\documentclass[%
reprint, twocolumn,
groupedaddress,
preprintnumbers,
bibnotes,
amsmath,amssymb,
aip,jcp, 
floatfix,
showkeys,
showpacs
]{revtex4-1}

\usepackage{import}
\usepackage{graphicx}
\usepackage{dcolumn}
\usepackage{bm}
\usepackage{hyperref}
\usepackage[table,dvipsnames,x11names]{xcolor}
\usepackage[normalem]{ulem}
\usepackage[utf8]{inputenc}
\usepackage[T1]{fontenc}

\usepackage{soul}
\usepackage{eufrak}

\usepackage{float}
\usepackage{tabularx,stackengine}

\usepackage{multirow}
\usepackage[capitalise]{cleveref}

\begin{document}
\renewcommand{\vec}[1]{\ensuremath{\mathbf{#1}}}
\newcommand{\mc}[3]{\multicolumn{#1}{#2}{#3}}

\definecolor{nice}{HTML}{C8D8EA}% light blue
\definecolor{alert}{HTML}{FCFF4B} % maximum yellow
\definecolor{morning}{HTML}{98AEAD} % morning blue
\definecolor{redish}{HTML}{FE7173} % 
\definecolor{greenish}{HTML}{50e25e} % 

\newcommand{\mim}{\ensuremath{\bigcirc\text{MI}\square}~}
\newcommand{\omiM}{\ensuremath{\bigcirc\text{MIM}}~}
\newcommand{\omiB}{\ensuremath{\bigcirc\text{MIB}}~}
\newcommand{\omiC}{\ensuremath{\bigcirc\text{MIC}}~}
\newcommand{\omiS}{\ensuremath{\bigcirc\text{MIS}}~}

%%%%%%%%%%%%%%%%%%%%%%%%%%%%%%%%%%%%%%%%%%%%%%%%%%%%%%%
%%%%%%%%%%%%%%%%%%%%%%%%%%%%%%%%%%%%%%%%%%%%%%%%%%%%%%%
%%%%%%%%%%%%%%%%%%%%%%%%%%%%%%%%%%%%%%%%%%%%%%%%%%%%%%%

\title{Inhibition of steel corrosion with imidazolium-based compounds - experimental and theoretical study}
\author{Dominik Legut}
\email[\textbf{Corresponding author:} ]{dominik.legut@vsb.cz}
\affiliation{IT4Innovations, V\v{S}B-Technical University of Ostrava, 17. listopadu 2172/15, 708 00 Ostrava-Poruba, Czech Republic}
\author{Andrzej Piotr Kądzielawa}
\affiliation{IT4Innovations, V\v{S}B-Technical University of Ostrava, 17. listopadu 2172/15, 708 00 Ostrava-Poruba, Czech Republic}
\affiliation{Instyut Fizyki Teoretycznej, Uniwersytet Jagielloński, ulica \L{}ojasiewicza 11, 30-348 Krak\'ow, Poland}
\author{Petr Pánek} 
\affiliation{Faculty of Materials Science and Technology, V\v{S}B-Technical University of Ostrava, 17. listopadu 2172/15, 708 00 Ostrava, Czech Republic}
\affiliation{Institute of Environmental Technology, V\v{S}B-Technical University of Ostrava, 17. listopadu 2172/15, 708 00 Ostrava, Czech Republic}
\author{Kristýna Marková}
\affiliation{Faculty of Materials Science and Technology, V\v{S}B-Technical University of Ostrava, 17. listopadu 2172/15, 708 00 Ostrava, Czech Republic}
\author{Petra Váňová}
\affiliation{Faculty of Materials Science and Technology, V\v{S}B-Technical University of Ostrava, 17. listopadu 2172/15, 708 00 Ostrava, Czech Republic}
\author{Kateřina Konečná}
\affiliation{Faculty of Materials Science and Technology, V\v{S}B-Technical University of Ostrava, 17. listopadu 2172/15, 708 00 Ostrava, Czech Republic}
\author{Šárka Langová}
\affiliation{Faculty of Materials Science and Technology, V\v{S}B-Technical University of Ostrava, 17. listopadu 2172/15, 708 00 Ostrava, Czech Republic}

\begin{abstract}
This work aims to investigate the corrosion inhibition of the mild steel in the 1 M HCl solution by 1-octyl-3-methylimidazolium hydrogen sulphate 1-butyl-3-methylimidazolium hydrogen sulphate, and 1-octyl-3-methylimidazolium chloride, using electrochemical, weight loss, and surface analysis methods as well as the full quantum-mechanical treatment. Polarization measurements prove that studied compounds are mixed-type inhibitors with a predominantly anodic reaction. The inhibition efficiency obtained from the polarization curves is about 80-92\% for all of the 1-octyl-3-methylimidazolium salts with a concentration higher than 0.005 mol/l, while it is much lower for 1-butyl-3-methylimidazolium hydrogen sulphate. The values measured in the weight loss experiments (after seven days) are to some extent higher (reaching up to 98\% efficiency). Furthermore, we have shown that the influence of the alkyl chain length on the inhibition efficiency is much larger than that of the anion type. %The adsorption process obeys the Langmuir isotherm. 
Furthermore, we obtain a realistic model of a single molecule on iron surface Fe(110) by applying the Density Functional Theory calculations. We use the state-of-the-art computational approach, including the meta-GGA strongly-constrained and appropriately normed semilocal density functional to model the electronic structure properties of both free and bounded-to-surface molecules of 1-butyl-, 1-hexyl-, and 1-octyl-3-methylimizadolium bromide, chloride, and hydrogen sulphate. From the calculations we extract, the HOMO/LUMO gap, hardness, electronegativity, and charge transfer of electrons from/to molecules-in-question. It supports the experimental findings and explains the influence of the alkyl chain length and the functional group on the inhibition process.
\end{abstract}

\date{\today}

\keywords{Imidazolium-based compounds; corrosion inhibition; polarization curve; electrochemical impedance spectroscopy; steel; density functional theory; electronic structure; ab-initio methods}
\pacs{}
\maketitle

%%%%%%%%%%%%%%%%%%%%%%%%%%%%%%%%%%%%%%%%%%
%%%%%╻┏┓╻╺┳╸┏━┓┏━┓╺┳┓╻ ╻┏━╸╺┳╸╻┏━┓┏┓╻%%%%%
%%%%%┃┃┗┫ ┃ ┣┳┛┃ ┃ ┃┃┃ ┃┃   ┃ ┃┃ ┃┃┗┫%%%%%
%%%%%╹╹ ╹ ╹ ╹┗╸┗━┛╺┻┛┗━┛┗━╸ ╹ ╹┗━┛╹ ╹%%%%%
%%%%%%%%%%%%%%%%%%%%%%%%%%%%%%%%%%%%%%%%%%

\section{Introduction}
\label{sec:introduction}

Many organic compounds are effective corrosion inhibitors. Recently, their subset, namely ionic liquids, has captivated the scientific community due to their remarkable properties – low vapour pressure, non-flammability, and thermal stability. They promote the formation of a chelate on a metal surface. Electrons are transferred between an organic compound containing nucleophile centres and metal. Atoms like oxygen, sulphur, and nitrogen, and unsaturated functional groups can share their free electron pairs.\cite{FANG2002179} The inhibition efficiency is higher if the organic compound can also accept the %free 
electrons from the surface of the metal.
One of the classes of such compounds is imidazolium derivatives that are widely studied as corrosion inhibitors. The inhibition efficiency usually increases with the increasing length of an alkyl chain.\cite{doi:10.1021/ie300977d,ZHANG201057} On the other hand, too large molecules can cause a steric hindrance and reduce the inhibition efficiency. Molecules with longer alkyl chains are usually more toxic and less decomposable, although these issues can be predicted employing quantum chemical calculations,\cite{SALAM2016393} a useful and economical tool to eliminate potentially problematic candidates before starting the experimental verification process.

Imidazolium and imidazolinium-based compounds are reported to exhibit corrosion resistance on aluminium,\cite{article15,ASHASSISORKHABI20064039} copper. \cite{ZHANG20043031,VASTAG2018526,QIANG201768,LI19991769,VERMA2017927,OTMACICCURKOVIC20092342,FATEH2020481,GODLEWSKA2013288}, and mild steel. \cite{OKAFOR2009850,GUO2017234,ZHANG20091881,DIAMANTI2016414,POURGHASEMIHANZA201696,GHANBARI20101205,CHONG2015219,VERMA2018100,LIKHANOVA20102088,DEYAB2017396,ALRASHED2019111015} Recently, other organic compounds were used as inhibitors on mild steels like Losartan potassium \cite{QIANG2021126863} and Ginkgo leaf extract \cite{QIANG20186} and tetrazole derivatives for the copper surface.\cite{QIANG202063}

Because of inadequacies of linear voltammetry and electrochemical impedance spectroscopy (EIS), other methods are also applied for the instantaneous determination of the corrosion rate. In recent years, the electrochemical frequency modulation technique has caught the attention of corrosion scientists as a rapid and non-destructive method allowing to determine the corrosion rate and polarization resistance without prior knowledge of the Tafel constants. A dual-frequency non-destructive potential wave perturbation around the corrosion potential is applied to corroding metal. This method was also applied to a corrosion study of the carbon steel using imidazolium-based compounds, and the results were compared with those obtained by other electrochemical methods.\cite{OBOT201883}

The inhibition mechanism is usually evaluated through the Gibbs energy of adsorption using various adsorption isotherms. The values of -20 kJ mol\textsuperscript{-1} or less negative indicate electrostatic interaction, and those being -40 kJ mol\textsuperscript{-1} or more negative indicate the chemisorption process. \cite{article37,ABDELREHIM2001268,QIANG2021126863} 

A strong dependence of the inhibition efficiency on the alkyl chain length was found in our previous work \cite{article39} on 1-alkyl-3-methylimidazolium bromides for the shorter chain length, breaking the trend for the largest 1-dodecyl-3-methylimidazolium bromide.\cite{article40} this study aims to find out whether the alkyl chain length or the functional group (Br, Cl, HSO$_4$) has a more significant effect on the inhibition process. The substitution of bromide by hydrogen sulphate could lead to the higher inhibition efficiency via the change of the highest occupied molecular orbital (HOMO) and charge distribution density according to our calculations, as well as previous modelling attempts.\cite{FANG2002179,KOWSARI201673,ZHENG2015168,SASIKUMAR2015105,YESUDASS2016252} These (quantum chemical) calculations are used to predict the interaction between the inhibitor and metal.\cite{KOWSARI201673,ZHENG2015168,SASIKUMAR2015105,YESUDASS2016252,ZHANG2005173,VERMA20191} The lower the value of the energy difference between HOMO and the lowest unoccupied molecular orbital (LUMO) with respect to the potential energy at infinity, the better the inhibition efficiency. Additionally, the quantities of ionization potential, electron affinity, global hardness, softness, and electronegativity are the key to determining the fraction of electrons transferred from donors (corroded substance) to acceptors (inhibitor), thus quantify the inhibition process.

\section{Details on calculations}
\label{sec:detailscalc}
\subsection{Density Functional Theory (DFT)}
We employ the Vienna ab-initio Software Package (VASP) for the computational part, \cite{VASP,VASP2,VASP3,VASP4} with the strongly constrained and appropriately normed semilocal density functional (SCAN)\cite{SCAN1,SCAN2} meta-GGA. Reciprocal grids of $7\times7\times1$, $7\times5\times1$, $6\times6\times1$, and $5\times5\times5$ for the iron surfaces Fe(100), Fe(110), Fe(111), and bulk bcc Fe are generated by the VASP-embedded routines (Mohnkorst-Pack), see Sec.~\ref{sssec:ironsurfaces}. For the calculations including the molecules, we utilize the $\Gamma$-point-only version of VASP. The exemplary Fe(110) surface in our calculations is represented by a trilayer of 252 iron atoms (three 24.35 by 20.09 \AA{} layers of 2.71 \AA{} thickness) and 30 \AA{} of vacuum. The design of each molecule is optimized in the box 24.35 $\times$ 20.09 $\times$ 38.12 \AA{} (for visualisation of HOMO and LUMO see \cref{app:visual}). To orient resultant structures to their principal axes, the Principal Component Analysis (PCA) \cite{PCA} is performed, and subsequently, the properties at consecutive distances from the surface are obtained (see Sec.~\ref{sssec:moleculessurface}).
To obtain the same point of reference for each resultant quantity, we determine the so-called \emph{workfunction} ($\Phi$), {\it, i.e.} the work required to transfer a trial charge to infinity as the value of the local potential of examined system at infinity (along the 'vacuum' axis). Shifting the energy scale by the value of $\Phi$ is equivalent to assuming that $\left. V_\text{local}(\vec{r}) \right|_{\|\vec{r}\| \rightarrow \infty} =  0$. In Sec.~,\ref{sssec:workfunction} we elaborate on our calculations as well as provide examples. The multiple linear regression taking into account the structural parameters\cite{doi:10.1002/zaac.201600230} with the cluster model focused on the geometry optimization \cite{article} is used here.

%%%%%%%%%%%%%%%%%%%%%%%%%%%%%%%%%%%%%%
%%%%%┏━╸╻ ╻┏━┓┏━╸┏━┓╻┏┳┓┏━╸┏┓╻╺┳╸%%%%%
%%%%%┣╸ ┏╋┛┣━┛┣╸ ┣┳┛┃┃┃┃┣╸ ┃┗┫ ┃ %%%%%
%%%%%┗━╸╹ ╹╹  ┗━╸╹┗╸╹╹ ╹┗━╸╹ ╹ ╹ %%%%%
%%%%%%%%%%%%%%%%%%%%%%%%%%%%%%%%%%%%%%
\section{Experimental}
\label{sec:experiment}
\subsection{Materials}
\label{ssec:materials}
A mild steel wire containing wt. 0.1\% C, 1.5\% Mn, 0.9\% Si, 0.03\% S, 0.04\% P was used for corrosion measurements. The specimens were ground using various grades of emery papers, which ended with the 2000 grade. They were cleaned with bidistilled water, degreased in acetone, and dried. The preparation and analysis of 1-octyl-3-methylimidazolium bromide as described in Ref. \cite{ABDELREHIM2001268} 1-octyl-3-methylimidazolium chloride was also prepared by microwave synthesis. 1-alkyl-3-methylimidazolium hydrogen sulphates were prepared by dissolving the corresponding 1-alkyl-3-methylimidazolium bromide in isopropyl alcohol. Pre-dried KHSO\textsubscript{4} was added to the solution, and the mixture was stirred at room temperature for 14 h. The precipitated KBr was removed by filtration, and the isopropyl alcohol was evaporated from the liquid phase under reduced pressure. The purity of the product was verified by determining residual bromides by the Mohr method (the purity was higher than 96\%). The yield of hydrogen sulphates was 80-85\%. The solutions were prepared by dilution of imidazolium salts in 1 M HCl. 

\subsection{Electrochemical measurements}
\label{ssec:measurements}
All the electrochemical measurements were carried out using Voltalab VM 40, Radiometer Analytical. A three-electrode configuration was used. A steel wire with 0.4 cm\textsuperscript{2} surface area functioned as the working electrode. A platinum coil and Ag/AgCl in 3M KCl were used as the auxiliary and reference electrodes. The experiments were carried out at temperature of 21$\pm 2^{\circ}$C without stirring. The working electrode was immersed into the tested solution for 45 minutes to attain a quasi-equilibrium state. The real equilibrium can hardly be reached. The potentiodynamic polarization curves were recorded in the range from -150 below to +150 mV above the open circuit potential (OCP) at a scan rate of 1 mV s\textsuperscript{-1} in the positive direction. The corrosion current density $i_{cor}$, corrosion potential $E_{cor}$ (mV), cathodic and anodic Tafel slopes $b_c$ and $b_a$ and polarization resistance $R_{pol}$ ($\Omega cm^2$) were determined. The inhibition efficiency $I E_{i_{cor}}$ (\%) was calculated from the following equation, where i\textsuperscript{0} and i\textsuperscript{i} are the corrosion current densities without and with the inhibitor:
\begin{align}
    I E_{i_{cor}} = \frac{i_{cor}^0 - i_{cor}^i}{i_{cor}^0} \times 100.
\end{align}
Similarly, the inhibition efficiency of the polarization resistance with ($R_{pol}$) and without inhibitor ($R_{pol}^0$)
calculated from the Tafel slopes 
\begin{align}
    R_{pol} \equiv \frac{b_a | b_b |}{\log(10) i_\text{cor} (b_a + |b_b|)}.
\end{align}
is determined simply by 
\begin{align}
     I E_{R_{pol}} = \frac{R_{pol} - R_{pol}^0}{R_{pol}} \times 100.
\end{align}
The  EIS measurements were carried out in a frequency range from 100 kHz to 100 mHz. A sine wave with 5 mV was used to perturb the system. An equivalent circuit with solution resistance $R_s$ in series with the parallel combination of the constant phase element (CPE) and polarization resistance $R_p$ was used. The polarization resistance includes all the metal/solution interface resistances - the charge transfer resistance, accumulation resistance, and diffusion layer resistance \cite{KOWSARI201673}. The inhibition efficiency was calculated from
\begin{align}
    IE_{EIS} = \frac{R_p - R_p^0}{R_p} \times 100,
\end{align}
where $R_p$ and $R_p^0$ are the polarization resistances of the solution with and without the inhibitor.

The use of CPE represents the more accurate fit in the case of deviation from an ideal capacitor due to different phenomena like surface roughness, inhibitor adsorption, porous layer formation, etc. The equation gives the impedance Z of CPE
\begin{align}
 \label{eq:ZCPE}
 Z_{CPE} = \left[ (\mathfrak{i} \omega )^n Y_0 \right]^{-1},    
\end{align}
where $\omega$ is the angular frequency (rad s\textsuperscript{-1}), and $\mathfrak{i}$ is an imaginary unit, the capacitance $Y_0$ is taken as the capacity of the so-called \emph{double layer capacitor}, and $n$ is an empirical exponent encompassing the imperfect nature of a physical system (unity in an ideal case) - including but not limited to the surface hardness.

\subsection{Weight loss measurements and surface analysis}
Gravimetric measurements were carried out with the wires prepared in the same way as the working electrode. The samples weighing about 0.22 g were immersed into the 1M HCl solution with or without the imidazolium salts for seven days, and the weight loss was determined. The inhibition efficiency was calculated from
\begin{align}
IE_{WL}   = \frac{WL^0-WL^i}{WL^0} \times 100
\end{align}
where $WL^0$ is the weight loss in 1M HCl solution, and WL$^i$ is the weight loss in the presence of the inhibitor.

The surface analysis was carried out using the scanning QUANTA 450 FEG EDX electron microscope, equipped with the EDX analyser. The snaps were acquired in the secondary electrons mode. 

%%%%%%%%%%%%%%%%%%%%%%%%%%%%%%%
%%%%%┏━┓┏━╸┏━┓╻ ╻╻  ╺┳╸┏━┓%%%%%
%%%%%┣┳┛┣╸ ┗━┓┃ ┃┃   ┃ ┗━┓%%%%%
%%%%%╹┗╸┗━╸┗━┛┗━┛┗━╸ ╹ ┗━┛%%%%%
%%%%%%%%%%%%%%%%%%%%%%%%%%%%%%%
\section{Results and discussion}
\label{sec:results}
\begin{figure}[t]
    \centering
\includegraphics[width=0.500\linewidth]{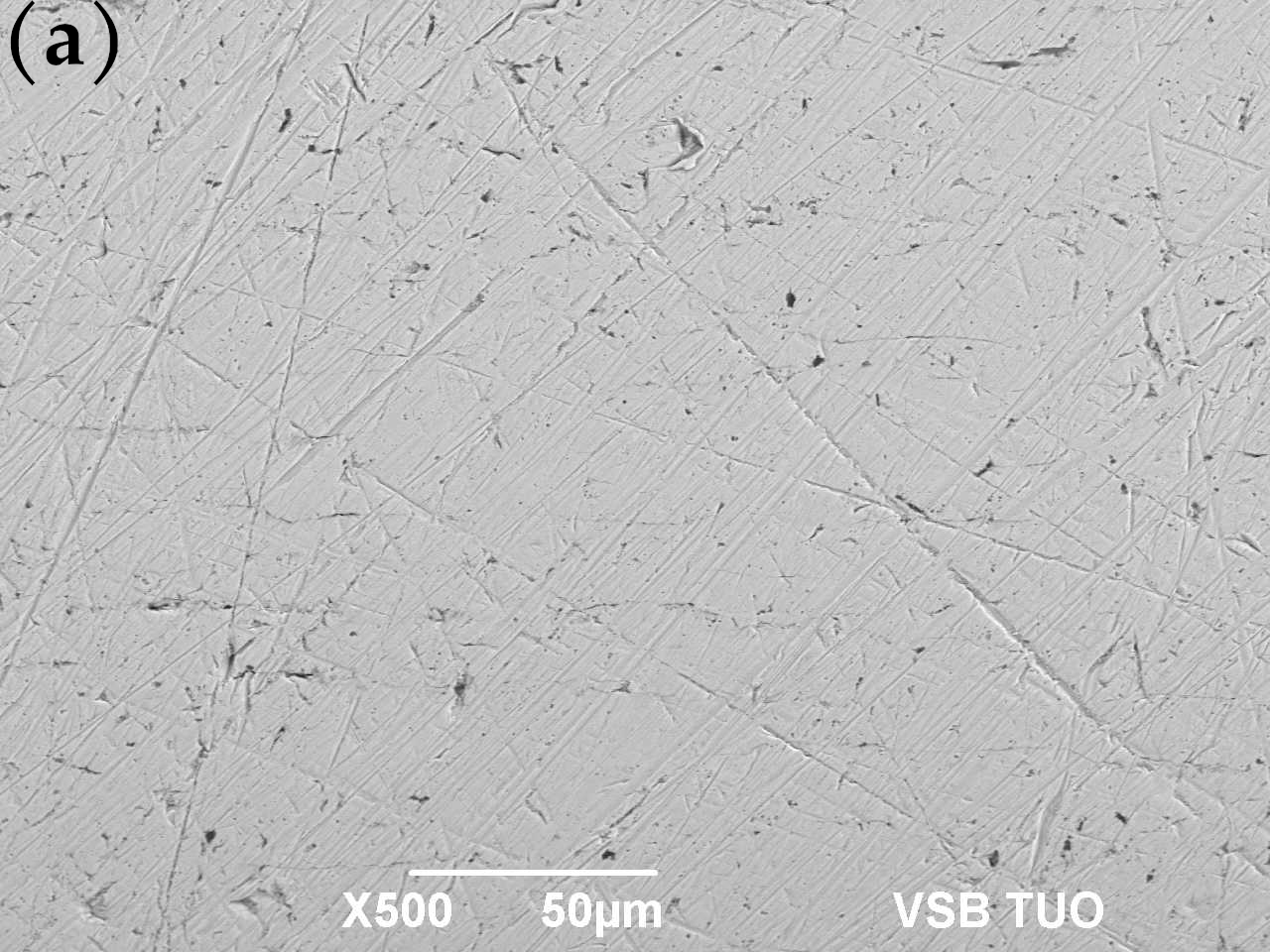}\includegraphics[width=0.500\linewidth]{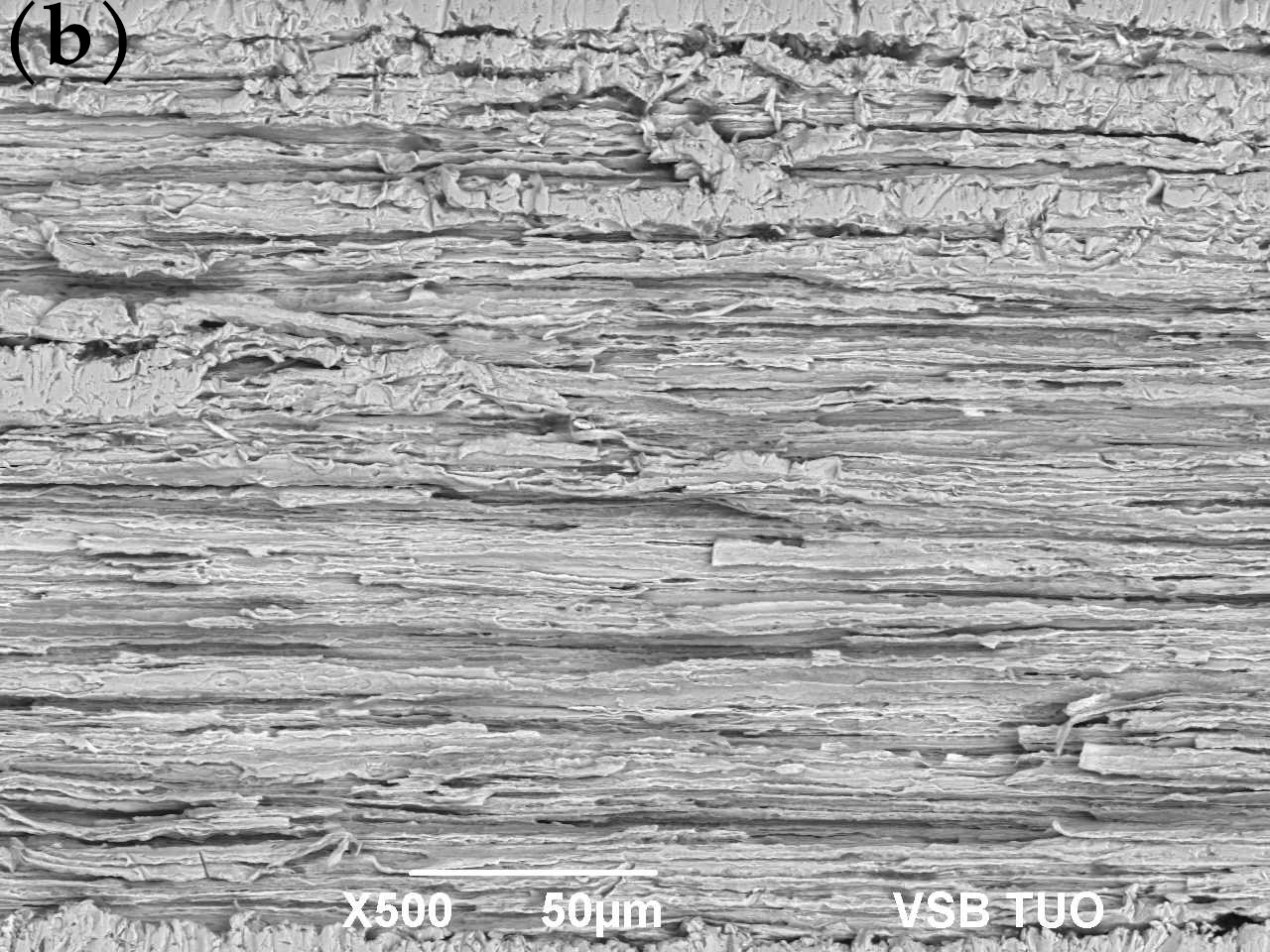}\vspace{-1pt}
\includegraphics[width=0.500\linewidth]{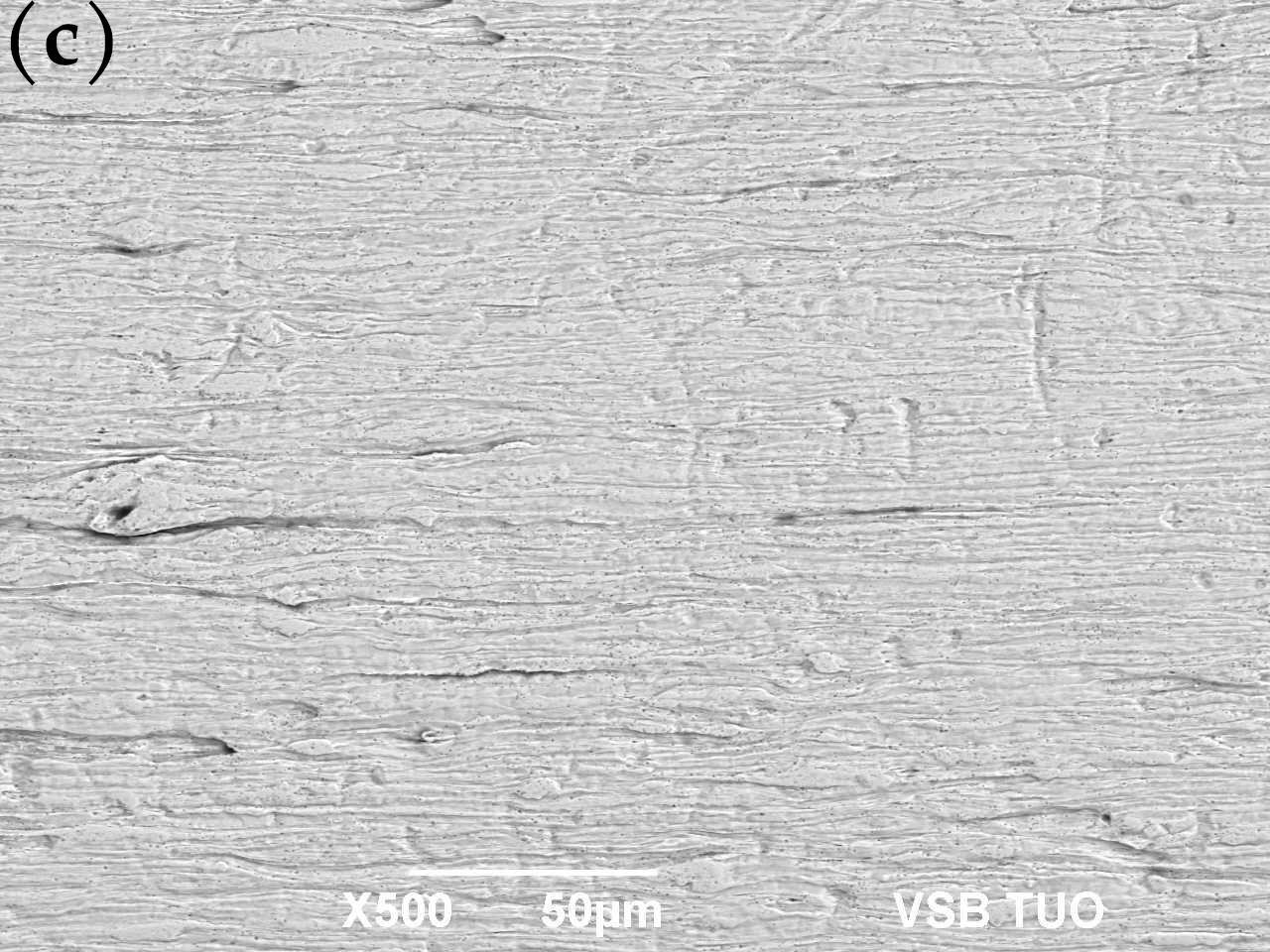}\includegraphics[width=0.500\linewidth]{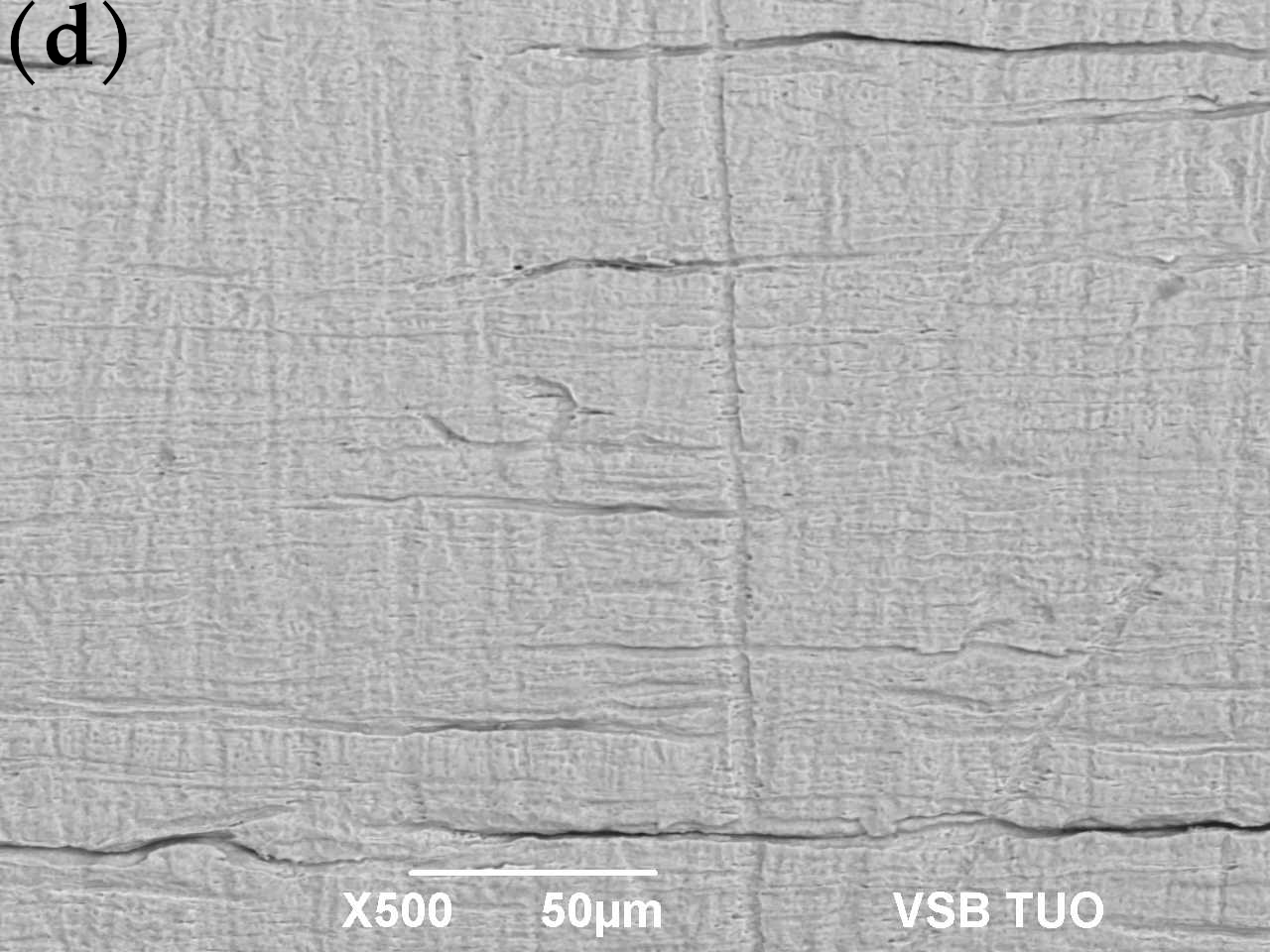}
    \caption{The surfaces of the wires used in the experiment obtained via the scanning electron microscope (SEM); the cases of the original mild steel wire (a) and wire after 7-day immersion in: 1M HCl solution (b), inhibited by 0.001M 1-octyl-3-methylimidazolium hydrogen sulphate solution (c), and inhibited by 0.001M 1-octyl-3-methylimidazolium chloride solution (d). }
    \label{fig:surfaces}
\end{figure}
\subsection{Potentiodynamic polarization curves and electrochemical impedance spectroscopy}
\label{ssec:spectroscopy}

\begin{figure*}[t]
    \centering
    \includegraphics[width=\linewidth]{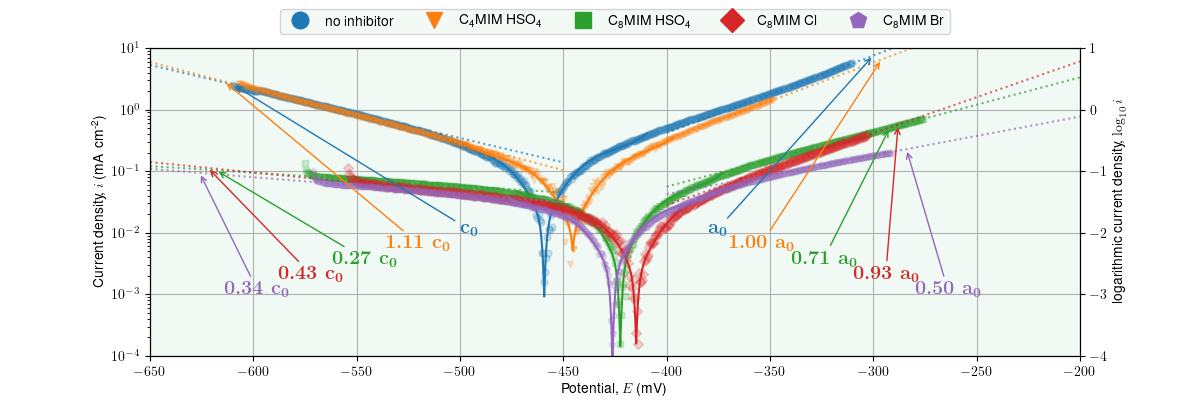}
    \caption{Potentiodynamic polarization curves of corrosion inhibition of mild steel in 1 M HCl in the absence and presence of 0.001M 1-octyl-3-methylimidazolium hydrogen sulphate, 1-octyl-3-methylimidazolium chloride, 1-octyl-3-methylimidazolium bromide, and 1-butyl-3-methylimidazolium hydrogen sulphate.
    The dotted lines represent the linear fit to the tail of the logarithmic current densities $\log i$. The numbers correspond to the ratio of the anodic ($a$) and cathodic ($c$) tails with respect to the values for a solution with no inhibitor (blue curve, $a_0$ and $c_0$).}
    \label{fig:expFig1}
\end{figure*}
\begin{figure*}[t]
    \centering
    \includegraphics[width=\linewidth]{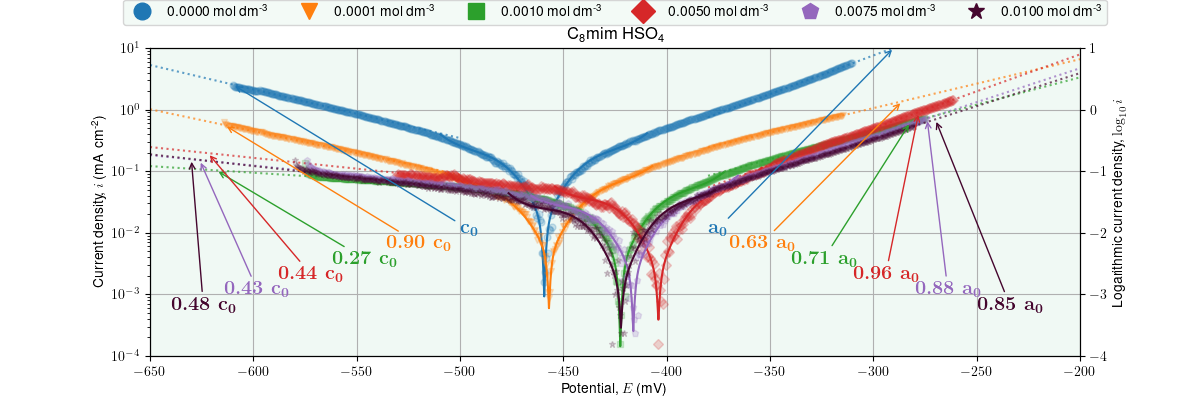}
    \caption{Potentiodynamic polarization curves of corrosion inhibition of mild steel in 1M HCl in the absence and presence of 1-octyl-3-methylimidazolium hydrogen sulphate. The numbers correspond to the ratio of the anodic ($a$) and cathodic ($c$) tails with respect to the values for a solution with no inhibitor (blue curve, $a_0$ and $c_0$). Solid lines are a guide to the eye.}
     
    \label{fig:expFig2}
\end{figure*}
\begin{figure*}[t]
    \centering
    \includegraphics[width=\linewidth]{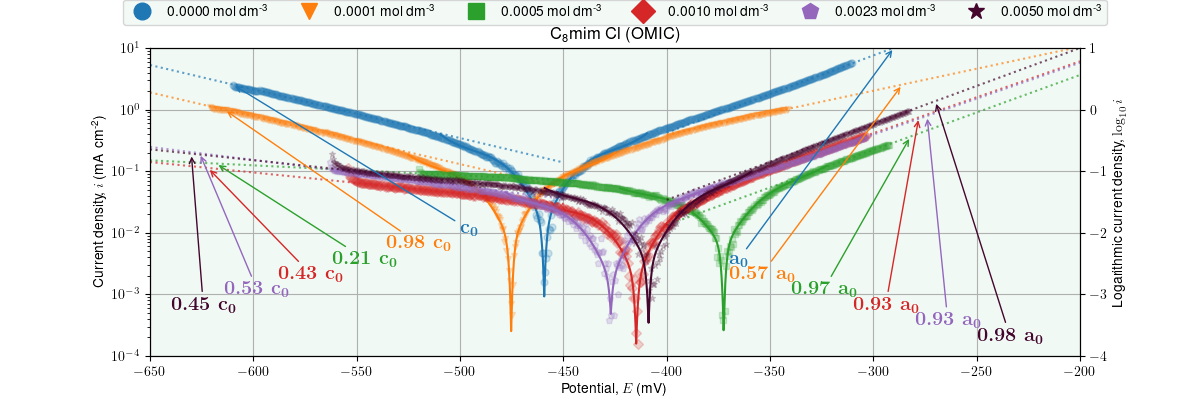}
    \caption{ The same as in Fig.\ref{fig:expFig2} but for C$_8$MImCl. Solid lines are a guide to the eye.
    }
   
    \label{fig:expFig3}
\end{figure*}
\begin{table*}[t] 
    \centering
    \label{tab:EIS}
    \caption{Electrochemical polarization parameters for mild steel in 1 M HCl solution in the absence and presence of 1-octyl-3-methylimidazolium chloride, 1-octyl-3-methylimidazolium hydrogen sulphate and 1-butyl-3-methylimidazolium hydrogen sulphate at 20$^\circ$C.}
    \resizebox{\linewidth}{!}{
    \begin{tabular}{l|rdd|d|dd|dd}
  &  \mc{1}{c}{$c$  ($mol/l$)  } &  \mc{1}{c}{$b_a$ ($mV/dec$) } &  \mc{1}{c|}{$-b_c$ ($mV/dec$) } &  \mc{1}{c|}{$E_{cor}$ ($mV$) } & \mc{1}{c}{i$_{cor}$ ($mA/cm^2$)} & \mc{1}{c|}{$IE_{i_{cor}}$ (\%)} &  \mc{1}{c}{$R_{pol}$  ($\Omega cm^2$) } &  \mc{1}{c}{  $IE_{R_{pol}}$ (\%) }  \\\hline
                               &$\varnothing$                &  80  &  126  &  -443  & 122.0 &     &  174 &    \\\hline
C$_8$MImHSO\textsubscript{4}   &  0.1 10\textsuperscript{-3} & 126  &  141  &  -466  &  50.3 &  59 &  574 & 70 \\
                               &  1.0 10\textsuperscript{-3} & 113  &  465  &  -418  &  38.5 &  68 & 1024 & 83 \\
                               &  5.0 10\textsuperscript{-3} &  83  &  285  &  -399  &  32.4 &  73 &  865 & 80 \\
                               &  7.5 10\textsuperscript{-3} &  91  &  296  &  -403  &  26.8 &  78 & 1125 & 84 \\
                               & 10.0 10\textsuperscript{-3} &  93  &  263  &  -408  &  22.8 &  81 & 1314 & 87 \\\hline
C$_8$MImCl                     &  0.1 10\textsuperscript{-3} & 140  &  129  &  -484  &  98.1 &  20 &  297 & 41 \\
                               &  1.0 10\textsuperscript{-3} &  86  &  295  &  -410  &  21.6 &  82 & 1336 & 87 \\
                               &  2.3 10\textsuperscript{-3} &  86  &  238  &  -405  &  23.1 &  81 & 1183 & 85 \\
                               &  5.0 10\textsuperscript{-3} &  81  &  284  &  -404  &  30.7 &  75 &  892 & 80 \\\hline
C$_4$MImHSO$_4$                &  1.0 10\textsuperscript{-3} &  80  &  114  &  -444  &  91.8 &  25 &  222 & 22     
    \end{tabular}
    }
    \label{tab:polarization}
\end{table*}
\begin{figure*}[t]
    \centering
    \includegraphics[width=0.49\linewidth,trim={2em 0.8em 1em 2em}, clip]{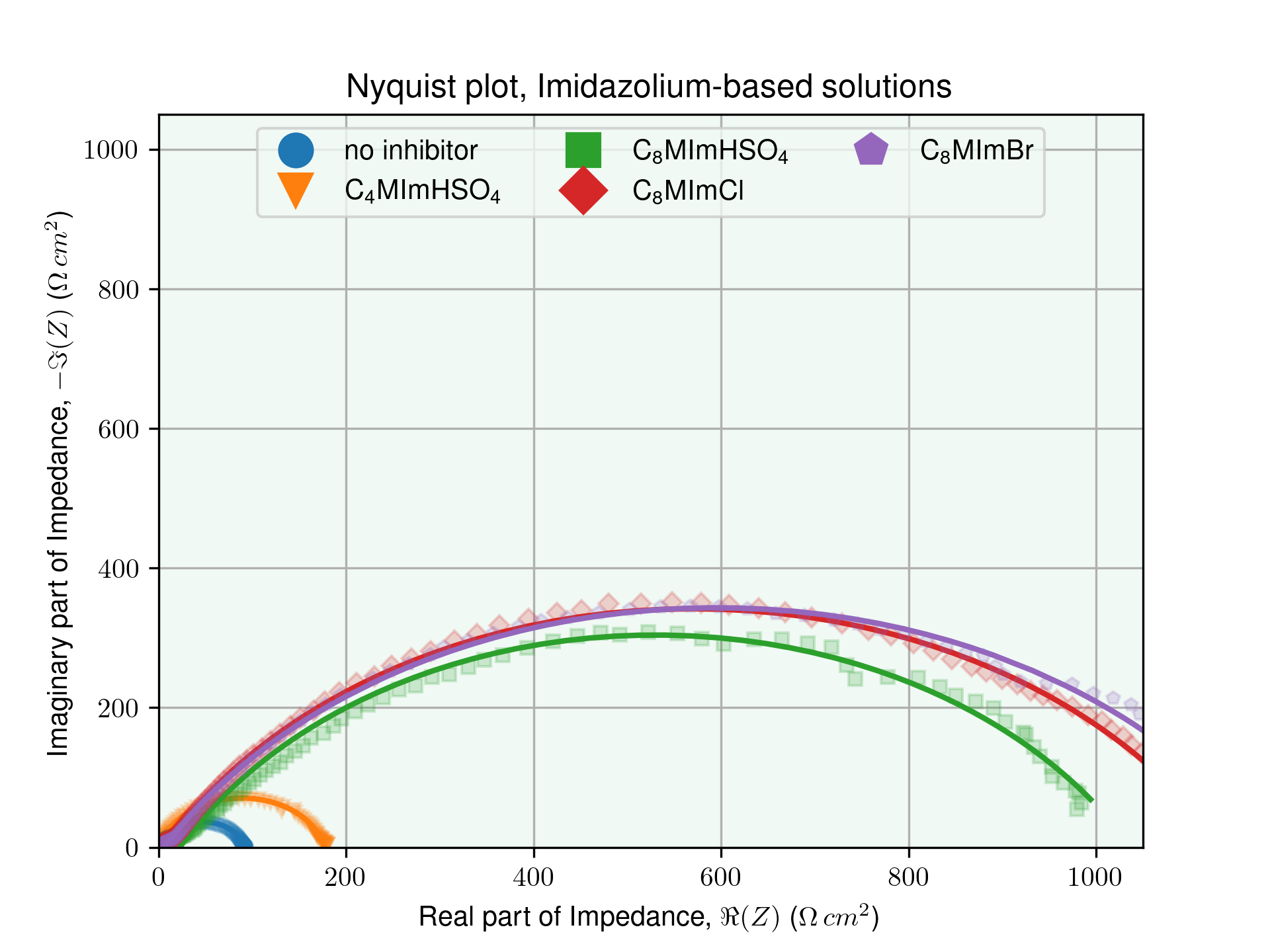}
    \includegraphics[width=0.49\linewidth,trim={2em 0.8em 1em 2em}, clip]{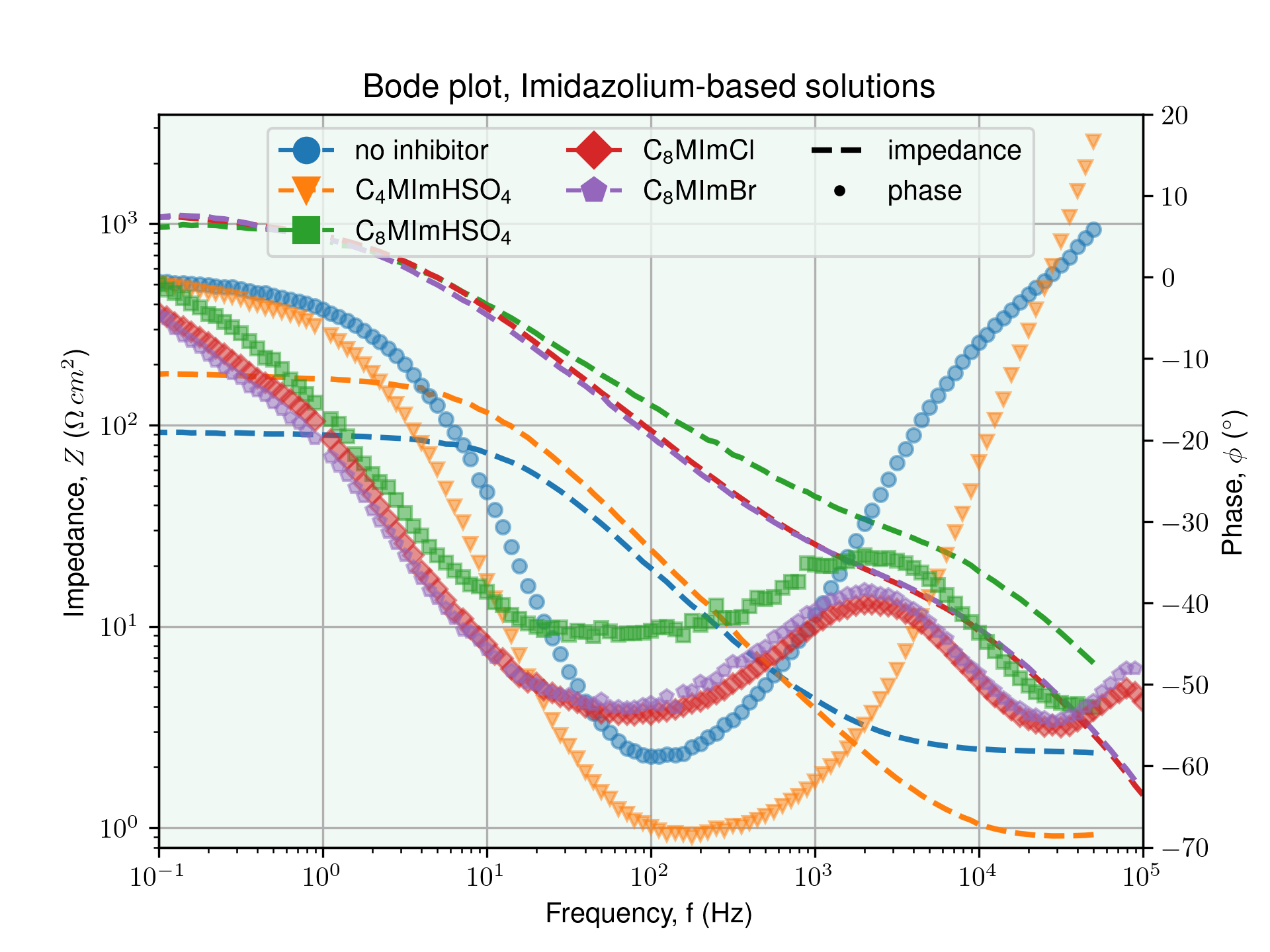}
    \caption{Nyquist (left) and Bode (right) plots of corrosion inhibition of mild steel in 1 M HCl in the absence and presence of different imidazolium-based compounds. Note the qualitative changes in the Bode plot: From a simple sharp minimum of the phase for the absence of inhibitor (blue dots) to either broadened valley (for C\textsubscript{8}MImHSO\textsubscript{4}) or two distinctive minima (\emph{e.g.}, C\textsubscript{8}MImBr). For details, see the main text.}
    \label{fig:impedanceBodeOthr}
\end{figure*}
\begin{figure*}[t]
    \centering
    \includegraphics[width=0.49\linewidth,trim={2em 0.8em 1em 2em}, clip]{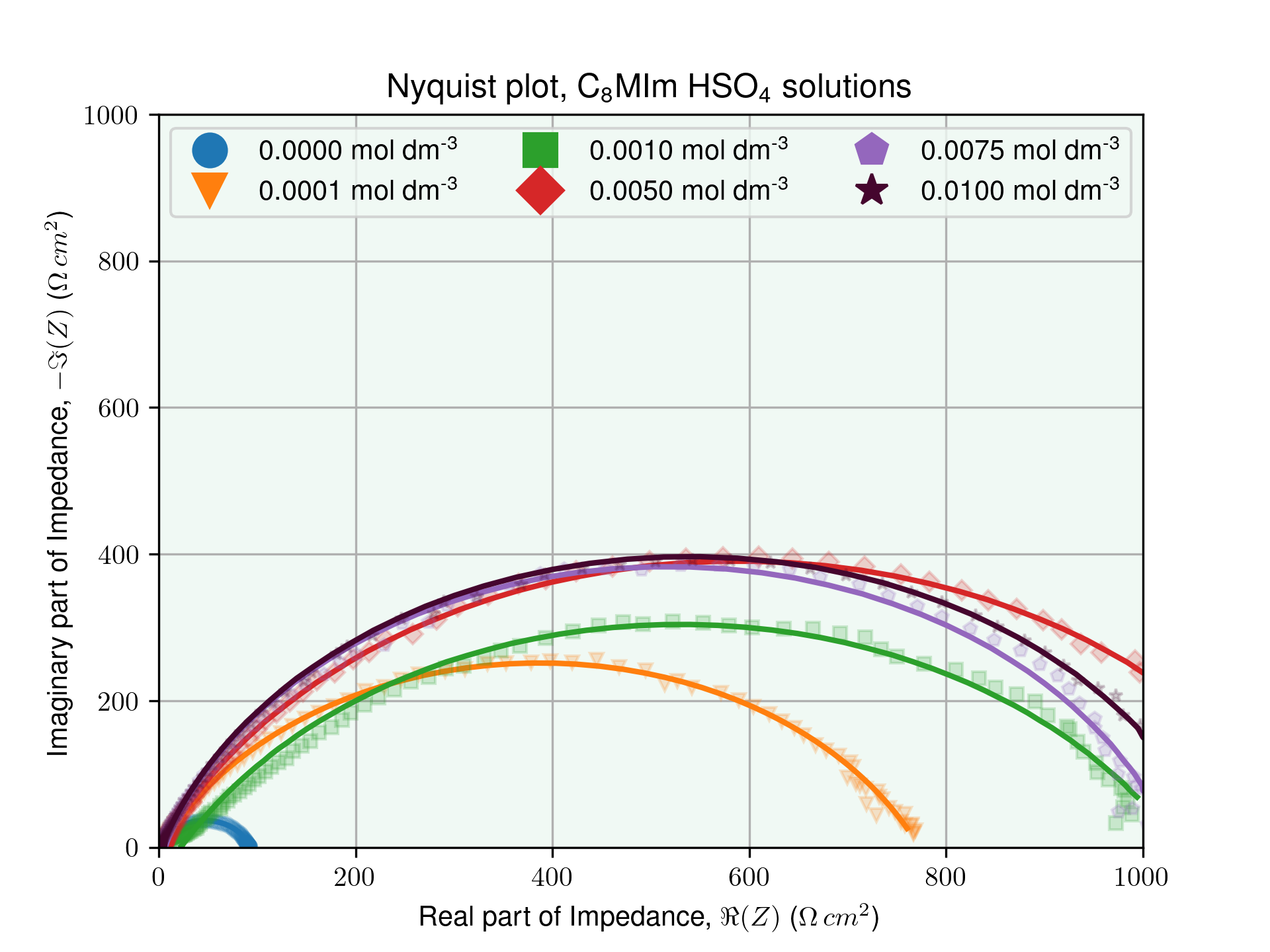}
    \includegraphics[width=0.49\linewidth,trim={2em 0.8em 1em 2em}, clip]{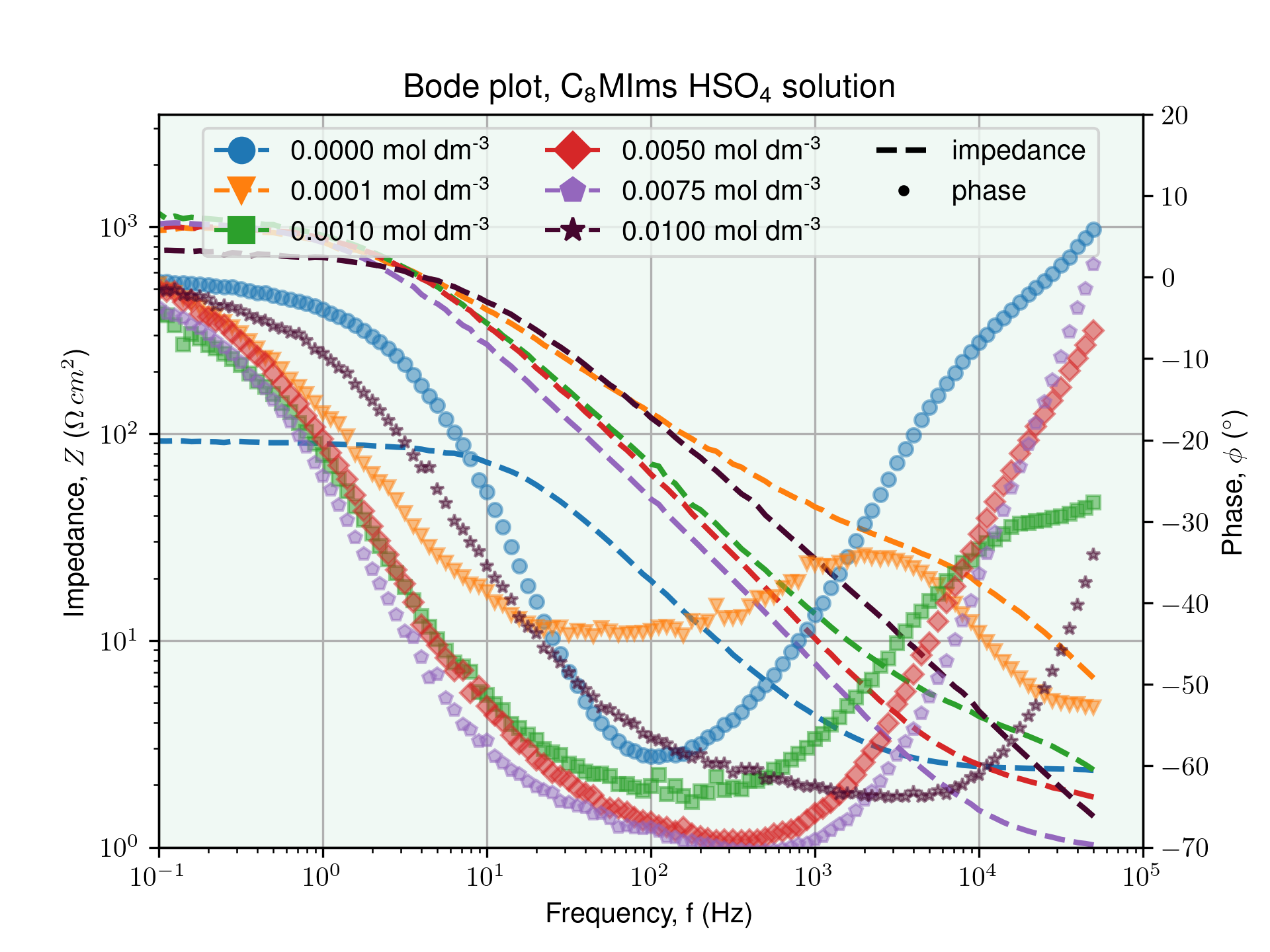}
    \caption{Nyquist (left) and Bode (right) plots of corrosion inhibition of mild steel in 1 M HCl in the absence and presence of 1-octyl-3-methylimidazolium hydrogen sulphate. For details, see the main text. }
    \label{fig:impedanceBodeOMIS}
\end{figure*}
\begin{figure*}[t]
    \centering
    \includegraphics[width=0.49\linewidth,trim={2em 0.8em 1em 2em}, clip]{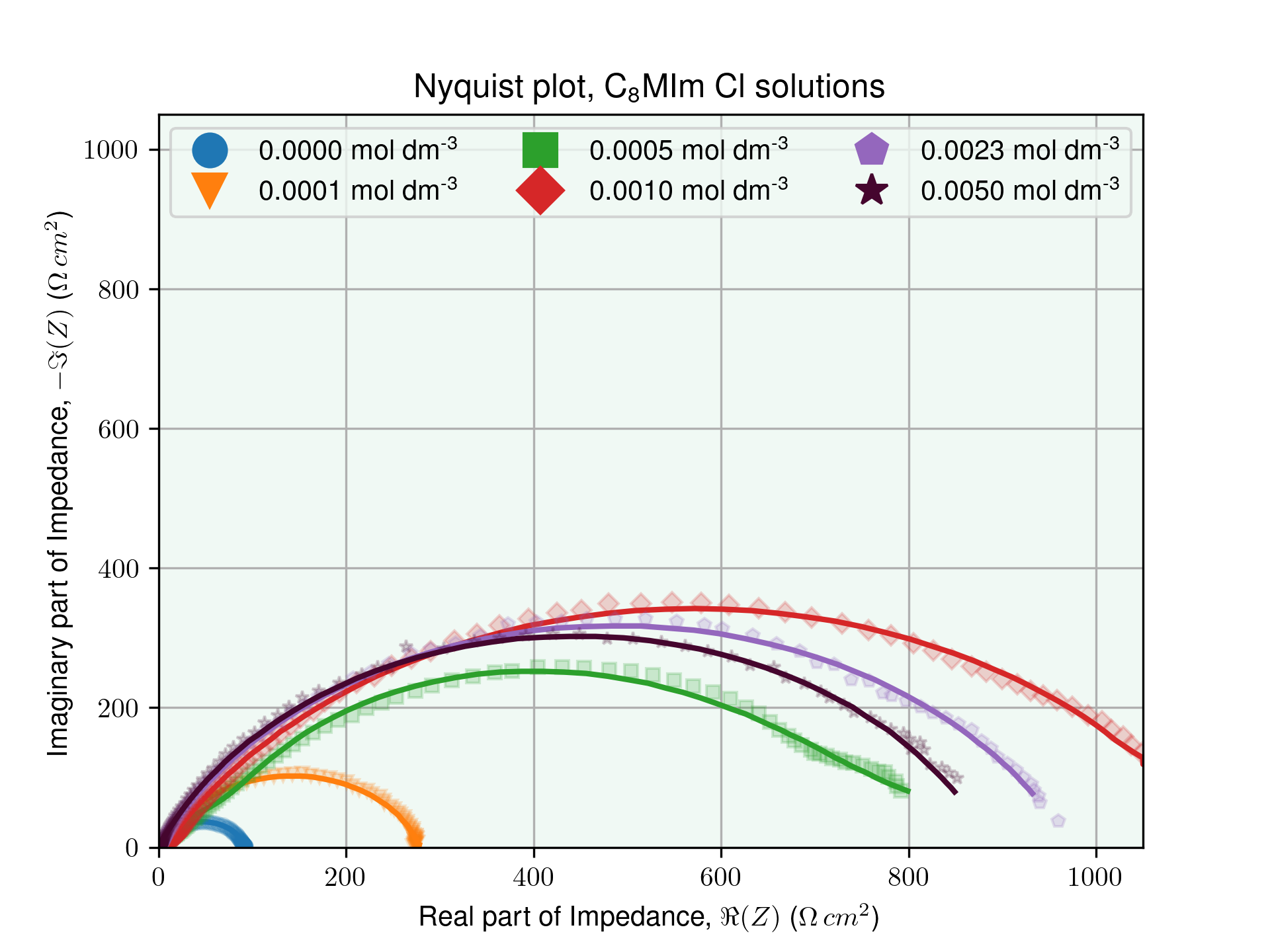}
    \includegraphics[width=0.49\linewidth,trim={2em 0.8em 1em 2em}, clip]{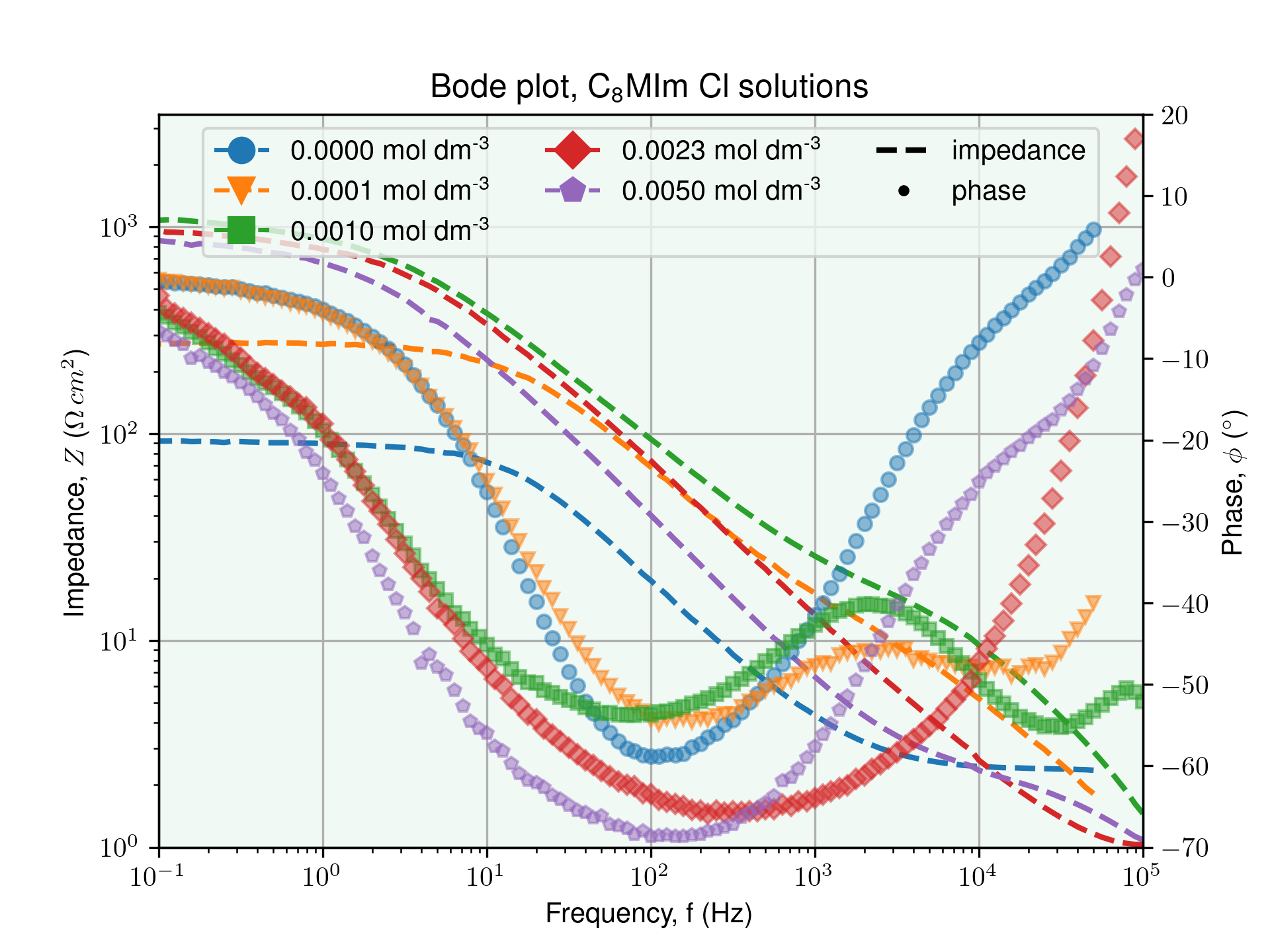}
    \caption{Nyquist (left) and Bode (right) plots of corrosion inhibition of mild steel in 1 M HCl in the absence and presence of 1-octyl-3-methylimidazolium chloride. For details, see the main text.} 
    \label{fig:impedanceBodeOMIC}
\end{figure*}
\begin{table*}[]
    \centering   
    \caption{Parameters of the Electrochemical Impedance Spectroscopy (EIS), namely, the polarization resistance (R\textsubscript{p}), capacity ($c_{dl}$), impedance-frequency exponent (n, \emph{cf.} \eqref{eq:ZCPE}), and the inhibition efficiency (IE\textsubscript{EIS}), concatenated for the case of mild steel in 1 M HCl solution both in the absence and presence of 1-octyl-3-methylimidazolium hydrogen sulphate, 1-octyl-3-methylimidazolium chloride, and 1-butyl-3-methylimidazolium hydrogen sulphate at 20$^\circ$C.}
    \label{tab:impedance}
    \begin{tabular}{l|rdddd}
             &  \mc{1}{c}{ c $(mol/l$)  } &  \mc{1}{c}{ R\textsubscript{p} ($\Omega cm^2$)  } &  \mc{1}{c}{ $C_{dl}$ ($\mu F cm^{-2}$)  } &  \mc{1}{c}{ $n$ } &  \mc{1}{c}{ IE\textsubscript{EIS} (\%) } \\\hline
            &  $\varnothing$              &    89  &  179  &  0.88  &   \\\hline
C\textsubscript{8}MImHSO\textsubscript{4}   &  0.1 10\textsuperscript{-3} &   771  &   69  &  0.74  &  88 \\
                                            &  1.0 10\textsuperscript{-3} &  1015  &  113  &  0.78  &  91 \\
                                            &  5.0 10\textsuperscript{-3} &  1173  &  113  &  0.75  &  92 \\
                                            &  7.5 10\textsuperscript{-3} &  1028  &   89  &  0.82  &  91 \\
                                            & 10.0 10\textsuperscript{-3} &  1072  &  116  &  0.81  &  92 \\\hline
C\textsubscript{8}MImCl                     &  0.1 10\textsuperscript{-3} &   273  &   70  &  0.82  &  67 \\
                                            &  1.0 10\textsuperscript{-3} &  1130  &  121  &  0.69  &  92 \\
                                            &  2.3 10\textsuperscript{-3} &   974  &  115  &  0.74  &  91 \\
                                            &  5.0 10\textsuperscript{-3} &   889  &  167  &  0.76  &  90 \\\hline
C\textsubscript{4}MImHSO\textsubscript{4}   &  1.0 10\textsuperscript{-3} &   176  &  154  &  0.86  &  49 \\
    \end{tabular}
\end{table*}
\begin{table}[]
    \centering   
    \caption{The Inhibition efficiency ($IE_{WL}$) of C\textsubscript{8}MImHSO\textsubscript{4} and C\textsubscript{8}MImCl after 7 days, from the weight loss (WL) measurements for different  concentrations of respective inhibitors.}
    \label{tab:weightloss}
    \begin{tabular}{l|rdd}
            &  \mc{1}{c}{ $c$ ($mol/l$)  } &  \mc{1}{c}{ $WL$ (\%)  } &  \mc{1}{c}{ $IE_{WL}$ (\%)} \\\hline\hline
           & \mc{1}{c}{$\varnothing$} & 54.8 & \\\hline
C\textsubscript{8}MImHSO\textsubscript{4} & 0.1 10\textsuperscript{-3}  & 18.2   &  67\\
           &  1.0 10\textsuperscript{-3}  &  1.4   &  97\\
           &  5.0 10\textsuperscript{-3}  &  1.0   &  98\\
           & 10.0 10\textsuperscript{-3}  &  1.3   &  98\\\hline
C\textsubscript{8}MImCl    & 0.1 10\textsuperscript{-3}  & 11.7   &  80\\
           &  1.0 10\textsuperscript{-3}  &  2.1  &  96\\      
           &  2.3 10\textsuperscript{-3}  &  1.2  &  98\\      
           &  5.0 10\textsuperscript{-3}  &  3.5  &  94\\      
    \end{tabular}
\end{table}

 Fig.~\ref{fig:expFig1} presents potentiodynamic polarization curves of corrosion inhibition of mild steel in 1M HCl in the absence and presence of 0.001M 1-octyl-3-methylimidazolium hydrogen sulphate (C$_8$MImHSO\textsubscript{4}), 1-octyl-3-methylimidazolium chloride (C$_8$MImCl), 1-octyl-3-methylimidazolium bromide (C$_8$MImBr), 1-butyl-3-methylimidazolium hydrogen sulphate (C$_4$MImHSO\textsubscript{4}),  
and 1-dodecyl-3-methylimidazolium bromide (C\textsubscript{12}MImBr) 
obtained in the previous study \cite{ABDELREHIM2001268}. The cathodic and anodic current densities in the presence of C$_8$MImHSO\textsubscript{4}, C$_8$MImCl, and C$_8$MImBr are much lower than C$_4$MImHSO\textsubscript{4} being just very close to the pure 1M HCl solution indicating that C$_4$MImHSO\textsubscript{4} is a rather limited inhibitor. The lowest current densities can be seen for C$_8$MImBr, but the differences are not significant. It seems that the inhibition efficiencies for all the salts with eight carbons chains are similar. The results are summarized in Tab.~\ref{tab:EIS} and indicate that the length of the alkyl chain plays a more significant role than the type of anion. These findings are also supported by quantum-mechanical modelling; see below. \\
Figs.~\ref{fig:expFig2} and~\ref{fig:expFig3} show the potentiodynamic polarization curves for various concentrations of C$_8$MImHSO\textsubscript{4} and C$_8$MImCl. The marked differences of current densities can be seen only between 0.0001 and 0.001 M solutions of both salts. The corrosion potentials are shifted to the positive direction compared with the HCl solution except for 0.0001 M C$_8$MImCl. The values of $i_{cor}$ differ less than 85 mV from the blank; hence both salts exhibit the traits of mixed-type inhibitors with a prevailing anodic effect \cite{QIANG20186,FERREIRA2004129,SASIKUMAR2015105,YAN20085953}. The presence of inhibitors causes the anodic and cathodic current density to fall. 
The electrochemical polarization parameters and the inhibition efficiency calculated from the corrosion current density are reported in Table~\ref{tab:polarization}. It is seen in Figs. \ref{fig:expFig2} and \ref{fig:expFig3} that the Tafel slopes are not totally parallel for both anodic and cathodic reactions for different inhibitor concentrations, hence a possible modification of the inhibiting mechanism might occur. The inhibition efficiencies are very similar for both salts with the same concentration. The highest concentration was not used for chloride because of the similar values of the corrosion current densities for 0.001 and 0.005 M solutions. The differences between C$_8$MImHSO\textsubscript{4} and C$_8$MImCl can be attributed to the various purity or subjectivity of the evaluation rather than to the various properties of the compounds. Only one concentration was applied for C$_4$MImHSO\textsubscript{4} since the inhibition efficiency was low.

The EIS measurements were evaluated using the EIS analyser software. In Figs~\ref{fig:impedanceBodeOthr}, \ref{fig:impedanceBodeOMIS}, and~\ref{fig:impedanceBodeOMIC} on the left, we present Nyquist plots of corrosion inhibition of mild steel in 1 M HCl in the absence and presence of 1-octyl-3-methylimidazolium hydrogen sulphate and 1-octyl-3-methylimidazolium chloride. The lines present the fitted spectrum, and the points indicate the experimental values. For each inhibitor, the spectra exhibit one depressed capacitive loop while Bode plots (right side of Figs~\ref{fig:impedanceBodeOthr}, \ref{fig:impedanceBodeOMIS}, and~\ref{fig:impedanceBodeOMIC}) reveal a new phase angle shift that indicates there are two-time constants in some cases (\emph{e.g.}, C\textsubscript{8}MIm HSO\textsubscript{4} with concentration $0.0001$ mol/l). Capacitive loops probably coincide, and the protective layer formation cannot be seen in Nyquist plots. The values obtained from EIS measurement are presented in Table~\ref{tab:impedance}. The solution resistances are not reported in thisTab.because of the considerable calculation error. The values were less than 2 $\Omega \, cm^2$. Except for the 0.0001 M solutions, the inhibition efficiencies of C\textsubscript{8}MIm HSO\textsubscript{4} and C\textsubscript{8}MImCl are almost independent on the inhibitor concentration and reach the values of about 90\%. The differences between the voltammetric and EIS measurements can be explained by the unequal time of measurement, and the OCP value can slightly change with time. As to C$_4$MImHSO$_4$, additional experiments would be necessary, but they were not carried out because of the low inhibition efficiency.

\subsection{Weight loss measurements and surface analysis}
The results of the weight loss measurements are shown in Table~\ref{tab:weightloss}. The experiments were not carried out with C$_4$MImHSO$_4$ because of the low inhibition efficiency found by the electrochemical measurements.
The inhibition efficiency slightly increased after seven days of immersion into the solutions. 
Figure~\ref{fig:surfaces} shows the mild steel surface after 7 days immersion into 1M HCl solution and into 0.001 M 1-octyl-3-methylimidazolium hydrogen sulphate and 0.001 M 1-octyl-3-methylimidazolium chloride solution. The surface in the presence of both inhibitors is much less damaged than in 1M HCl solution. 

%%%%%%%%%%%%%%%%%%%%%%%%%%%%%%%%%%%%%%%%%%%%%%%%%%%%%
%%%%%╺┳╸┏━╸┏━┓┏━┓┏━╸╺┳╸╻┏━╸┏━┓╻     ┏━┓┏━┓┏━┓╺┳╸%%%%%
%%%%% ┃ ┣╸ ┃ ┃┣┳┛┣╸  ┃ ┃┃  ┣━┫┃     ┣━┛┣━┫┣┳┛ ┃ %%%%%
%%%%% ╹ ┗━╸┗━┛╹┗╸┗━╸ ╹ ╹┗━╸╹ ╹┗━╸   ╹  ╹ ╹╹┗╸ ╹ %%%%%
%%%%%%%%%%%%%%%%%%%%%%%%%%%%%%%%%%%%%%%%%%%%%%%%%%%%%
\subsection{Theoretical modelling}
\label{ssec:modeling}
We performed quantum-mechanical calculations to elucidate the inhibition reaction and quantify it by several descriptors, starting from the clean iron surface and adsorbed molecules. For the naming conventions, we refer the reader to Appendix~\ref{app:naming}.

\subsubsection{iron surfaces}
\label{sssec:ironsurfaces}
In order to investigate electronic structure properties of various adsorbed molecules we start with modeling of the bcc iron (space group Im-3m, $a=b=c=2.87$ \AA{}, $\alpha=\beta=\gamma=90^\circ$), then calculate its electronic properties in: \emph{(i)} the bulk, \emph{(ii)} the Fe(100) surface, \emph{(iii)} the Fe(110) surface, and \emph{(iv)} the Fe(111) surface.

The most stable iron surface is the one with the lowest surface energy $\gamma$ \cite{Israelachvili2011}
\begin{align}
    \gamma = \frac{E_\text{slab} - N \cdot E_\text{bulk}}{2A},
\end{align}
where $E_\text{slab}$ is the energy of a slab of $N$ atoms with area $A$ exposed to vacuum, and $E_\text{bulk}$ is the energy (per atom) of the bcc iron. As illustrated in Table~\ref{tab:surface_energy}, the surface energy is minimized for Fe(110), which is consistent both with the experimental data \cite{tyson} and the precedent calculations \cite{sahputra}.
\begin{table}
    \centering
    \caption{The surface energy $\gamma$ for slabs of $N$ atoms and energy $E_G$, with the area $A$ exposed to vacuum.}
    \begin{tabular}{c|rdddd}
     \mc{2}{r}{$N$} &  \mc{1}{c}{\,\,$\tfrac{E_G}{N}$ (eV)\,\,} &  \mc{1}{c}{\,\, $A$ (\AA{}\textsuperscript{2})\,\,} &  \mc{1}{c}{\,\, $\gamma$ (eV \AA{}\textsuperscript{-2})\,\,} &  \mc{1}{c}{ \,\, $\gamma$ (J m\textsuperscript{-2})\,\,}\\\hline\hline
bulk&54&-8.34069&&& \\
Fe(100)&108&-8.12508&74.132&0.157&2.52 \\
Fe(111)&108&-7.92548&128.401&0.175&2.80 \\
Fe(110)&72&-8.04459&69.893&0.153&2.44 \\\hline
\mc{5}{c}{W.R. Tyson and W.A. Miller \cite{tyson}} & 2.41 \\
\mc{5}{c}{I. H. Sahputra \textit{et al.} \cite{sahputra}} & 2.44
    \end{tabular}
    \label{tab:surface_energy}
\end{table}

\subsubsection{Free molecules}
\label{sssec:homolumogap}
As a starting point of our modelling, we optimize the free molecules and obtain their chemical properties, i.e., the energy of the Highest Occupied Molecular Orbital (E\textsubscript{HOMO}) and the Lowest Unoccupied Molecular Orbital (E\textsubscript{LUMO}), the hardness
\begin{align}
    \eta \equiv \frac{1}{2} \left( - E_\text{HOMO} + E_\text{LUMO} \right),
\end{align}
and electronegativity
\begin{align}
    \chi \equiv -\frac{1}{2} \left( E_\text{HOMO} + E_\text{LUMO} \right).
\end{align}
The quantitative results are summarized in Tab.~\ref{tab:HOMOLUMO}. The groups --Br, --Cl, and --HSO\textsubscript{4} change the electronegativity by ~300-400\% and hardness $\eta$ only by 50-100\% compared to the molecules without any of the aforementioned group. The effect of the length of the alkyl chain (C\textsubscript{n}H\textsubscript{2n+1}) is small - both quantities increase slightly (~5\%) with the increasing $n$.
{
\setlength{\tabcolsep}{0pt}
\begin{table}[t]
    \centering
    \caption{The electronic properties of single molecules (\mim): the HOMO/LUMO energies, hardness $\eta$, electronegativity $\chi$. For definitions, see main text. The grey-colored cells correspond to the 'harder' spin-orbital. }
    \begin{tabular}{l|dd|dd}
           & \mc{1}{p{40pt}}{~$E_\text{HOMO}$} & \mc{1}{p{40pt}|}{~$E_\text{LUMO}$} & \mc{1}{p{40pt}}{~$\eta$} & \mc{1}{p{40pt}}{~$\chi$} \\\hline\hline
\rowcolor{morning}BMIM $\sigma = \downarrow$     \,&  -5.33  &  -0.86  &   2.23  &   3.09  \\
\rowcolor{morning}HMIM $\sigma = \downarrow$     \,&  -5.32  &  -0.82  &   2.24  &   3.07  \\
\rowcolor{morning}OMIM $\sigma = \downarrow$     \,&  -5.31  &  -0.83  &   2.24  &   3.07  \\\hline
BMIM $\sigma = \uparrow$       \,&  -2.13  &  -0.22  &   0.96  &   1.17  \\
HMIM $\sigma = \uparrow$       \,&  -2.09  &  -0.21  &   0.94  &   1.15  \\
OMIM $\sigma = \uparrow$       \,&  -2.10  &  -0.22  &   0.94  &   1.15  \\\hline
BMIB                             &  -4.47  &  -1.67  &   1.40  &   3.07  \\
HMIB                             &  -4.52  &  -1.67  &   1.42  &   3.10  \\
OMIB                             &  -4.56  &  -1.67  &   1.44  &   3.11  \\\hline
BMIC                             &  -4.70  &  -1.49  &   1.60  &   3.09  \\
HMIC                             &  -4.83  &  -1.56  &   1.63  &   3.19  \\
OMIC                             &  -4.81  &  -1.55  &   1.64  &   3.19  \\\hline
BMIS                             &  -5.32  &  -1.82  &   1.75  &   3.57  \\
HMIS                             &  -5.37  &  -1.86  &   1.76  &   3.61  \\
OMIS                             &  -5.39  &  -1.82  &   1.79  &   3.62  \\
    \end{tabular}
    \label{tab:HOMOLUMO}
\end{table}
}

A note concerning the electronic nature of the bonding of 1-alkyl-3-methyl-imidazolium with an extra functional group - either Halogens (Bromide \& Chloride) or hydrogen sulphate (HSO\textsubscript{4}) is to be made here. By performing the Bader charge analysis,\cite{bader} we obtain the resultant charge enclosed in the specific volume (by creating the potential-weighed Voronoi diagram) for each atom. The results are presented in Tab.~\ref{tab:charge}. Note that the additional group attracts an electron from 1-butyl-3-methyl-imidazolium, indicating an ionic-like bond between the group and BMIM. The transfer of an electron to the functional group warrants the prospective electron transfer from the iron surface to an imidazolium-based compound. The results for 1-hexyl- and 1-octyl- are qualitatively the same and, as such, are omitted here.

\begin{table}[t]
    \centering
    \caption{The effective charge (in units of $|e|$) on each atom of  1-butyl-3-methyl-imidazolium without (first column) and with functional group (Bromide, Chloride, and hydrogen sulphate, respectively) obtain by the Bader charge analysis \cite{bader} (hence fractional values). The numbering of carbon and nitrogen atoms corresponds to the schematic figure on the left.}
    \label{tab:charge}
    \vspace{0.5cm}\includegraphics[width=0.20\linewidth]{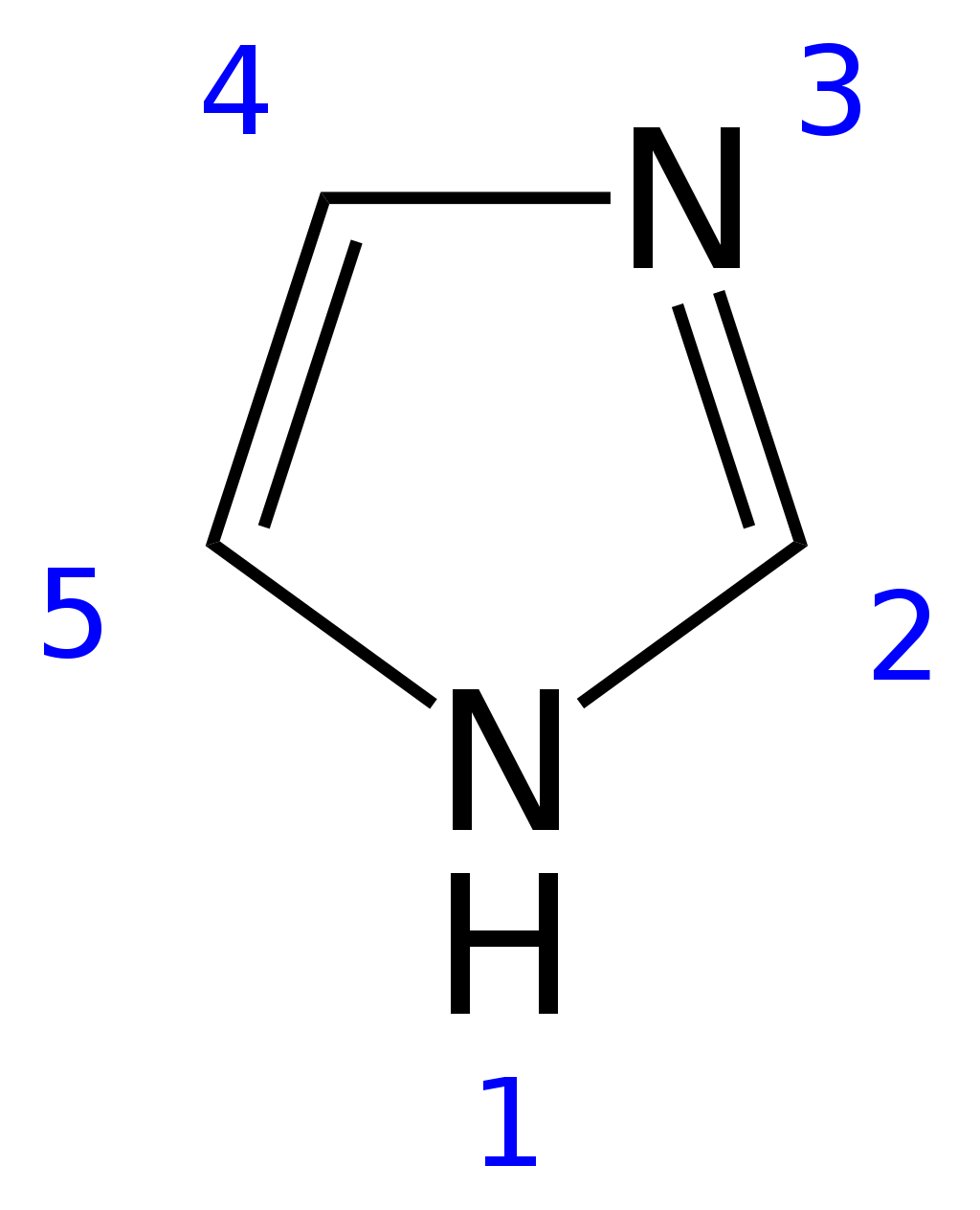}\hspace{-0.02\linewidth}
    \resizebox{0.8\linewidth}{!}{
    \begin{tabular}{l|dddd}
\vspace{-2cm}\\
                 & \text{BMIM}     & \text{BMIB}     & \text{BMIC}     & \text{BMIS}    \\ \hline \hline
Halogen                     &\varnothing  &    -0.8  &    -0.8  &\varnothing\\ \hline
C(\textcolor{blue}{1})      &        0.3  &     0.2  &     0.2  &     0.2   \\ \hline
C(\textcolor{blue}{2})      &        0.7  &     1.0  &     1.0  &     1.1   \\
C(\textcolor{blue}{3})      &        0.4  &     0.3  &     0.3  &     0.3   \\
C(\textcolor{blue}{4})      &        0.4  &     0.4  &     0.4  &     0.3   \\
C(\textcolor{blue}{5})      &        0.4  &     0.4  &     0.4  &     0.5   \\
N(\textcolor{blue}{1,3})    &       -1.3  &    -1.4  &    -1.4  &    -1.4   \\ \hline
S       &\varnothing & \varnothing &  \varnothing &\mc{1}{c}{\multirow{3}{*}{$\left. \begin{array}{r} 3.9 \\ -1.4 \\ -0.6 \\ \end{array}\right|$}}   \\
O       &\varnothing & \varnothing &  \varnothing & \\
OH      &\varnothing & \varnothing &  \varnothing & \\\cline{5-5}
HSO\textsubscript{4}        &            &             &              & \mc{1}{c}{$\left. \begin{array}{r} -0.9 \\ \end{array}\right|$}
    \end{tabular}
    }
    \vspace{-0.5cm}
\end{table}

\subsubsection{A molecule on iron (110) surface}
\label{sssec:moleculessurface}

The effect of the HOMO and LUMO energy levels modified by the presence of the molecules close to the Fe(110) surface are presented here. It allows us to examine the impact of the iron surface on the HOMO-LUMO gap, hardness $\eta$, and electronegativity $\chi$. Furthermore, we obtain the electronic structure for different distances of the \mim~molecules from the surface ({\it, i.e.}, the distance between the molecular centre of mass (CM) and the last layer of the Fe(110) surface) orienting the principal axis of the molecule (via the principal component analysis \cite{PCA}).

\begin{figure*}[t]
    \centering
    \includegraphics[width=0.49\linewidth,trim={0.0em 0.5em 2.5em 0.0em}]{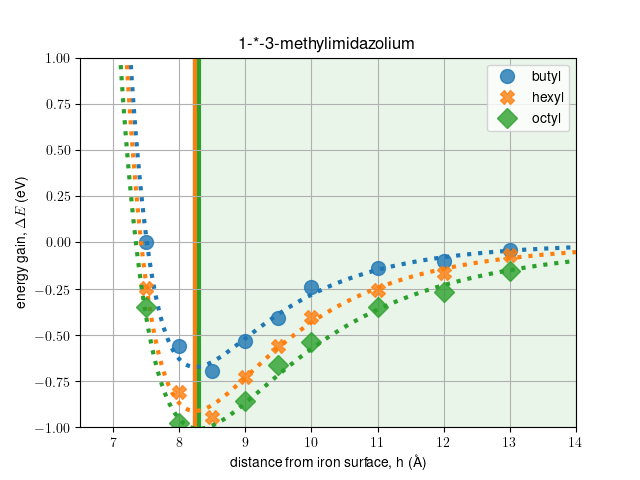}
    \includegraphics[width=0.49\linewidth,trim={2.5em 0.5em 0.0em 0.0em}]{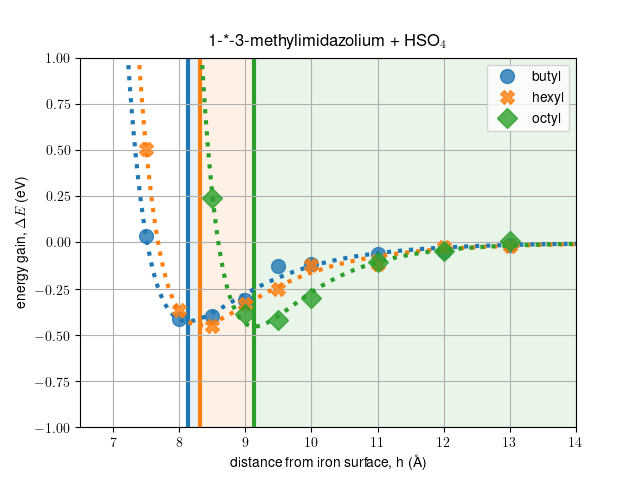}\\
    \includegraphics[width=0.49\linewidth,trim={0.0em 0.0em 2.5em 0.5em}]{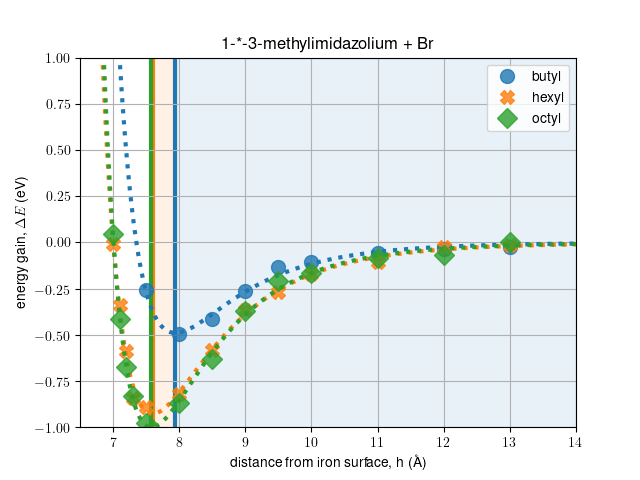}
    \includegraphics[width=0.49\linewidth,trim={2.5em 0.0em 0.0em 0.0em}]{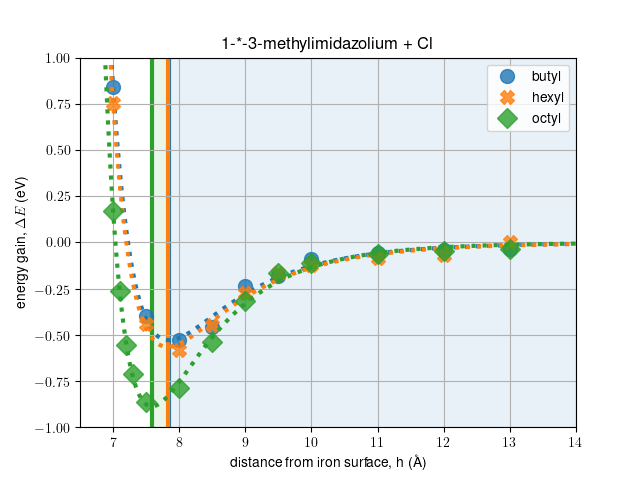}
    \caption{Relative energy $\Delta E \equiv E - E_0$ of molecules:: $\circ$MIM (top left), $\circ$MIS (top right), $\circ$MIB (bottom left), and $\circ$MIC (bottom right), versus the distance between their respective center of mass and the Fe(110) surface ($h$) along \textbf{c}. The vertical lines mark the equilibrium distance $h_0$.}
    \label{fig:distance2Fe}
\end{figure*}

We define the adsorption energy as\cite{WEI2021109555,aenm.201804000} 
\begin{align}
    E_\text{ads} \equiv E_\text{molecule on surface} - \left( E_\text{free molecule} + E_\text{clean surface} \right).
\end{align}
In Fig.~,\ref{fig:distance2Fe} we show the adsorption energy ($E_\text{ads}$) versus the distance $h$ between the molecule and Fe surface. 

In the next subsection (\ref{sssec:electronicProperties}), we test the formula for the charge expected transfer
\begin{align}
    \label{eq:charge-transfer}\notag
    \Delta N =  N^e_\text{surface} - N^e_\text{free} &\approx \frac{\chi_\text{Fe} - \chi_\text{\mim}}{2(\eta_\text{Fe} + \eta_\text{\mim})} \\
    &\approx \frac{-E_F - \chi_\text{\mim}}{2\eta_\text{\mim}},
\end{align}
(cf. Dohare, \textit{et al.} \cite{dohare}), both for the energy levels of the free-molecules with pure (non-interacting) Fe surface, as well as for the interacting system of \mim on Fe(110).

\subsubsection*{Workfunction}
\label{sssec:workfunction}
\begin{figure*}[t]
    \centering
    \includegraphics[width=0.49\textwidth]{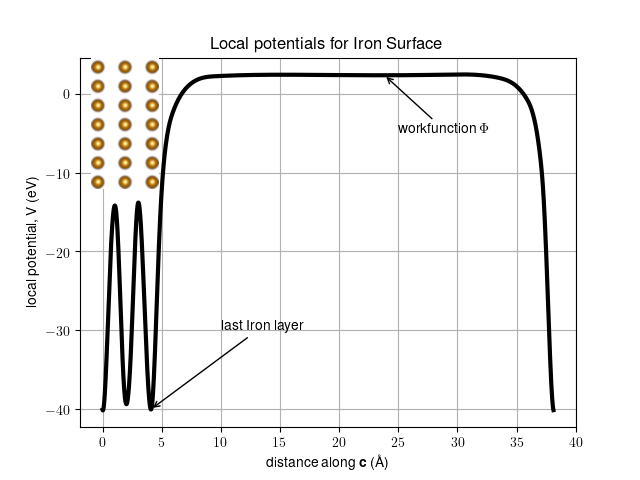}
    \includegraphics[width=0.49\textwidth]{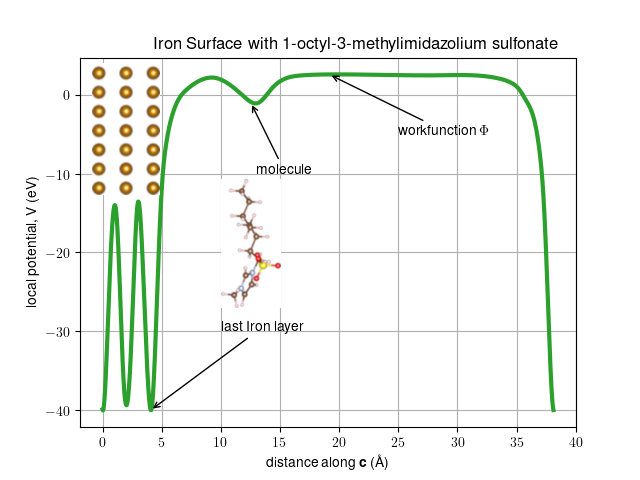}
    \caption{The local potential (averaged along $\mathbf{a}$ and $\mathbf{b}$ directions) of pure Fe(110) surface (left) and the surface with OMIS molecule at an exemplary distance (right) versus the direction $\mathbf{c}$. We define the proximity of the molecule to the surface as the difference between the last minimum corresponding to the iron layer and the minimum corresponding to the molecule.}
    \label{fig:workfunctionFe}
    \label{fig:workfunction}
\end{figure*}

 We estimate the so-called \emph{workfunction} $\Phi$, as a negative value of the local potential of the system in question at infinity along the 'vacuum' axis $\vec{c}$ (i.e., averaged along $\vec{a}$ and $\vec{b}$). In Fig.~\ref{fig:workfunction}, we illustrate this process on the exemplary cases of pure Fe(110) surface and BMIB on Fe(110). Ultimately, we add the value of $\Phi$ from the energy scale per case, resulting in comparable data.

\subsubsection{Electronic properties of molecules on iron surface}
\label{sssec:electronicProperties}

\begin{figure}[t]
    \centering
    \includegraphics[width=\linewidth]{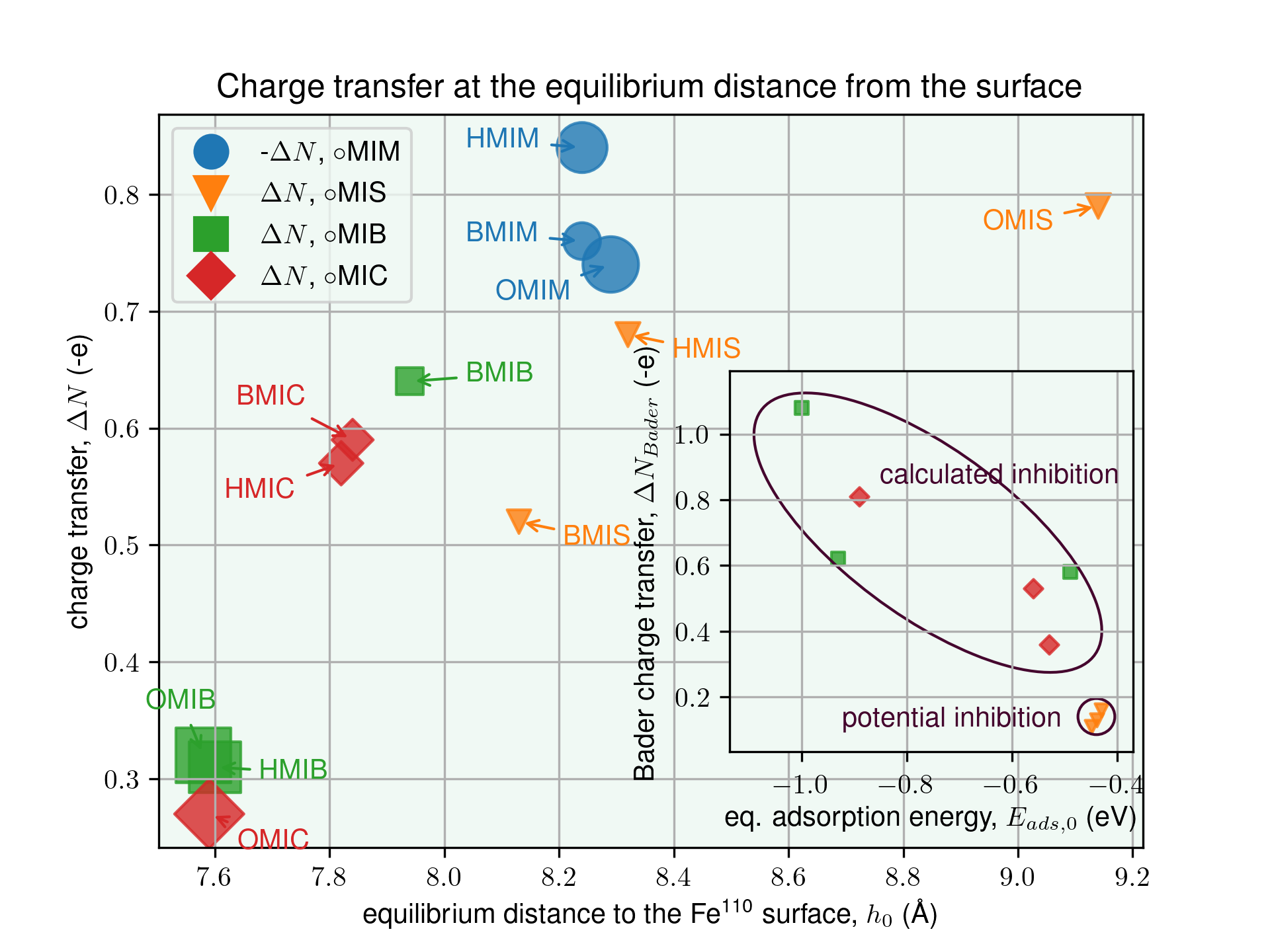}
    \caption{Charge transfer $\Delta N$ at the equilibrium position (i.e. minimizing the adsorption energy $E_\text{ads}$) versus the equilibrium distance $h$ and (inset) the adsorption energy value at equilibrium. The size of the points corresponds to the minimal values of $E_\text{ads}$. Note the clear correlation between the equilibrium distance and the charge transfer for \omiB, \omiC, and \omiS, with no dependence for \omiM (donor) molecules.}
    \label{fig:chargetransfer}
\end{figure}
{
\setlength{\tabcolsep}{0pt}
\begin{table*}[t]
    \centering
    \caption{Fermi energy $E_f$ and the chemical properties at the electrons on the molecule at the optimal distance from iron surface, namely, the HOMO/LUMO energies, hardness $\eta$, electronegativity $\chi$, and charge transfer $\Delta N$ (via \eqref{eq:charge-transfer}) for molecule on-surface and an excpected charge transfer $\Delta N_\text{free}$ extrapolated from the properties of the free molecules (cf. Tab.~\ref{tab:HOMOLUMO}) - the Fermi energy $E_f = -2.4750 eV$ - was taken as its value for the Fe(110) surface. }
    \begin{tabular}{l|ddddd|cc|c}
      &    \mc{1}{p{50pt}}{$E_f$  (eV) }& \mc{1}{p{50pt}}{$E_\text{HOMO}$  (eV)}& \mc{1}{p{50pt}}{$E_\text{LUMO}$  (eV)}& \mc{1}{p{50pt}}{$\eta$  (eV)}& \mc{1}{p{50pt}|}{$\chi$  (eV)}& \mc{1}{p{50pt}}{$\Delta N$ (-e)} & \mc{1}{p{50pt}}{$\Delta N_\text{free}$ (-e)} & \mc{1}{p{50pt}}{$\Delta N_\text{Bader}$ (-e)} \\\hline\hline
BMIM  &    -2.20   &   -8.40  &   -4.30   &    2.05  &    6.35  & \cellcolor{alert}\bf  -0.76  & \cellcolor{alert}\bf  -0.68 & -0.76 \\
HMIM  &    -2.18   &   -8.58  &   -4.46   &    2.06  &    6.52  & \cellcolor{alert}\bf  -0.84  & \cellcolor{alert}\bf  -0.70 & -0.68 \\
OMIM  &    -2.18   &   -8.11  &   -4.29   &    1.91  &    6.20  & \cellcolor{alert}\bf  -0.74  & \cellcolor{alert}\bf  -0.70 & -0.66 \\ \hline\hline
BMIB  &    -2.41   &   -5.55  &   -1.86   &    1.84  &    3.71  & \cellcolor{redish} 0.64  &   0.21 & \cellcolor{greenish}\bf 0.36 \\
HMIB  &    -2.32   &   -6.48  &   -1.75   &    2.36  &    4.12  & \cellcolor{redish} 0.31  &   0.22 & \cellcolor{greenish}\bf 0.53 \\
OMIB  &    -2.29   &   -6.46  &   -1.76   &    2.35  &    4.11  & \cellcolor{redish} 0.32  &   0.22 & \cellcolor{greenish}\bf 0.81 \\\hline
BMIC  &    -2.41   &   -5.70  &   -1.77   &    1.97  &    3.73  & \cellcolor{redish} 0.59  &   0.20 & \cellcolor{greenish}\bf 0.58 \\
HMIC  &    -2.40   &   -5.74  &   -1.77   &    1.98  &    3.75  & \cellcolor{redish} 0.57  &   0.22 & \cellcolor{greenish}\bf 0.62 \\
OMIC  &    -2.30   &   -6.66  &   -1.74   &    2.46  &    4.20  & \cellcolor{redish} 0.27  &   0.22 & \cellcolor{greenish}\bf 1.08 \\\hline
BMIS  &    -2.43   &   -5.87  &   -1.85   &    2.01  &    3.86  & \cellcolor{greenish}\bf 0.52  &   0.31 & \cellcolor{redish} 0.16 \\
HMIS  &    -2.46   &   -5.60  &   -1.81   &    1.89  &    3.70  & \cellcolor{greenish}\bf 0.68  &   0.32 & \cellcolor{redish} 0.13 \\
OMIS  &    -2.49   &   -5.46  &   -1.72   &    1.87  &    3.59  & \cellcolor{greenish}\bf 0.79  &   0.32 & \cellcolor{redish} 0.11 \\
    \end{tabular}
    \label{tab:HOMOLUMO-Fe}
\end{table*}
}

\begin{table}[t]
    \centering
    \caption{Comparison of the energy scales: Fermi energy $E_f$, on-surface HOMO energy, free-molecule HOMO energy, on-surface LUMO energy, and free-molecule LUMO energy. All values in eV.}
    \resizebox{\linewidth}{!}{
    \begin{tabular}{l|dd||dc}
&   \mc{1}{c}{$E_\text{HOMO}$ on Fe(110)} &  \mc{1}{c||}{$E_\text{HOMO}$ free} & \mc{1}{c}{$E_\text{LUMO}$ on Fe(110)} & \mc{1}{c}{$E_\text{LUMO}$ free} \\\hline\hline
BMIB  &   -5.55  &  -4.47  &   -1.86   &  -1.67  \\  
HMIB  &   -6.48  &  -4.52  &   -1.75   &  -1.67  \\  
OMIB  &   -6.46  &  -4.56  &   -1.76   &  -1.67  \\\hline
BMIC  &   -5.70  &  -4.70  &   -1.77   &  -1.49  \\  
HMIC  &   -5.74  &  -4.83  &   -1.77   &  -1.56  \\  
OMIC  &   -6.66  &  -4.81  &   -1.74   &  -1.55  \\\hline
BMIS  &   -5.87  &  -5.32  &   -1.85   &  -1.82  \\
HMIS  &   -5.60  &  -5.37  &   -1.81   &  -1.86  \\
OMIS  &   -5.46  &  -5.39  &   -1.72   &  -1.82  \\
    \end{tabular}
    }
    \label{tab:HOMOLUMO-Fe-comp-1}
\end{table}

\begin{table*}[t]
    \centering
    \caption{Comparison of on-surface and free-molecule hardness' $\eta$, electronegativities $\chi$, and charge transfers $\Delta N$ (via \eqref{eq:charge-transfer}).}
    \begin{tabular}{l|cc||cc||dc}
   & \mc{1}{c}{$\eta$ (eV) on Fe(110)} & \mc{1}{c||}{$\eta$ (eV) free} & \mc{1}{c}{$\chi$ (eV) on Fe(110)} & \mc{1}{c||}{$\chi$ (eV) free} & \mc{1}{c}{$\Delta N$ on Fe(110)} & \mc{1}{c}{$\Delta N$ free }\\\hline\hline
BMIB  &     1.84   &   1.40   &   3.71  &   3.07    &    0.64 &  0.21 \\
HMIB  &     2.36   &   1.42   &   4.12  &   3.10    &    0.31 &  0.22 \\
OMIB  &     2.35   &   1.44   &   4.11  &   3.11    &    0.32 &  0.22 \\\hline
BMIC  &     1.97   &   1.60   &   3.73  &   3.09    &    0.59 &  0.20 \\
HMIC  &     1.98   &   1.63   &   3.75  &   3.19    &    0.57 &  0.22 \\
OMIC  &     2.46   &   1.64   &   4.20  &   3.19    &    0.27 &  0.22 \\\hline
BMIS  &     2.01   &   1.75   &   3.86  &   3.57    &    0.52 &  0.31 \\
HMIS  &     1.89   &   1.76   &   3.70  &   3.61    &    0.68 &  0.32 \\
OMIS  &     1.87   &   1.79   &   3.59  &   3.62    &    0.79 &  0.32 \\
    \end{tabular}
    \label{tab:HOMOLUMO-Fe-comp-2}
\end{table*}
The main aim of the modelling described above is to determine the charge transfer to the adsorbed molecule to estimate the corrosion inhibition efficiency. We compare the results \emph{(i)} obtained via the separate calculations (i.e., assuming that the interaction with surface does not significantly change the chemistry of \mim~molecules); \emph{(ii)} extending \eqref{eq:charge-transfer} to the system of \mim~interacting with Fe(110) surface; \emph{(iii)} via the Bader charge analysis\cite{bader} of the latter model. In Table~,\ref{tab:HOMOLUMO-Fe} we calculate the chemical quantities as described in Sec.~\ref{sssec:homolumogap}, and in subsequent Tabs.~\ref{tab:HOMOLUMO-Fe-comp-1}~and~\ref{tab:HOMOLUMO-Fe-comp-2} we compare the impact of the iron surface on the \mim~molecules. Please note that the changes of the Fermi energy of the surface correspond to the fraction of charge accepted on the molecule in our periodic model (with a large but ultimately constant number of electrons). At first, the simple (non-functionalized) imidazolium molecule is not an acceptor but a donor of electrons to the Fe(110) surface; see the charge transfer for \omiM in Table~\ref{tab:HOMOLUMO-Fe}. However, for the imidazolium molecule functionalized with either Br, Cl, or HSO\textsubscript{4} group, the charge transfer is towards these molecules, see \omiC, \omiB, and \omiS inTab.\ref{tab:HOMOLUMO-Fe}.

A crucial remark is to be made here. Equation \eqref{eq:charge-transfer} \emph{does not} represent the actual charge transfer but rather \emph{the aptitude} of a molecule to accept some additional charge. This means that the results in \cref{tab:HOMOLUMO-Fe} need to be supplemented with the analysis of the molecular charge, i.e. with the Bader charge analysis of the resultant spacial charge distribution. Moreover, the chemical properties of \omiM molecules in proximity of Fe(110) surface are a clear indicator of a charge transfer \emph{from molecule to the surface} (i.e., its ionization). Note that the drastic changes of both hardness ($\eta_\text{free} \approx 1 \rightarrow \eta_\text{surf} \approx 2$) and electronegativity  ($\eta_\text{free} \approx 1 \rightarrow \eta_\text{surf} \approx 6$) combined with lowering the HOMO-LUMO energies are what one expects is such a case.

As presented in \cref{tab:HOMOLUMO-Fe}, the compounds are divided into three distinctive groups: \emph{(i)} 'pure' \omiM molecules, \emph{(ii)} Halogen enriched \omiB and \omiC, and \emph{(iii)} containing the hydrogen sulphate (see \cref{fig:chargetransfer}). In \emph{(i)} the molecules neither accept the electron from the surface (in fact they are donors as $\Delta N_\text{Bader} < 0$) nor have the capacity to accept charge. Halogen-enriched \emph{(ii)} molecules do not indicate smaller possibility of accepting a new electron as the alkyl chain becomes longer as an extra participation is \emph{already included} in its electronic structure ($\Delta N_\text{Bader} \gg 0$). Finally, in \emph{(iii)} we do not observe significant charge being accepted on the \omiS molecules as $\Delta N_\text{Bader}$ decreases with the alkyl chain-length. The subgroups \emph{(ii)} and \emph{(iii)} are distinctively marked in the Inset of Fig.~\ref{fig:chargetransfer}. Nevertheless, the result is compensated by growing (and significantly larger than other compounds) capacity for becoming an acceptor ($\Delta N \gg 0$). These discrepancies are caused (most probably) by the finite size of the iron surface (chemical potential $\mu$ being low but still $> 0$), causing the Fe-originated charge also to be limited. Technically, the finite-size scaling for the slab volume is to answer this question - a task feasible using the next generation HPC clusters with increased memory size and decreased node communication latency, as our calculations here reached the limits of the state-of-the-art supercomputers.\cite{euroexa} However, our results show important trends in inhibitor behaviour for the discussed compounds (cf. conclusions in Sec.~\ref{sec:conlustion}).

\section{Conclusions}
\label{sec:conlustion}
In this paper we present a complementary picture of experimental and theoretical results showing the dependence of the efficiency of corrosion inhibition on the length of the alkyl chain of 1-alkyl-methylimidazolium chlorides, and 1-alkyl-methylimidazolium hydrogen sulphates. We not only show the qualitative differences between these two groups but also provide a comprehensive computational tool for labeling future candidate compounds as an extension to the already available approaches.

The inhibition efficiency of the imidazolium salts with various anions and various alkyl chain lengths was investigated through potentiodynamic polarization curves like corrosion current densities, EIS, weight loss measurements, and surface analysis. 1-octyl-3-methylimidazolium hydrogen sulphate and 1-octyl-3-methylimidazolium chloride proved to be mixed-type inhibitors with the predominant anodic inhibitive effect for mild steel. The inhibition efficiency of both ionic liquids is about 90\% for the concentration of 0.001 M and higher in 1M HCl solutions. The results were compared with those obtained for 1-octyl-3-methylimidazolium bromide \cite{article39}. The differences between 1-octyl-3-methylimidazolium salts were not significant.
On the other hand, the inhibition efficiency of 1-butyl-3-methylimidazolium hydrogen sulphate was much lower. It can be suggested that the alkyl chain length influences the inhibitor behaviour more than the type of anion for the same type of ionic liquid. The inhibition efficiency values obtained from the weight loss measurements were slightly higher than those from the electrochemical measurement. The reason probably consists of the slow formation of the protective film. 

The quantum chemical calculations support the experimental findings when accounted for both the accepted and potentially accepted charge. While the total charge transfer ($\Delta N_\text{tot} \equiv \Delta N + \Delta N_\text{Bader}$ oscillates around 1 for \omiC and \omiB (increasing $\sim 10 \%$ with doubling of the alkyl chain, its value for the  \omiS compounds jumps by $\sim 35 \%$ (BMIS $\rightarrow$~OMIS, see all in  \cref{tab:HOMOLUMO-Fe}). These trends are clearly not unravelled when the molecule is being assessed separately from the surface ($\Delta N_\text{free} \approx \text{const}$ for \omiB, \omiC, and \omiS), as it was shown in the previous works.\cite{dohare} It is important to note that although this separate approach does not predict the qualitative behaviour of the inhibition efficiency vs the chain length, it does, however, predict whether the molecule will become acceptor or not.

\section{Acknowledgements}
\label{sec:acknowledgements}
This work was supported by the Technology Agency of the Czech Republic, project No. FV10089, \emph{Synthesis of ionic liquids in microwave reactor}, by the ERDF in the IT4Innovations national supercomputing centre - path to exascale project (CZ.02.1.01/0.0/0.0/16\_013/0001791) within the OPRDE, and by the project e-INFRA CZ (ID:90140) by the Ministry of Education, Youth and Sports of the Czech Republic.

\section{References}
\bibliography{clanekbibtex} 

\clearpage
\onecolumngrid
\appendix
\section{Formulae and the naming convention}
\label{app:naming}
\begin{figure}
    \centering
    \includegraphics[width=0.9\textwidth]{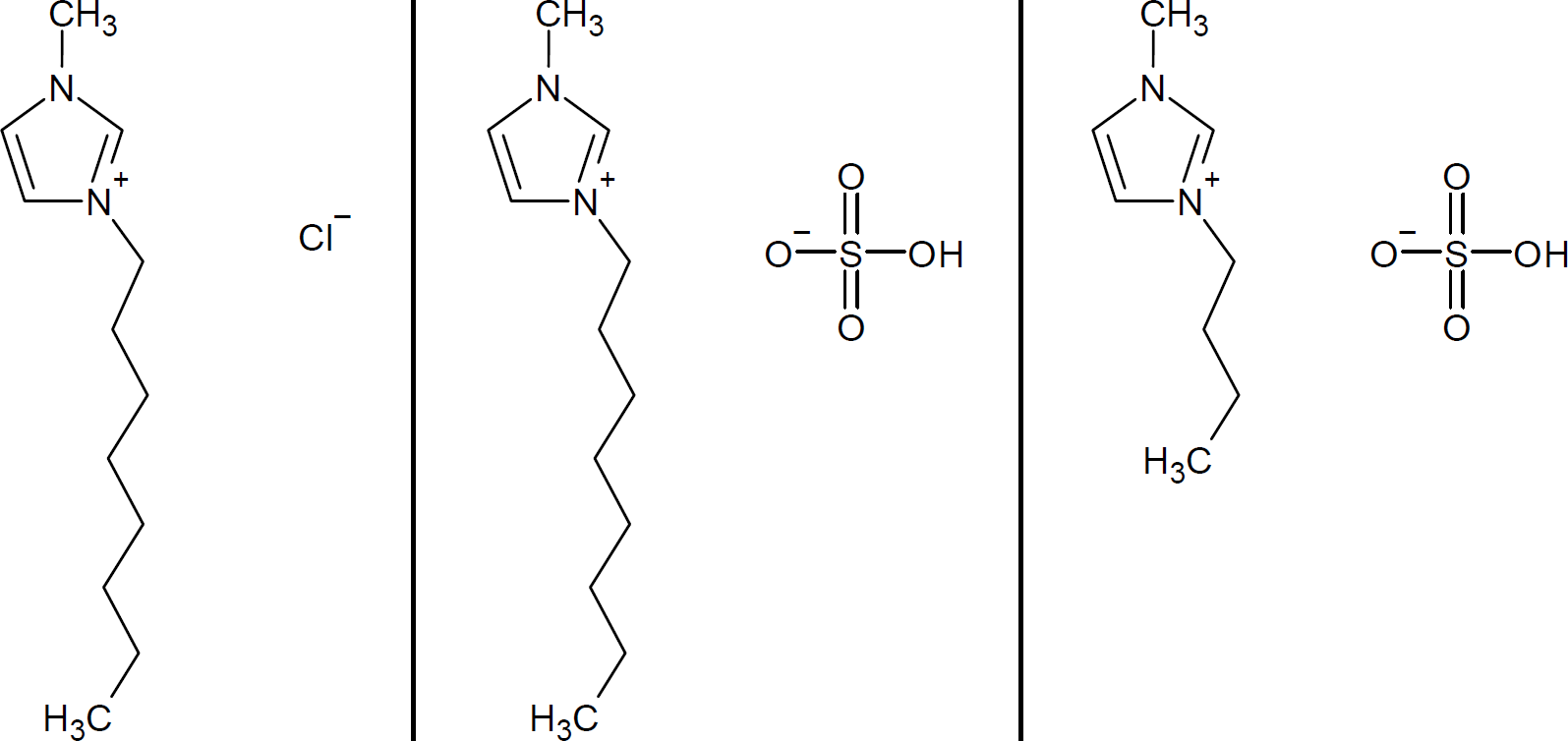}
    \caption{Skeletal formulae of the experimentally studied molecules, \emph{i.e.}, the 1-octyl-3-methylimidazolium chloride (left), 1-octyl-3-methylimidazolium hydrogen sulphate (center), and 1-butyl-3-methylimidazolium hydrogen sulphate (right).}
    \label{fig:formulae}
\end{figure}

In this paper we study the compounds derived from imidazole, \emph{i.e.}, C\textsubscript{3}H\textsubscript{4}N\textsubscript{2} with a skeletal formula as in Fig.~\ref{fig:skeletal} with a methyl (CH\textsubscript{3}) group at the position 3 and an alkyl chain (see below) at the position 1. The compounds studied both experimentally and theoretically, \emph{i.e.}, 1-octyl-3-methylimidazolium chloride, 1-octyl-3-methylimidazolium hydrogen sulphate, and 1-butyl-3-methylimidazolium hydrogen sulphate, have their full skeletal formulae displayed in Fig.~\ref{fig:formulae}.
\begin{figure}[h!]
    \centering\vspace{-0.3cm}
    \includegraphics[width=0.10\linewidth]{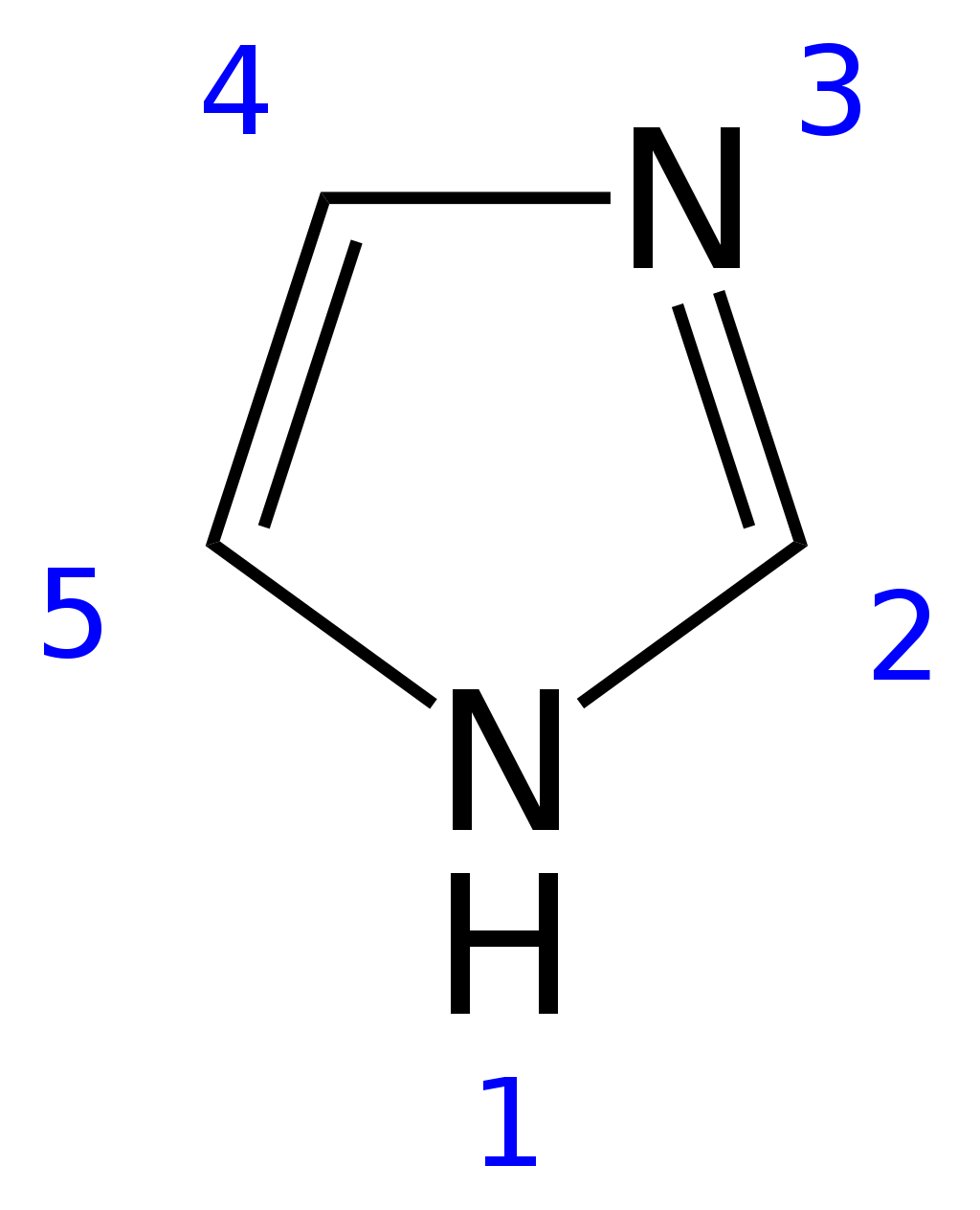}
    \vspace{-0.35cm}
    \caption{A skeletal formula of an imidazolium molecule with a numbering convention of substituent groups used in this paper.}
    \label{fig:skeletal}\vspace{-0.3cm}
\end{figure}

For simplicity, we refer to the molecules in question, \emph{i.e.}, 1-$\bigcirc$-3-\textbf{m}ethyl\textbf{i}midazolium $\square$, as \mim, where we take the alkyl substituent group $\bigcirc$ as \\
\begin{center}
    \begin{tabular}{lc|c}
    \,\,\,alkyl\,\,\,& \,\,\,formula\,\,\, & \,\,\,symbol ($\bigcirc$)\,\,\,\\\hline
    butyl & $-C_4H_{9 }$ & \textbf{B} \\
    hexyl & $-C_6H_{13}$ & \textbf{H} \\
    octyl & $-C_8H_{17}$ & \textbf{O} \\
    \end{tabular}
\end{center}
Furthermore, we take the functional group $\square$ as
\begin{center}
    \begin{tabular}{lc|c}
    \,\,\,\,\,\,group name & \,\,\,formula\,\,\, & \,\,\,symbol ($\square$)\,\,\,\\\hline
    $$\O$$ & $ $ & \textbf{M} \\
    Bromide & $Br$ & \textbf{B} \\
    Chloride & $Cl$ & \textbf{C} \\
    Hydrogen sulphate & $-HSO_4$ & \textbf{S} \\
    \end{tabular}
\end{center}
Hence, 1-octyl-3-methylimidazolium is being referred as \textbf{OMIM}, whereas 1-butyl-3-methylimidazolium bromide as \textbf{BMIB}.

\subsection{Visualization of HOMO and LUMO}
\label{app:visual}
\begin{figure*}[h!]
     \includegraphics[angle=90,origin=c,width=0.24\linewidth]{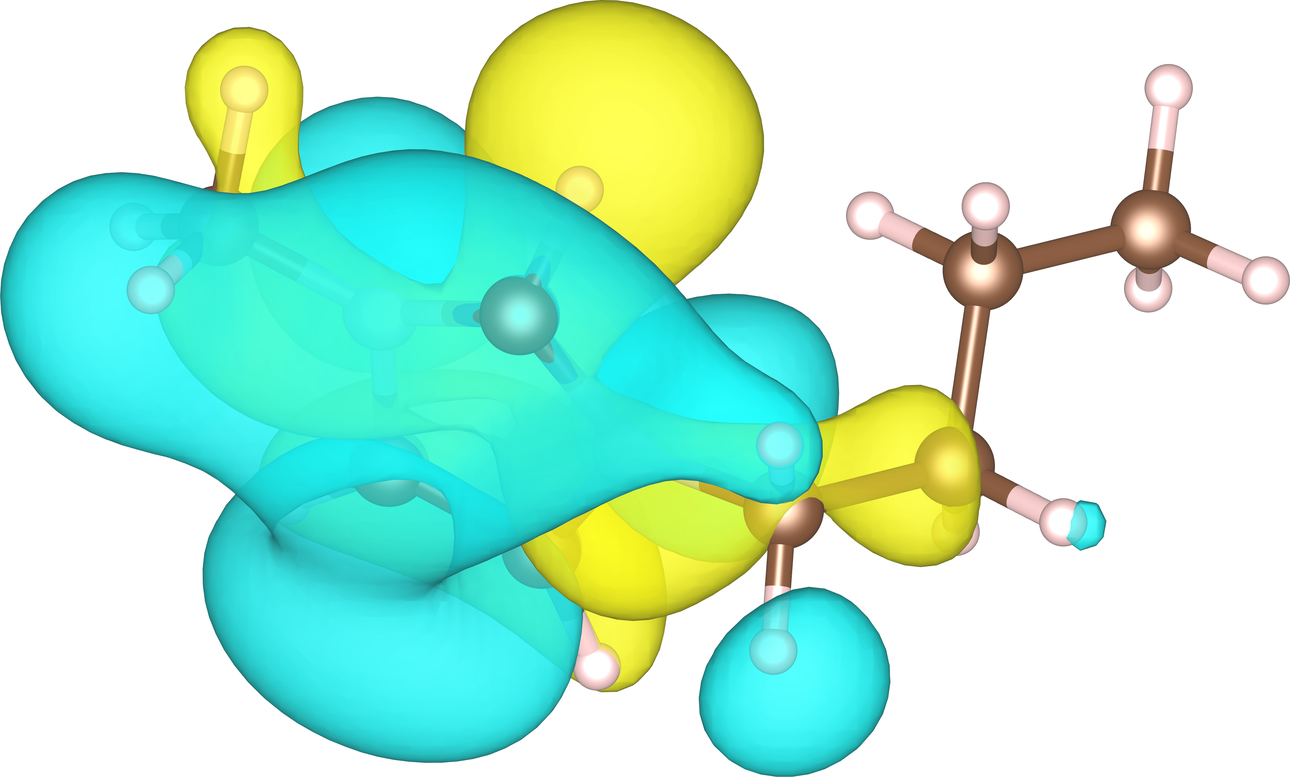}
     \includegraphics[angle=90,origin=c,width=0.24\linewidth]{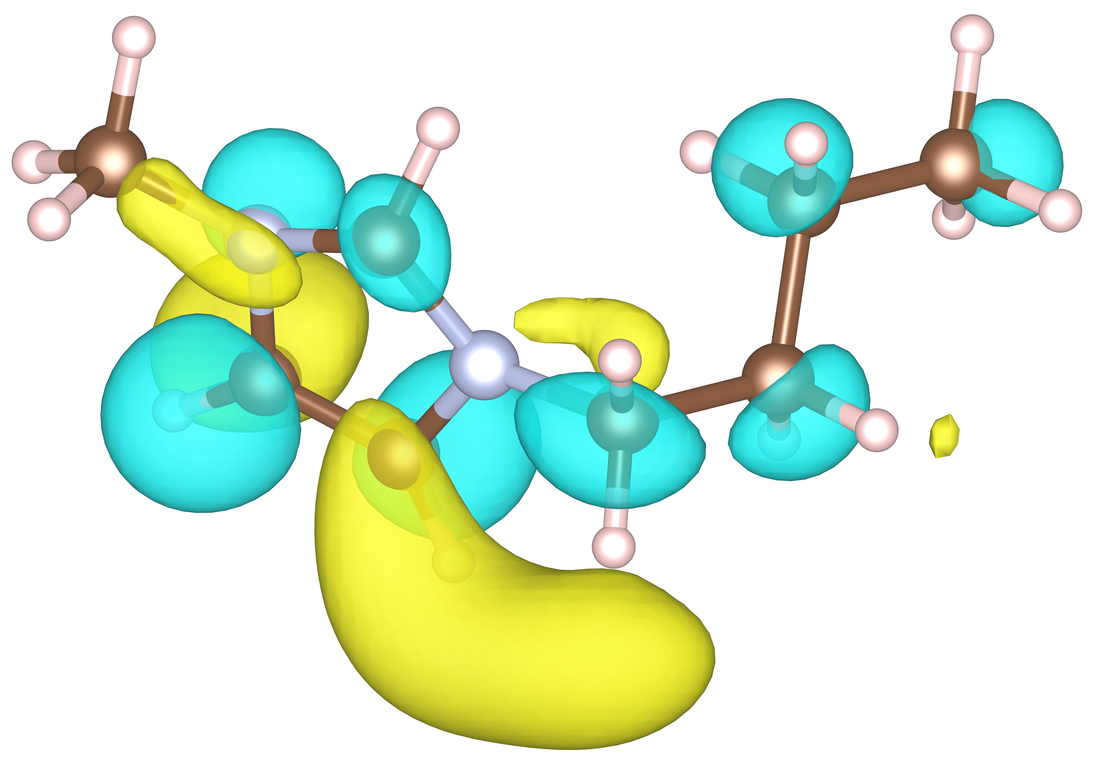}
     \includegraphics[angle=90,origin=c,width=0.24\linewidth]{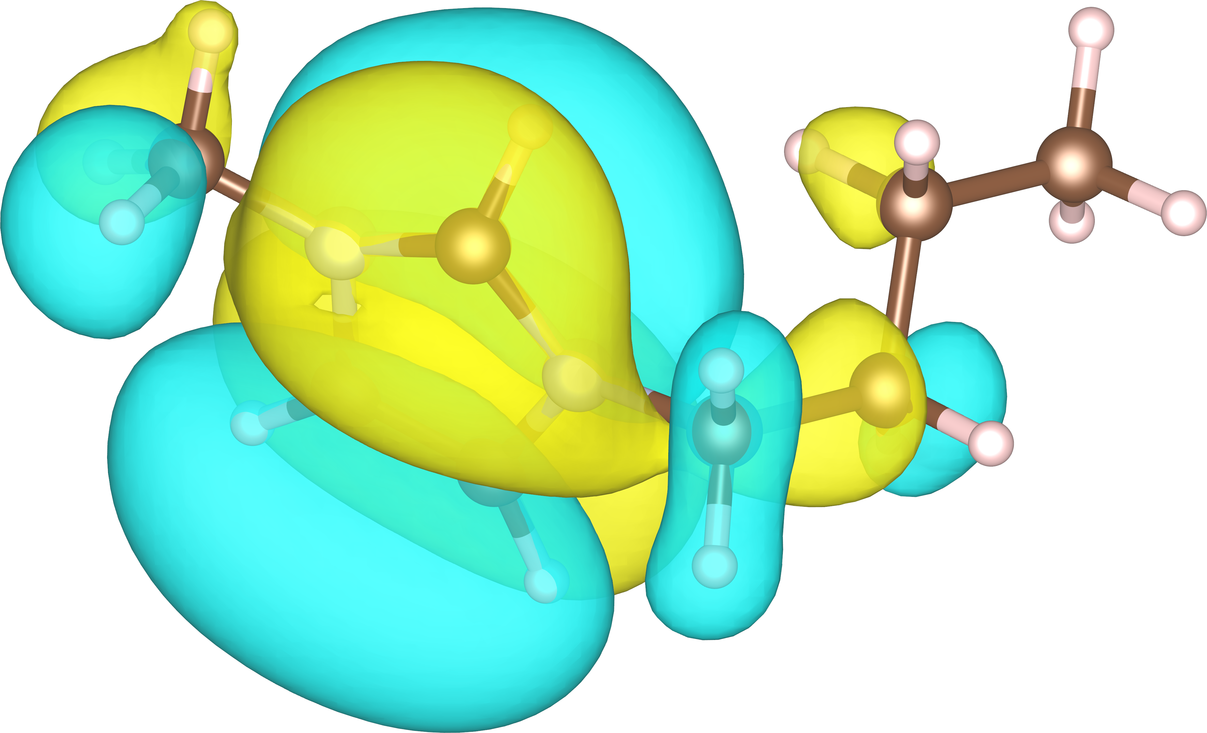}
     \includegraphics[angle=90,origin=c,width=0.24\linewidth]{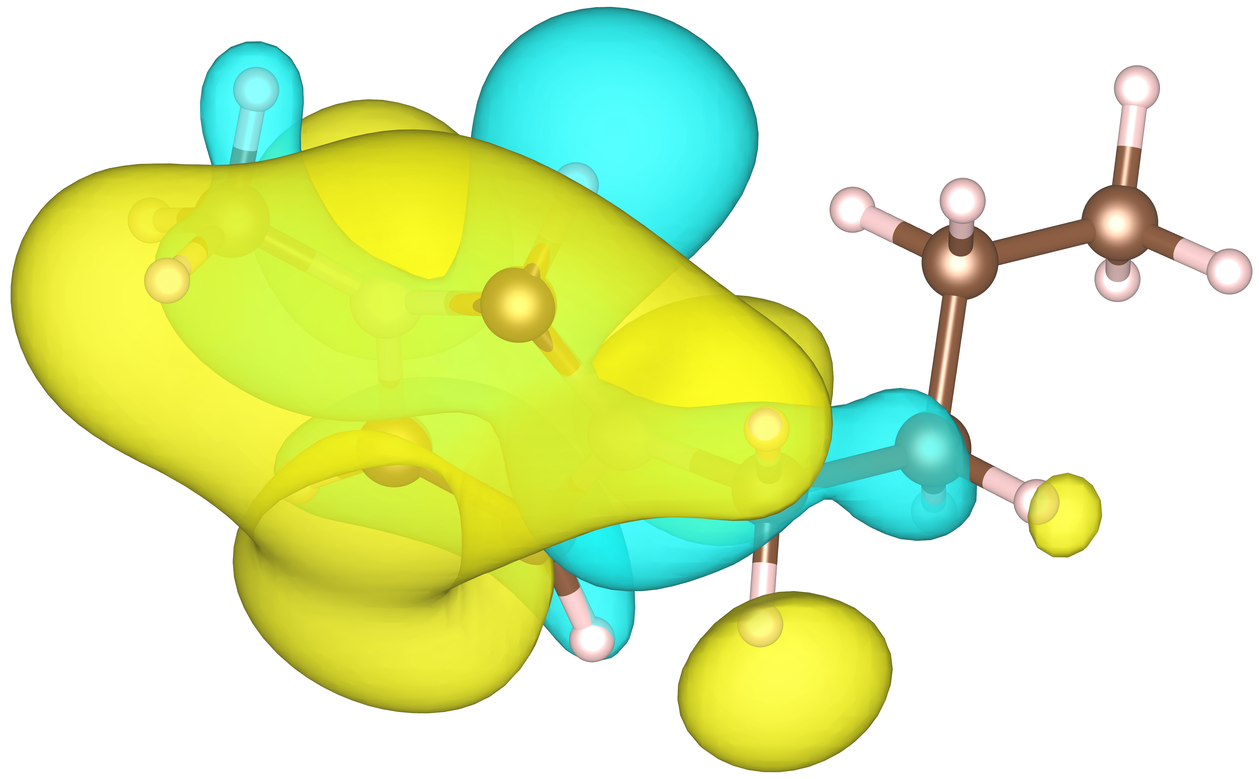}
    \hspace{-\textwidth}
    \vspace{-120pt}
    \resizebox{\textwidth}{!}{ \Large\bf
    \begin{tabular}{llll}
     \hspace{0.11\textwidth}(a) \hspace{0.11\textwidth} & \hspace{0.11\textwidth} (b) \hspace{0.11\textwidth} & \hspace{0.11\textwidth} (c)  \hspace{0.11\textwidth} & \hspace{0.11\textwidth} (d) \hspace{0.11\textwidth} \\
     &&& \\ &&& \\ &&& \\ &&& \\ &&& \\ &&& \\ &&& \\ &&& \\ &&& \\ &&& \\ &&& \\ &&& \\ &&& \\ &&& \\ &&& \\ &&& \\ &&& \\ &&& \\ &&& \\ &&& \\
    \end{tabular}}
    \caption{1-butyl-3-methylimidazolium:  The highest occupied molecular orbital (HOMO) \textbf{(a), (c)} and the lowest unoccupied molecular orbital (LUMO) \textbf{(b), (d)} wavefunctions for spin $\uparrow$ and $\downarrow$ respectively.}
    \label{fig:BMIM-HOMOLUMO}
\end{figure*}

\begin{figure*}
    \centering
     \includegraphics[angle=90,origin=c,width=0.24\linewidth]{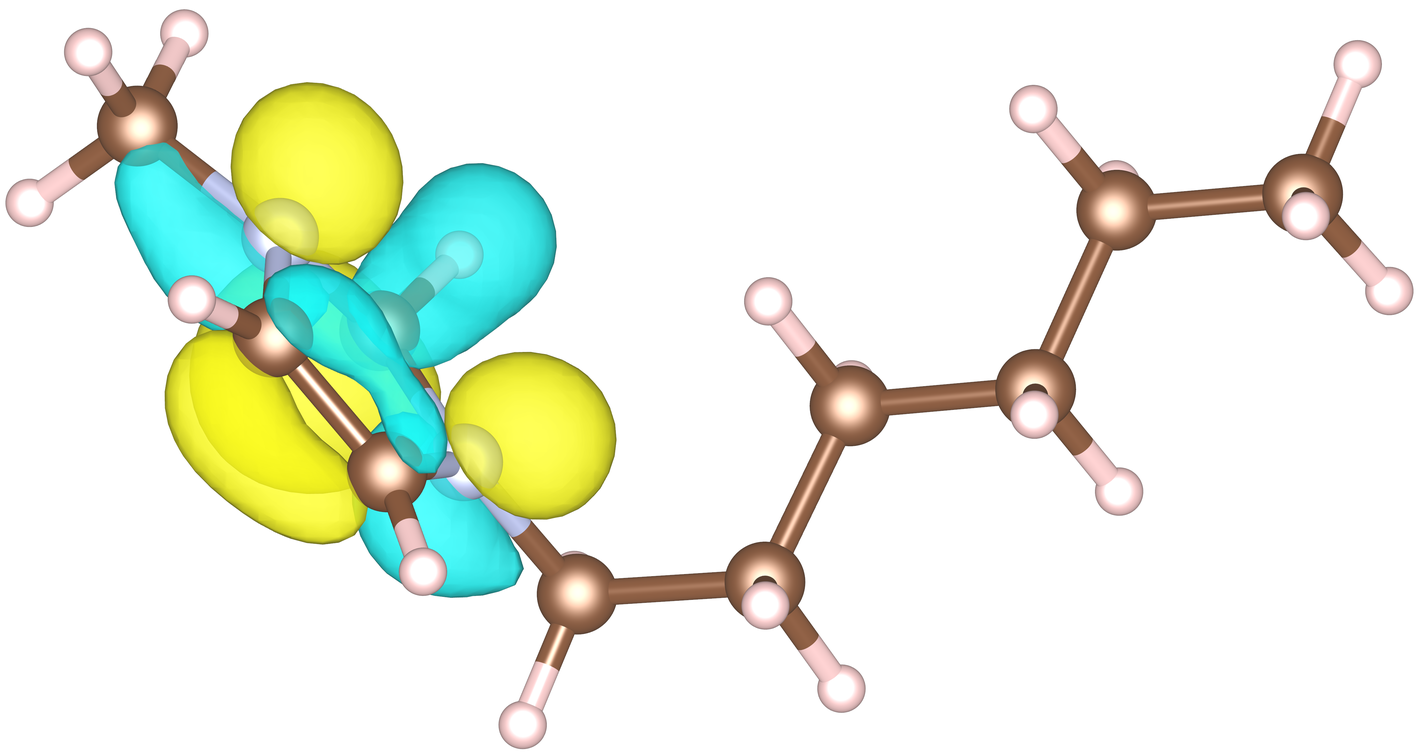}
     \includegraphics[angle=90,origin=c,width=0.24\linewidth]{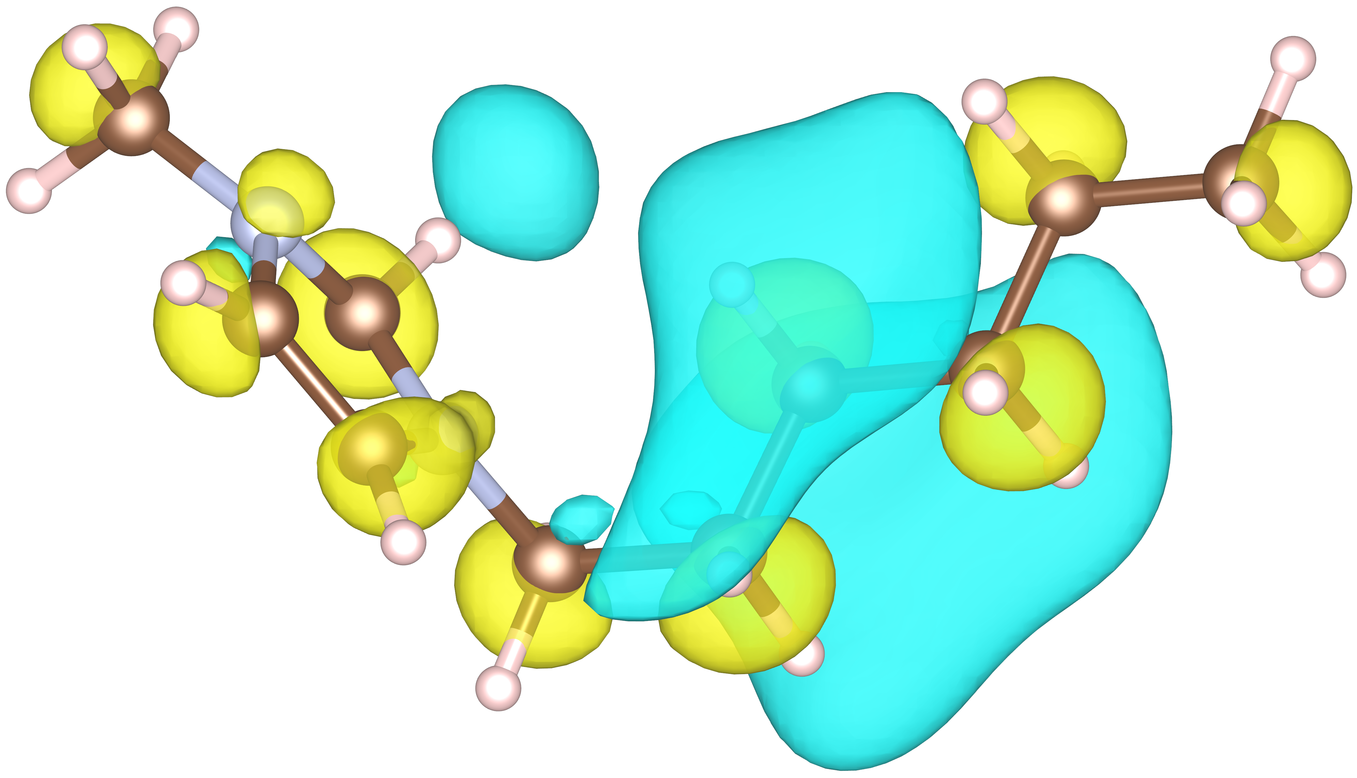}
     \includegraphics[angle=90,origin=c,width=0.24\linewidth]{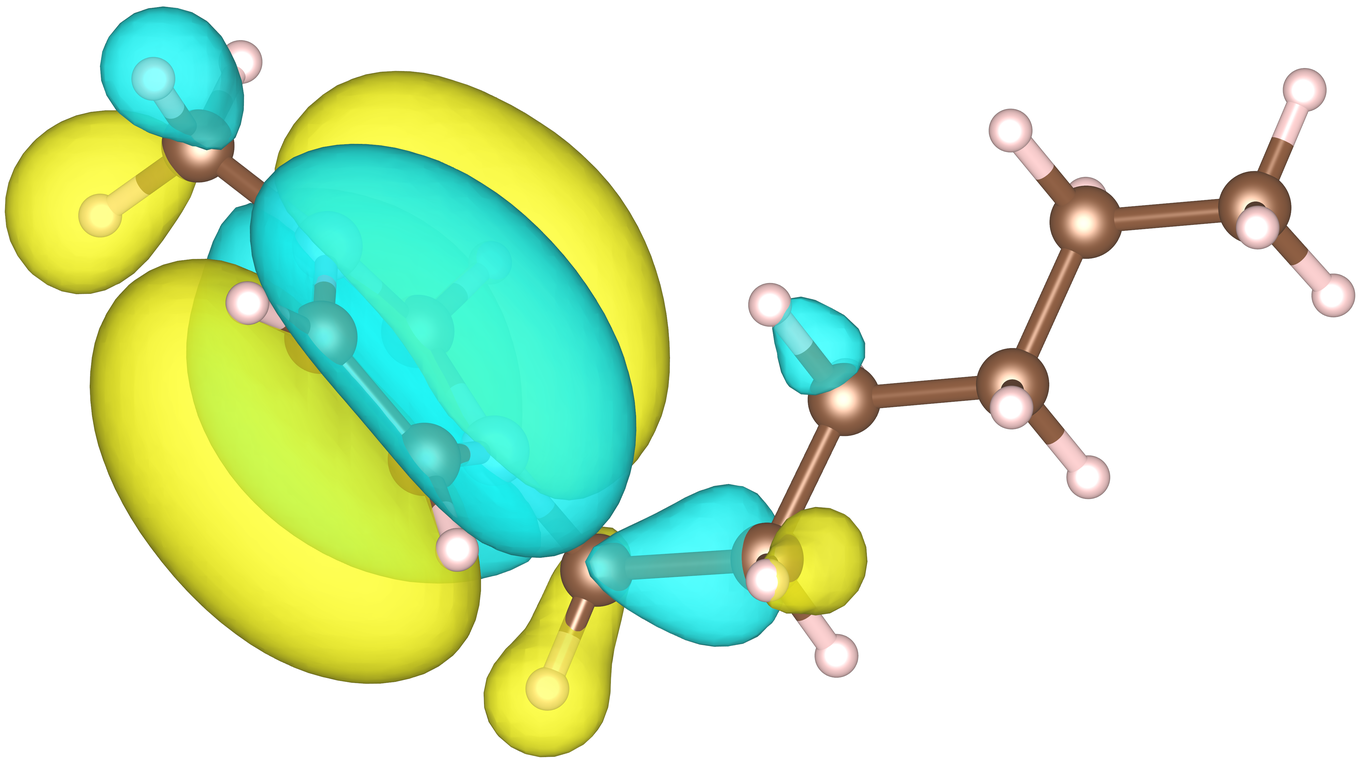}
     \includegraphics[angle=90,origin=c,width=0.24\linewidth]{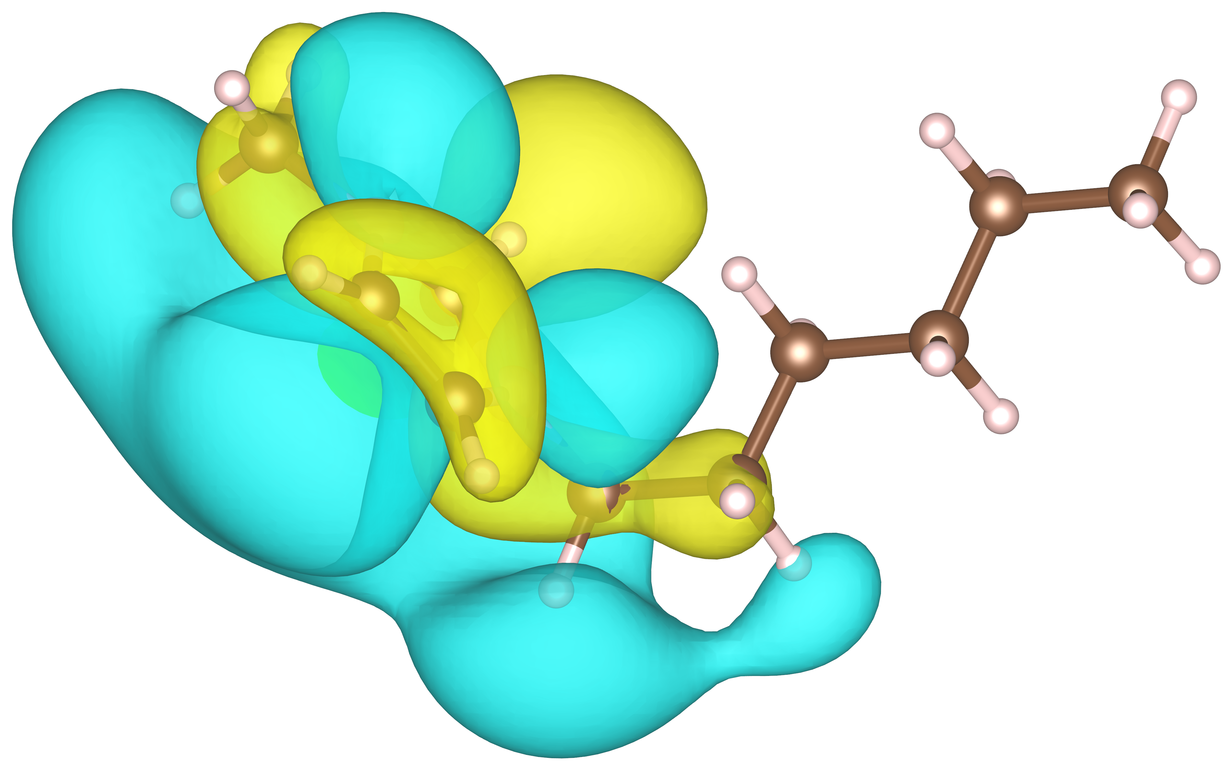}
    \hspace{-\textwidth}
    \vspace{-120pt}
    \resizebox{\textwidth}{!}{ \Large\bf
    \begin{tabular}{llll}
     \hspace{0.11\textwidth}(a) \hspace{0.11\textwidth} & \hspace{0.11\textwidth} (b) \hspace{0.11\textwidth} & \hspace{0.11\textwidth} (c)  \hspace{0.11\textwidth} & \hspace{0.11\textwidth} (d) \hspace{0.11\textwidth} \\
     &&& \\ &&& \\ &&& \\ &&& \\ &&& \\ &&& \\ &&& \\ &&& \\ &&& \\ &&& \\ &&& \\ &&& \\ &&& \\ &&& \\ &&& \\ &&& \\ &&& \\ &&& \\ &&& \\ &&& \\
    \end{tabular}}
    \caption{1-hexyl-3-methylimidazolium:  The highest occupied molecular orbital (HOMO) \textbf{(a), (c)} and the lowest unoccupied molecular orbital (LUMO) \textbf{(b), (d)} wavefunctions for spin $\uparrow$ and $\downarrow$ respectively.}
    \label{fig:HMIM-HOMOLUMO}
\end{figure*}

\begin{figure*}
    \centering
     \includegraphics[angle=90,origin=c,width=0.24\linewidth]{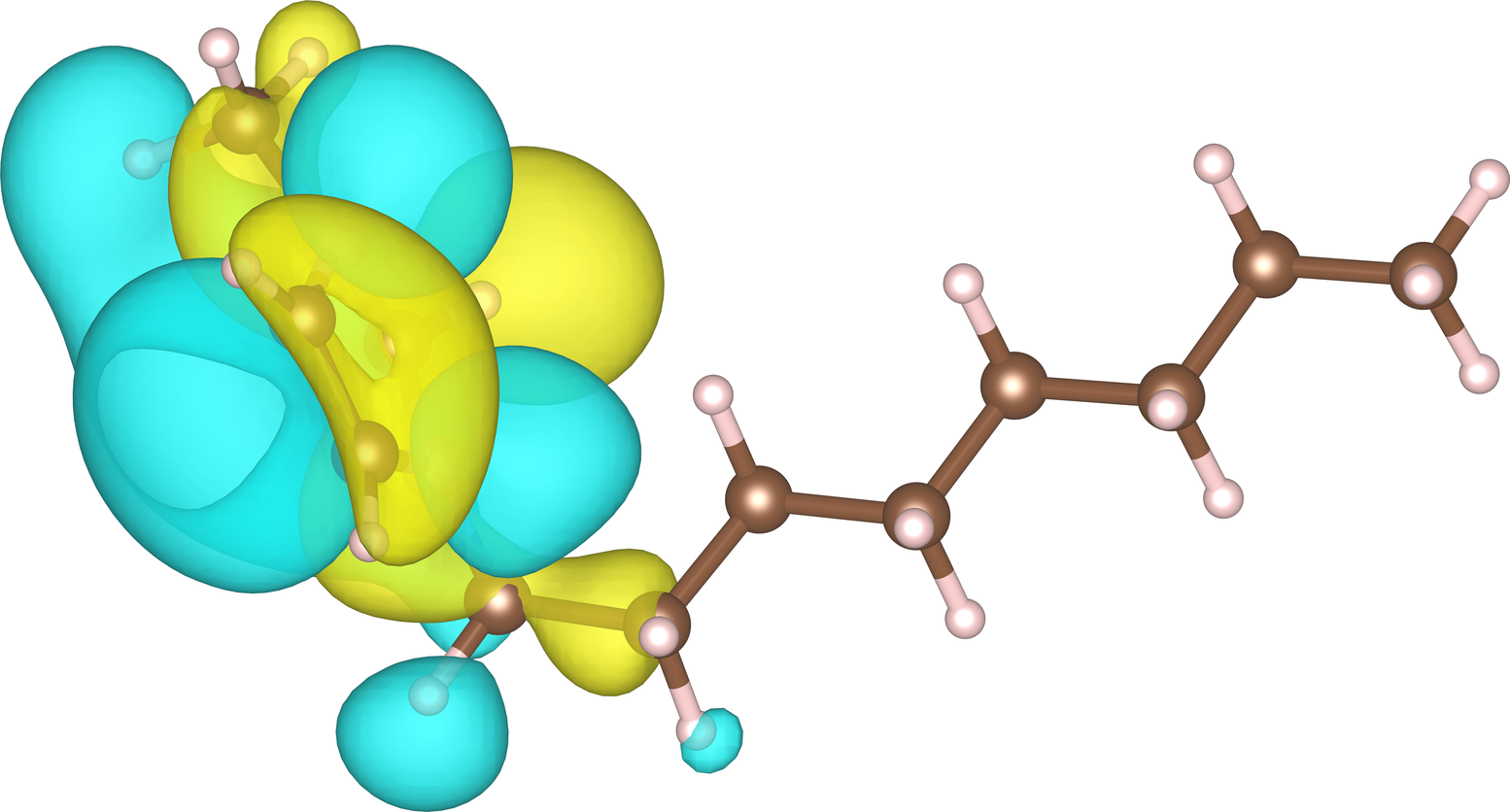}
     \includegraphics[angle=90,origin=c,width=0.24\linewidth]{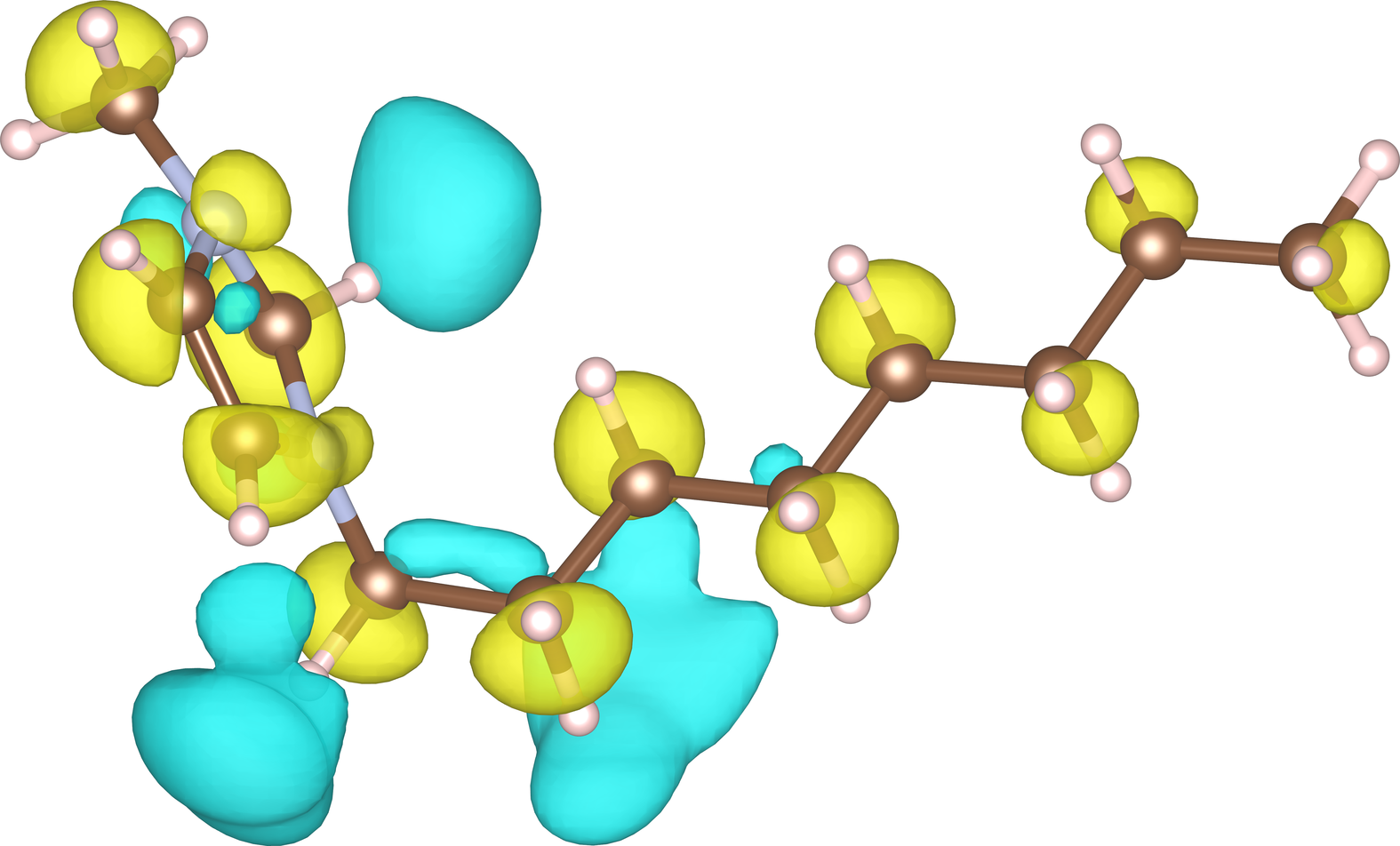}
     \includegraphics[angle=90,origin=c,width=0.24\linewidth]{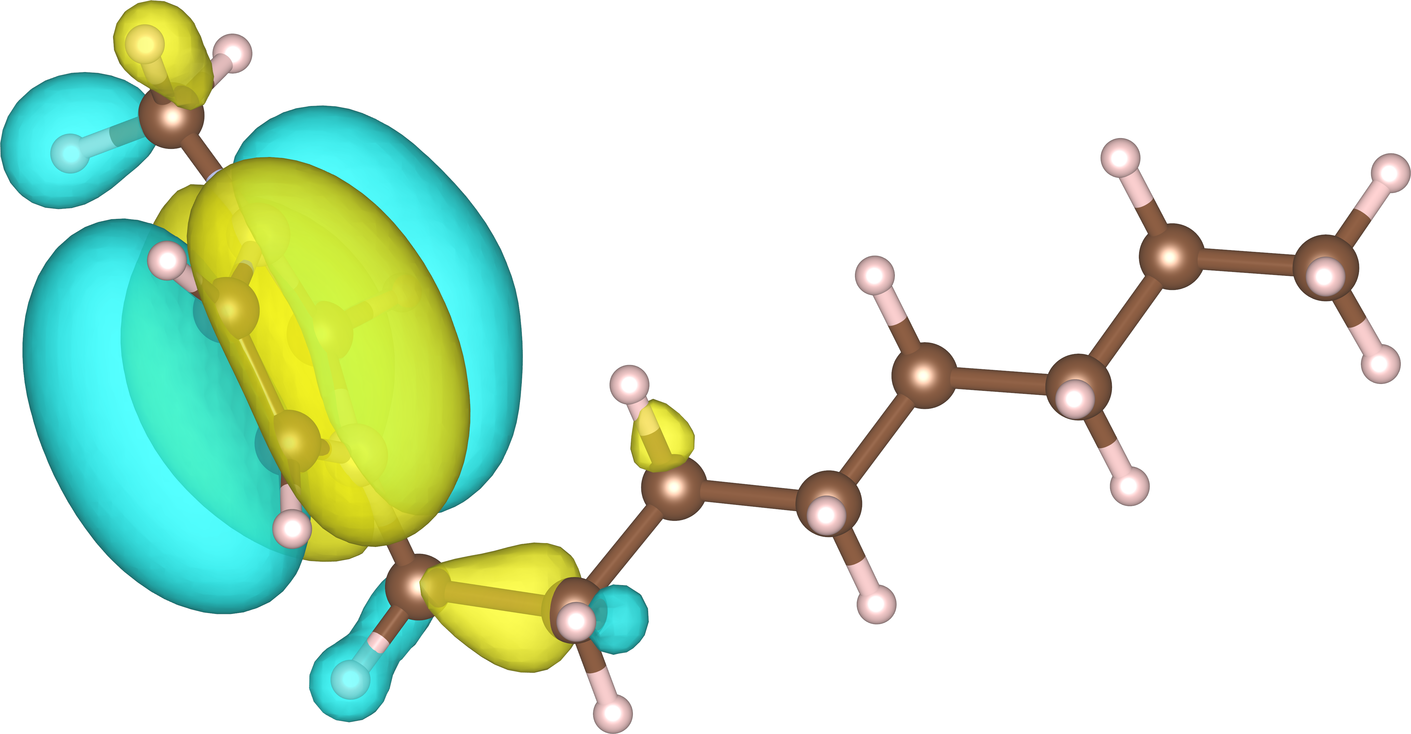}
     \includegraphics[angle=90,origin=c,width=0.24\linewidth]{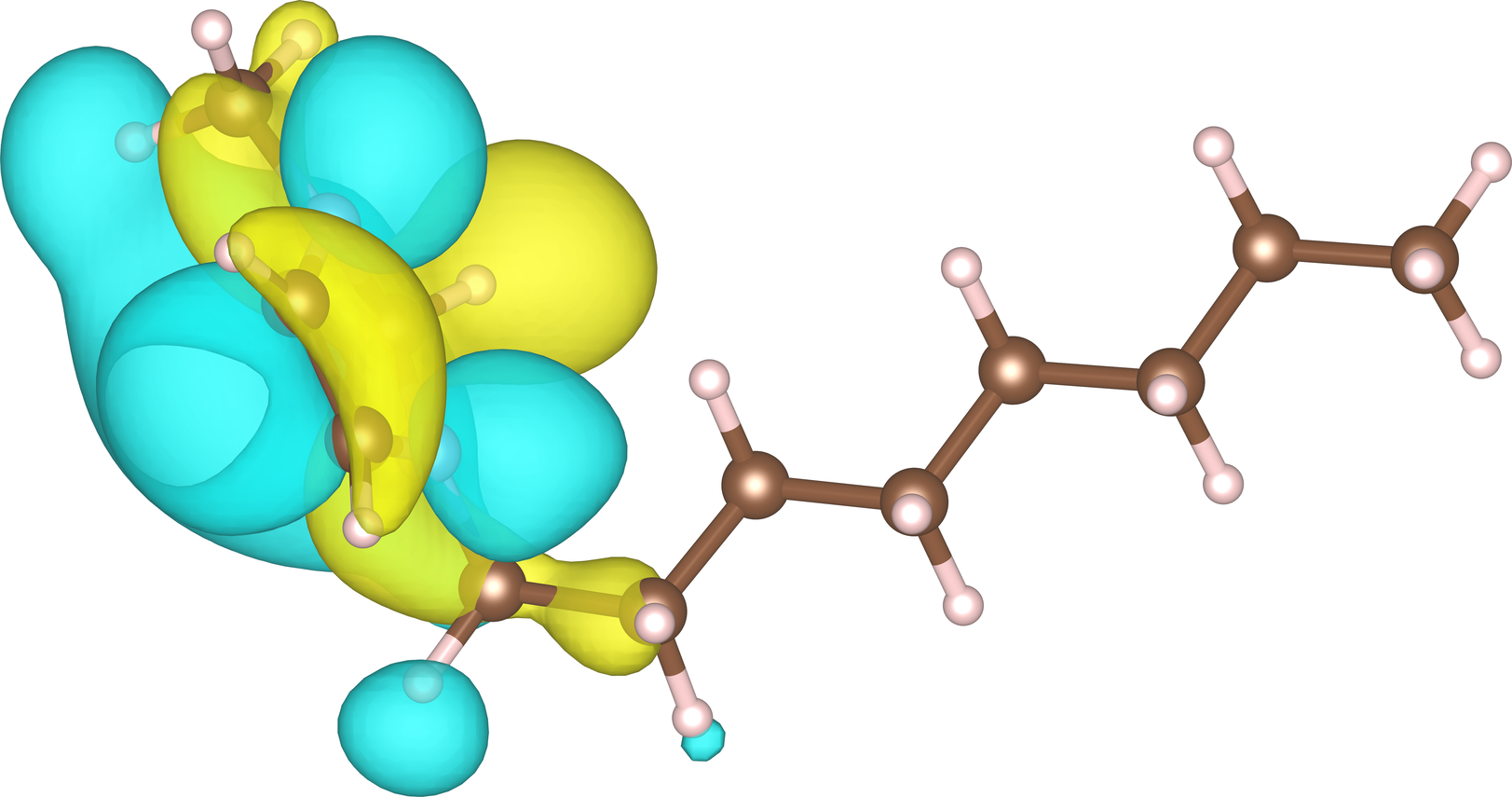}
    \hspace{-\textwidth}
    \vspace{-120pt}
    \resizebox{\textwidth}{!}{ \Large\bf
    \begin{tabular}{llll}
     \hspace{0.11\textwidth}(a) \hspace{0.11\textwidth} & \hspace{0.11\textwidth} (b) \hspace{0.11\textwidth} & \hspace{0.11\textwidth} (c)  \hspace{0.11\textwidth} & \hspace{0.11\textwidth} (d) \hspace{0.11\textwidth} \\
     &&& \\ &&& \\ &&& \\ &&& \\ &&& \\ &&& \\ &&& \\ &&& \\ &&& \\ &&& \\ &&& \\ &&& \\ &&& \\ &&& \\ &&& \\ &&& \\ &&& \\ &&& \\ &&& \\ &&& \\
    \end{tabular}}
    \caption{1-octyl-3-methylimidazolium:  The highest occupied molecular orbital (HOMO) \textbf{(a), (c)} and the lowest unoccupied molecular orbital (LUMO) \textbf{(b), (d)} wavefunctions for spin $\uparrow$ and $\downarrow$ respectively.}
    \label{fig:OMIM-HOMOLUMO}
\end{figure*}

\begin{figure*}
    \centering
     \includegraphics[angle=90,origin=c,width=0.32\linewidth]{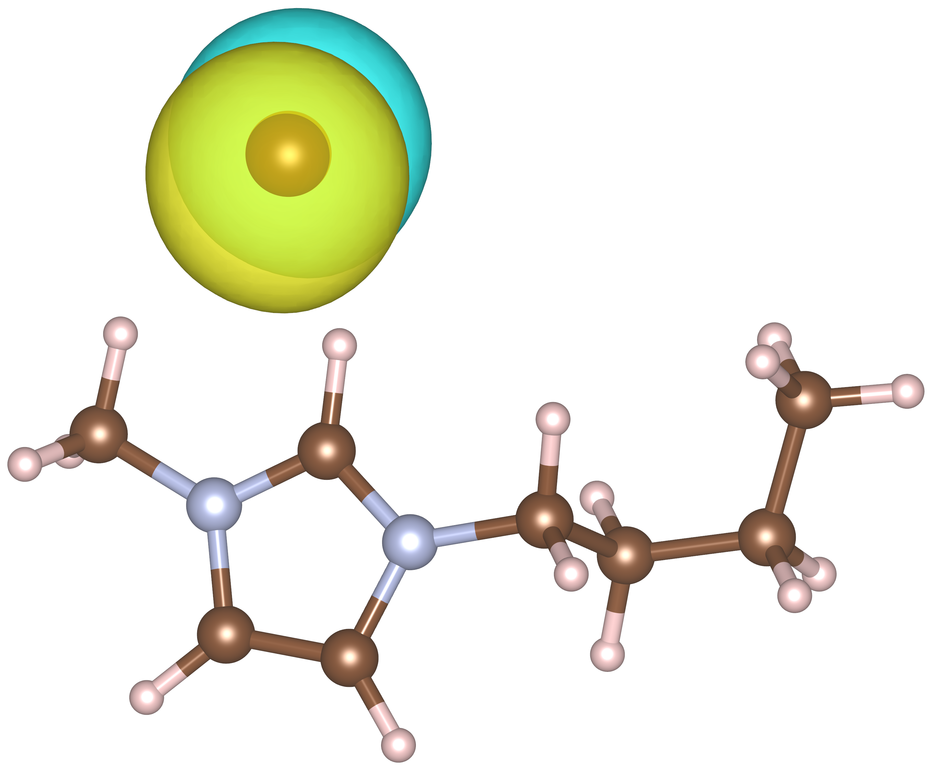}
     \includegraphics[angle=90,origin=c,width=0.32\linewidth]{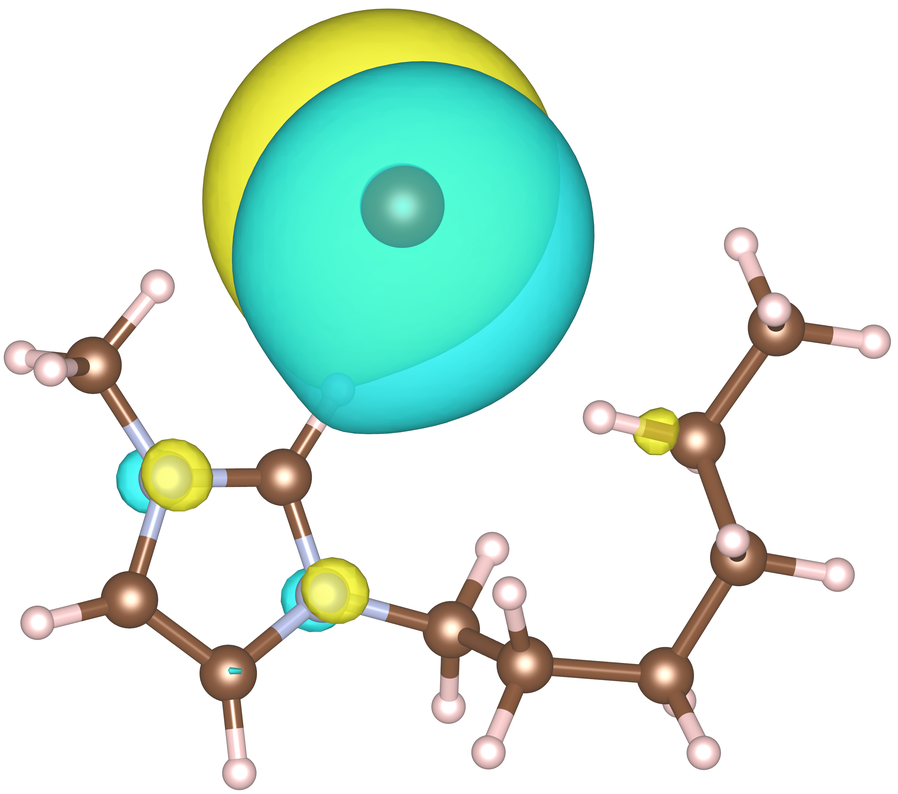}
     \includegraphics[angle=90,origin=c,width=0.32\linewidth]{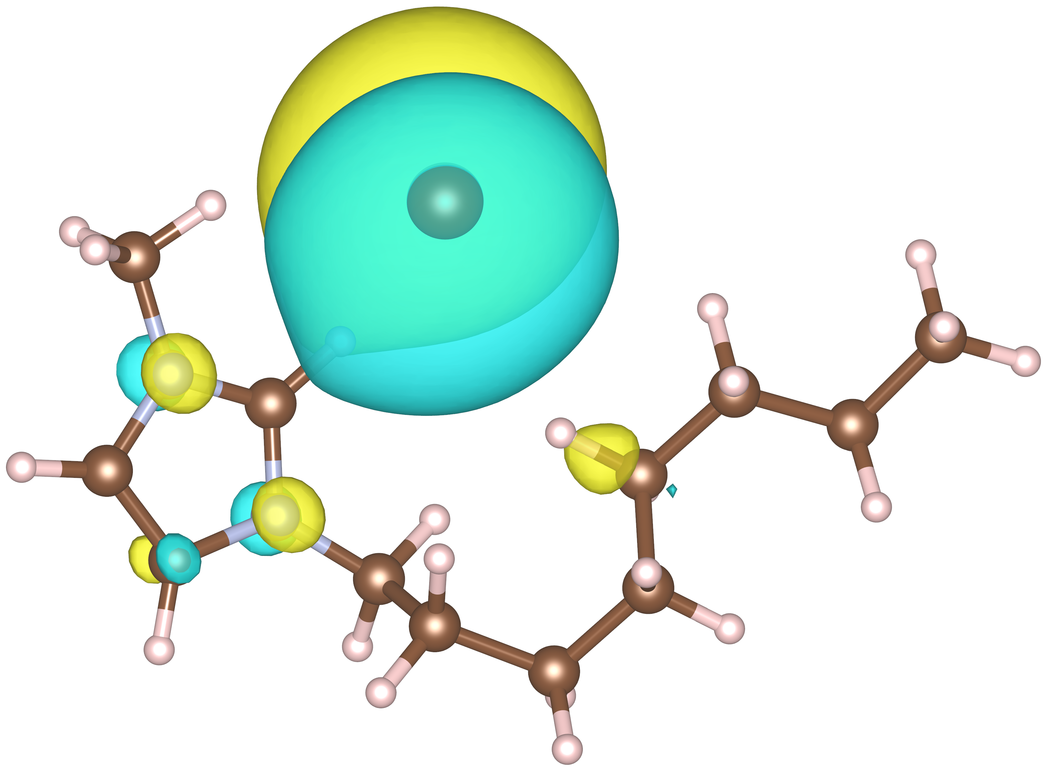}
    \hspace{-\textwidth}
    \vspace{-120pt}
    \resizebox{\textwidth}{!}{ \Large\bf
    \begin{tabular}{lll}
     \hspace{0.11\textwidth}(a) \hspace{0.11\textwidth} & \hspace{0.11\textwidth} (c) \hspace{0.11\textwidth} & \hspace{0.11\textwidth} (e)  \hspace{0.11\textwidth} \\
     && \\ && \\ && \\ && \\ && \\ && \\ && \\ && \\ && \\ && \\ && \\ && \\ && \\ && \\ && \\ && \\ && \\ && \\ && \\
    \end{tabular}}\vspace{-70pt}
     \includegraphics[angle=90,origin=c,width=0.32\linewidth]{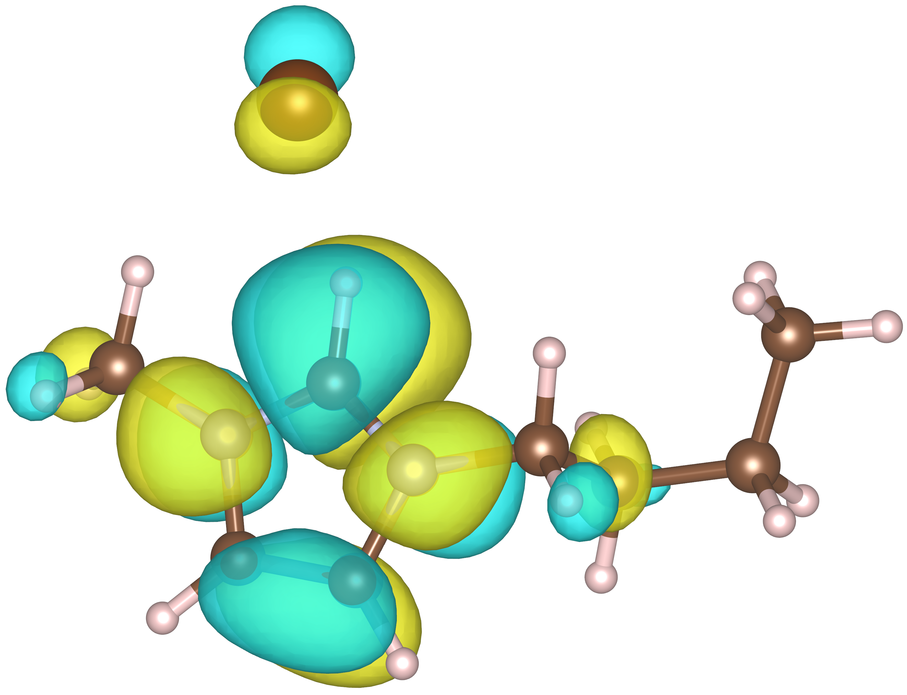}
     \includegraphics[angle=90,origin=c,width=0.32\linewidth]{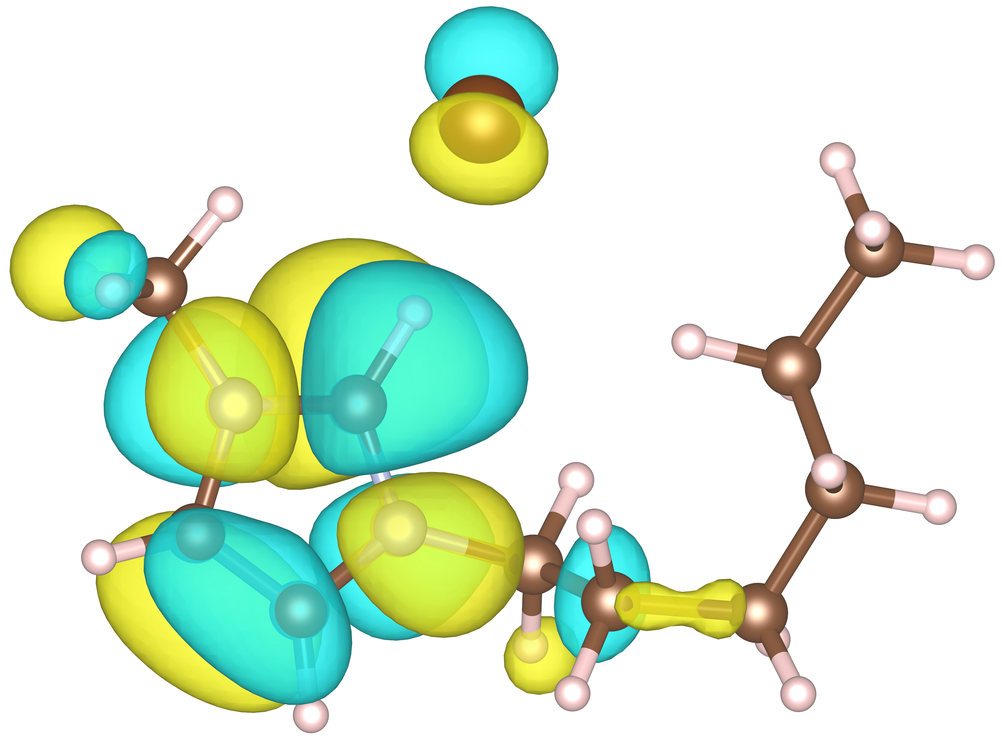}
     \includegraphics[angle=90,origin=c,width=0.32\linewidth]{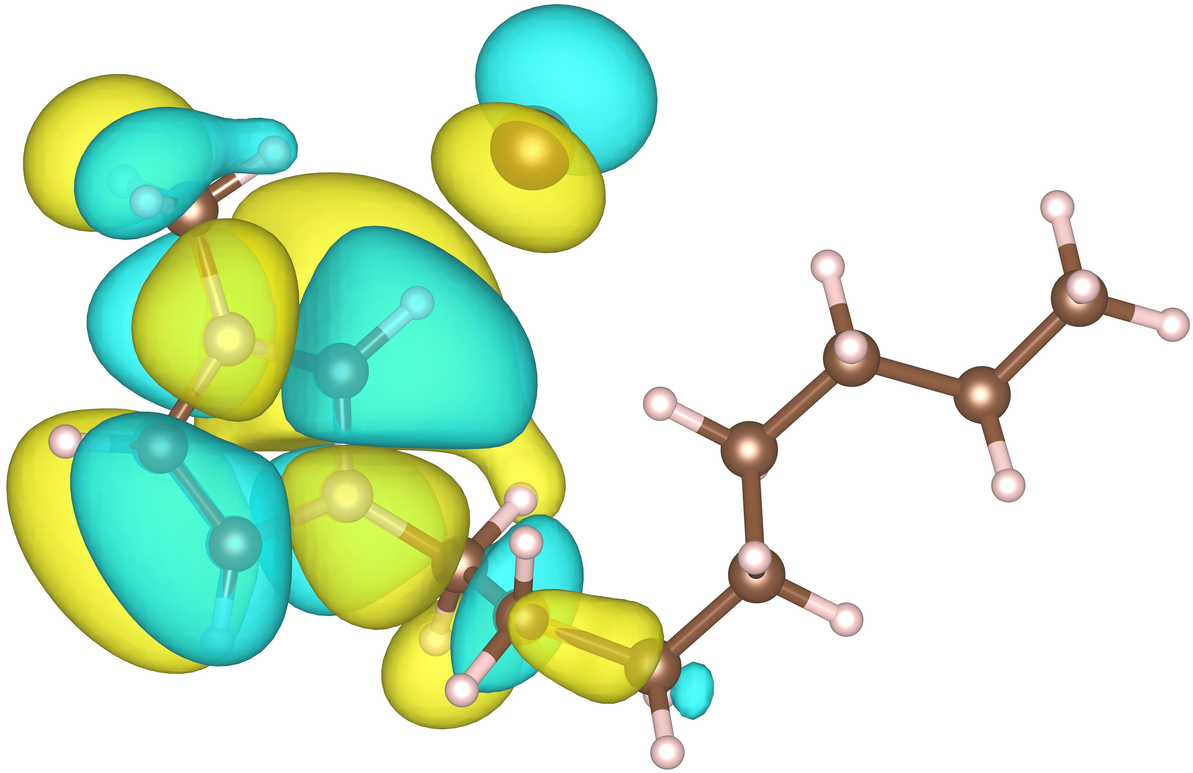}
    \hspace{-\textwidth}
    \vspace{-120pt}
    \resizebox{\textwidth}{!}{ \Large\bf
    \begin{tabular}{lll}
     \hspace{0.11\textwidth}(b) \hspace{0.11\textwidth} & \hspace{0.11\textwidth} (d) \hspace{0.11\textwidth} & \hspace{0.11\textwidth} (f)  \hspace{0.11\textwidth} \\
     && \\ && \\ && \\ && \\ && \\ && \\ && \\ && \\ && \\ && \\ && \\ && \\ && \\ && \\ && \\ && \\ && \\ && \\ && \\
    \end{tabular}}\vspace{-50pt}
    \caption{The highest occupied molecular orbital (HOMO) (\textbf{top}) and the lowest unoccupied molecular orbital (LUMO) (\textbf{bottom}) wavefunctions of 
    1-butyl-3-methylimidazolium bromide \textbf{(a), (b)};
    1-hexyl-3-methylimidazolium bromide \textbf{(c), (d)};
    1-octyl-3-methylimidazolium bromide \textbf{(e), (f)}.
    }
    \label{fig:MIB-HOMOLUMO}
\end{figure*}

\begin{figure*}
    \centering
     \includegraphics[angle=90,origin=c,width=0.32\linewidth]{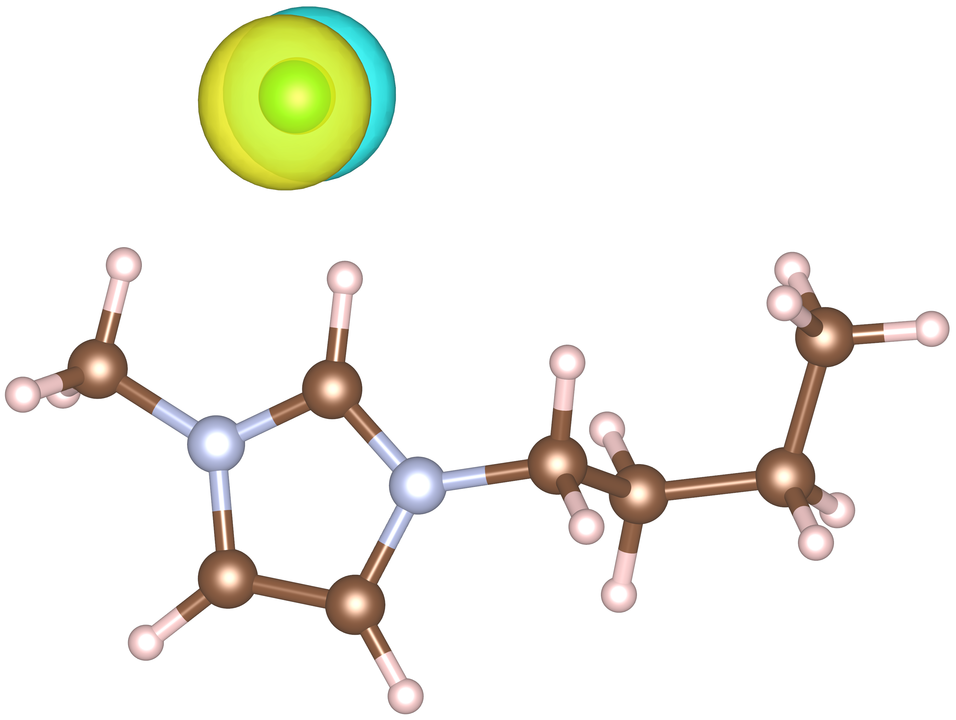}
     \includegraphics[angle=90,origin=c,width=0.32\linewidth]{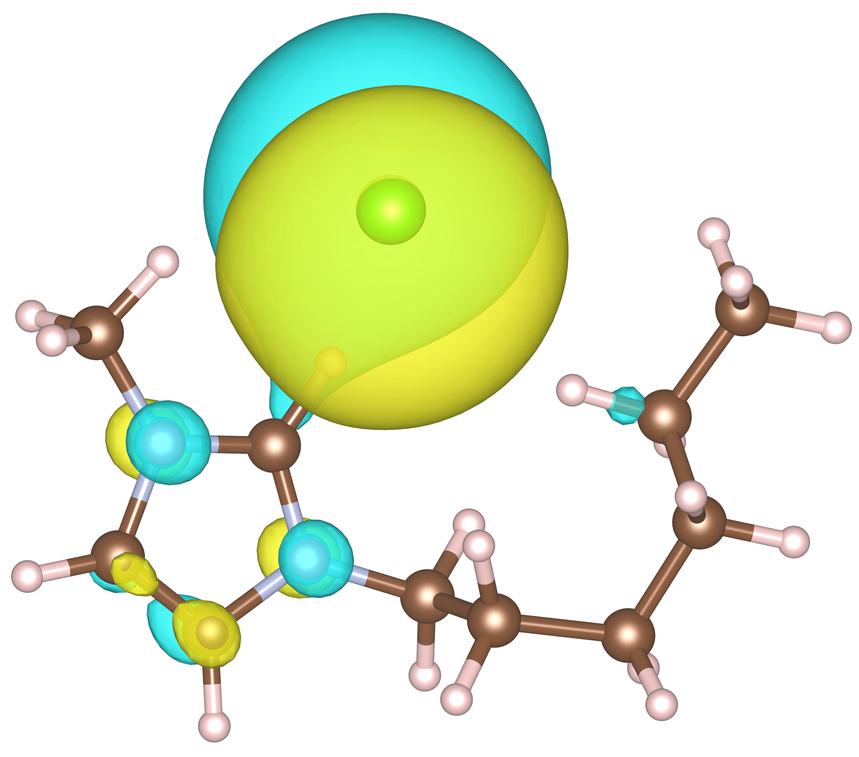}
     \includegraphics[angle=90,origin=c,width=0.32\linewidth]{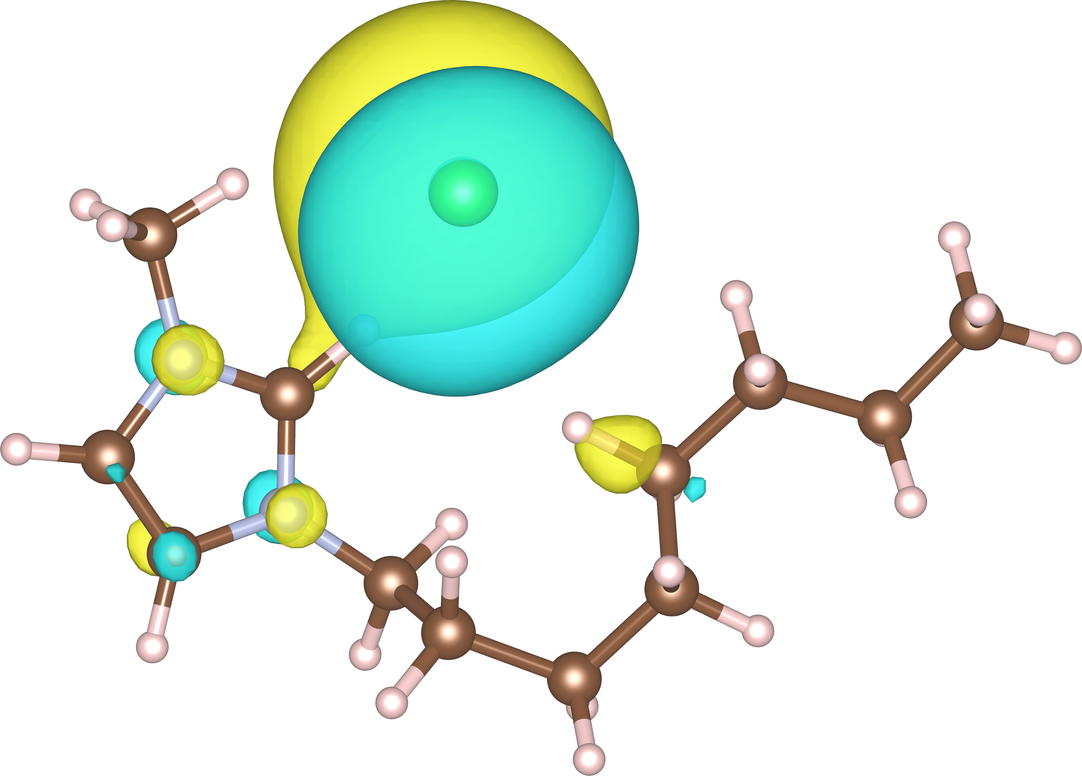}
    \hspace{-\textwidth}
    \vspace{-120pt}
    \resizebox{\textwidth}{!}{ \Large\bf
    \begin{tabular}{lll}
     \hspace{0.11\textwidth}(a) \hspace{0.11\textwidth} & \hspace{0.11\textwidth} (c) \hspace{0.11\textwidth} & \hspace{0.11\textwidth} (e)  \hspace{0.11\textwidth} \\
     && \\ && \\ && \\ && \\ && \\ && \\ && \\ && \\ && \\ && \\ && \\ && \\ && \\ && \\ && \\ && \\ && \\ && \\ && \\
    \end{tabular}}\vspace{-70pt}
     \includegraphics[angle=90,origin=c,width=0.32\linewidth]{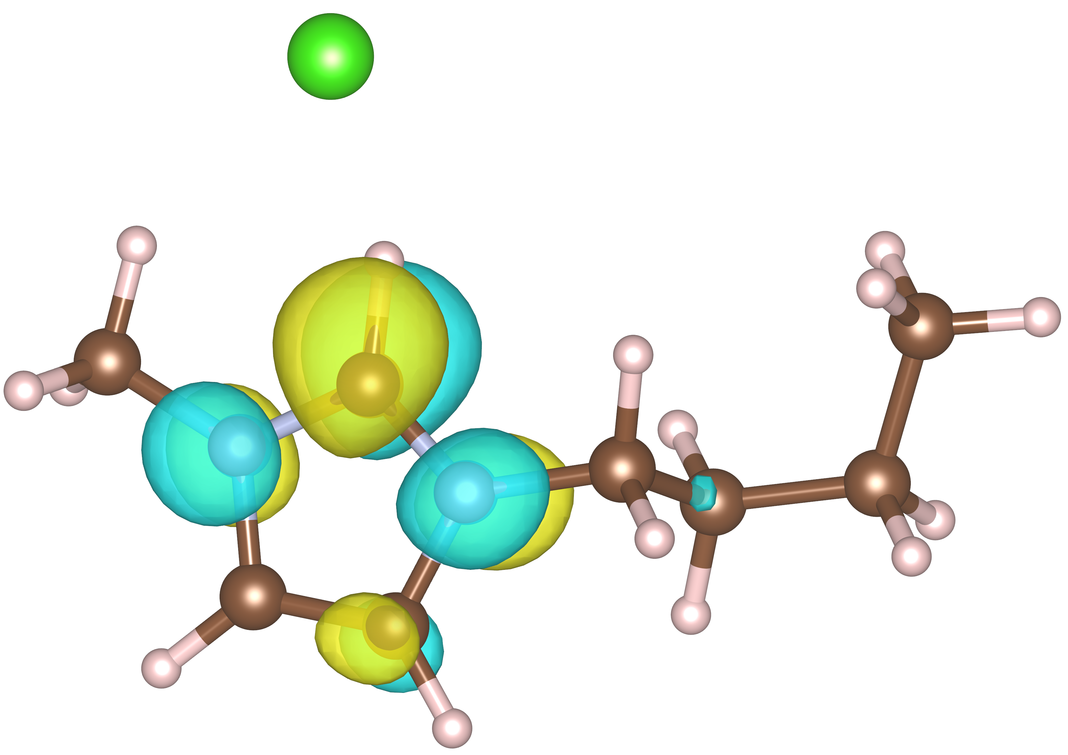}
     \includegraphics[angle=90,origin=c,width=0.32\linewidth]{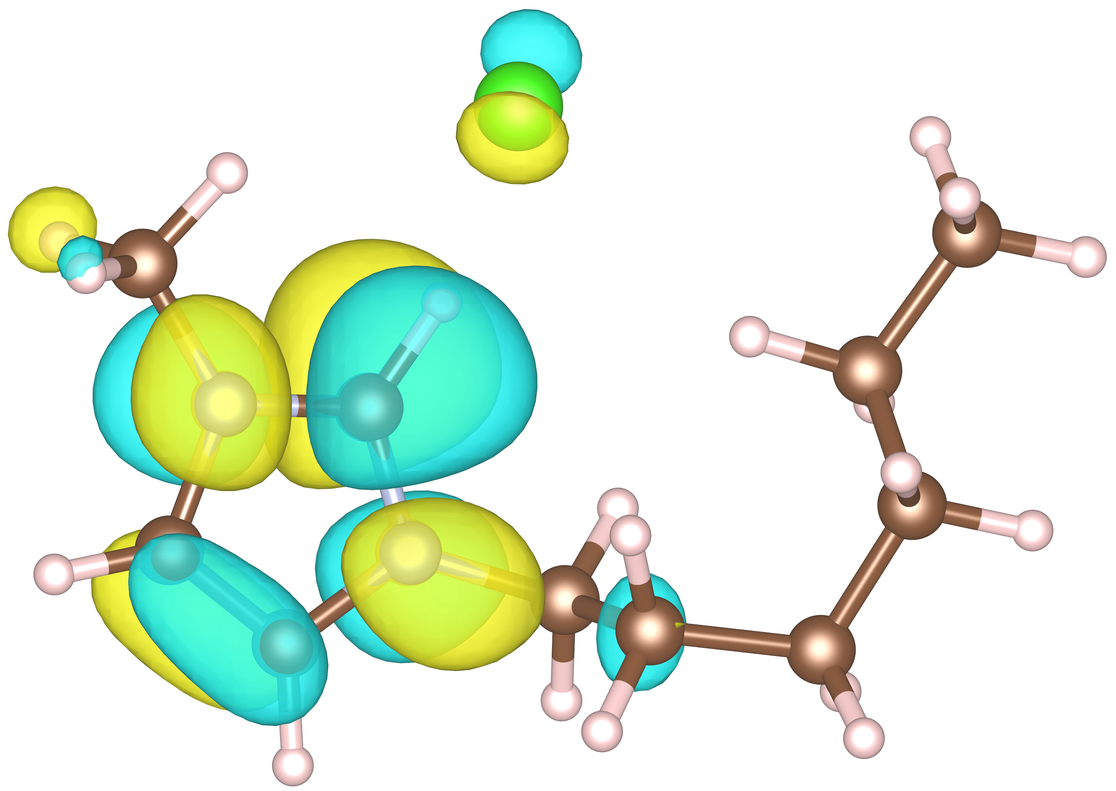}
     \includegraphics[angle=90,origin=c,width=0.32\linewidth]{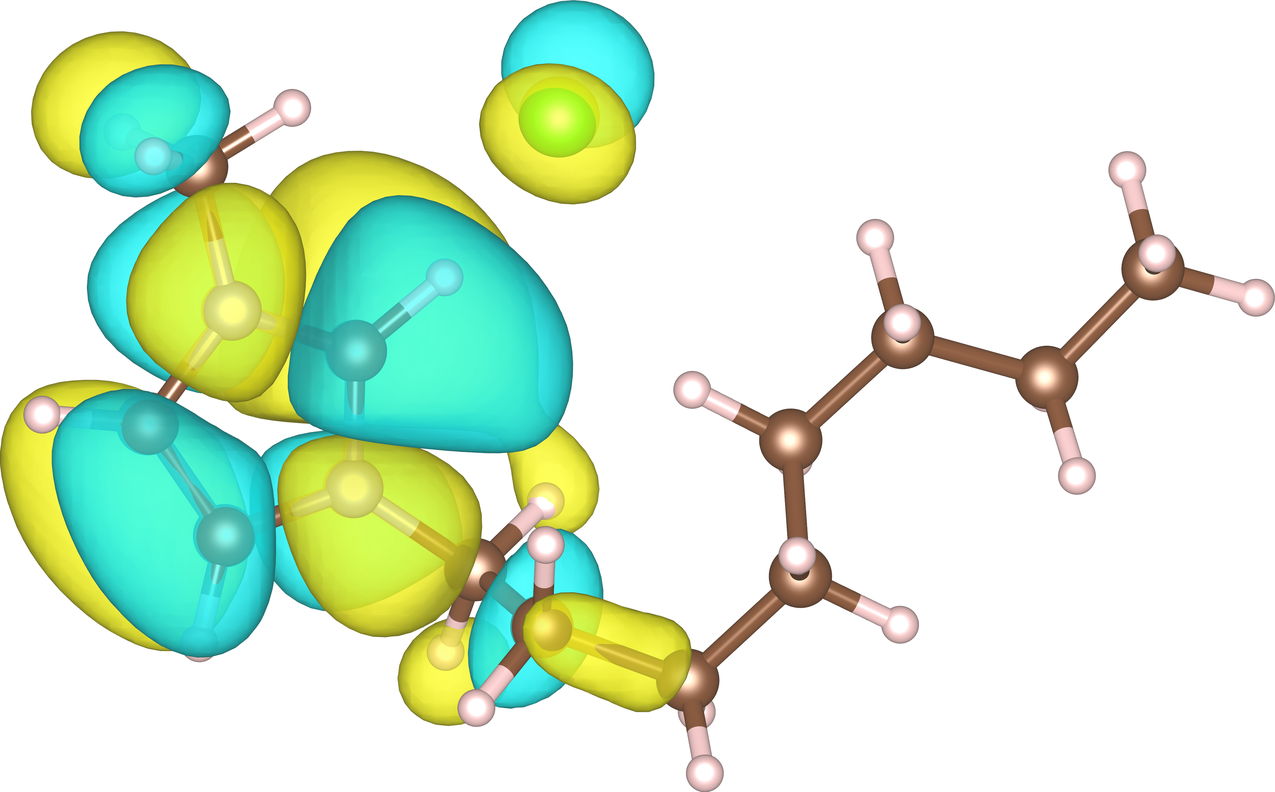}
    \hspace{-\textwidth}
    \vspace{-120pt}
    \resizebox{\textwidth}{!}{ \Large\bf
    \begin{tabular}{lll}
     \hspace{0.11\textwidth}(b) \hspace{0.11\textwidth} & \hspace{0.11\textwidth} (d) \hspace{0.11\textwidth} & \hspace{0.11\textwidth} (f)  \hspace{0.11\textwidth} \\
     && \\ && \\ && \\ && \\ && \\ && \\ && \\ && \\ && \\ && \\ && \\ && \\ && \\ && \\ && \\ && \\ && \\ && \\ && \\
    \end{tabular}}\vspace{-50pt}
    \caption{The highest occupied molecular orbital (HOMO) (\textbf{top}) and the lowest unoccupied molecular orbital (LUMO) (\textbf{bottom}) wavefunctions of 
    1-butyl-3-methylimidazolium  chloride \textbf{(a), (b)};
    1-hexyl-3-methylimidazolium  chloride \textbf{(c), (d)};
    1-octyl-3-methylimidazolium  chloride \textbf{(e), (f)}.
    }
    \label{fig:MIC-HOMOLUMO}
\end{figure*}

\begin{figure*}
    \centering
     \includegraphics[angle=90,origin=c,width=0.32\linewidth]{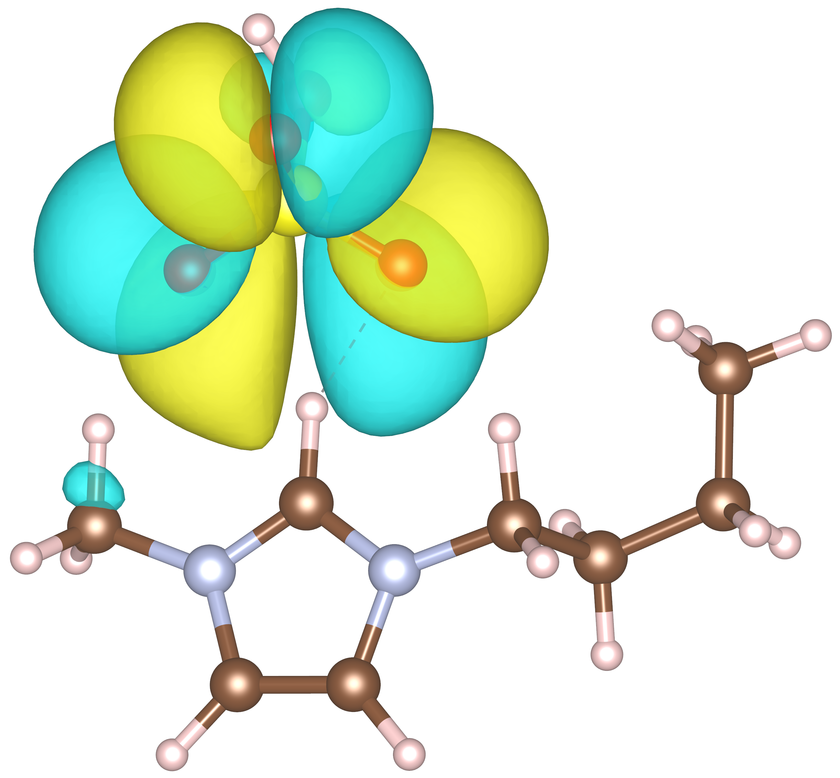}
     \includegraphics[angle=90,origin=c,width=0.32\linewidth]{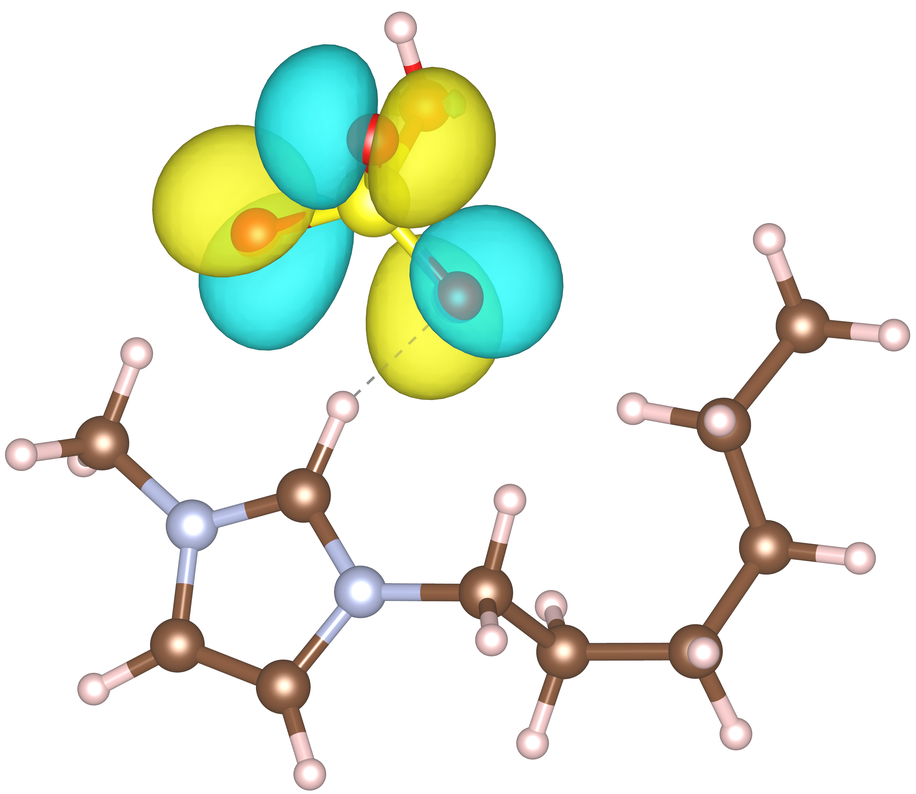}
     \includegraphics[angle=90,origin=c,width=0.32\linewidth]{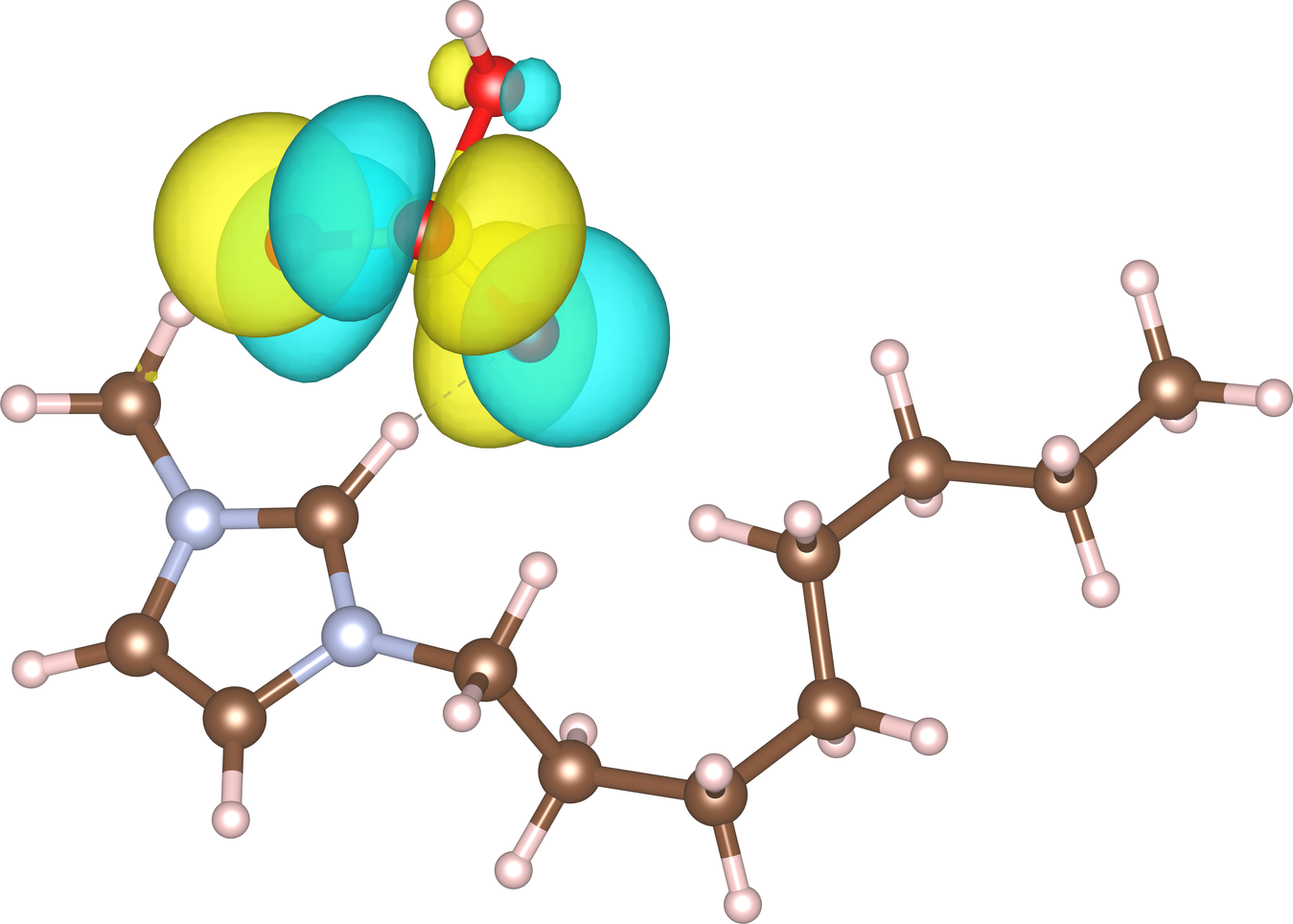}
    \hspace{-\textwidth}
    \vspace{-120pt}
    \resizebox{\textwidth}{!}{ \Large\bf
    \begin{tabular}{lll}
     \hspace{0.11\textwidth}(a) \hspace{0.11\textwidth} & \hspace{0.11\textwidth} (c) \hspace{0.11\textwidth} & \hspace{0.11\textwidth} (e)  \hspace{0.11\textwidth} \\
     && \\ && \\ && \\ && \\ && \\ && \\ && \\ && \\ && \\ && \\ && \\ && \\ && \\ && \\ && \\ && \\ && \\ && \\ && \\
    \end{tabular}}\vspace{-70pt}
     \includegraphics[angle=90,origin=c,width=0.32\linewidth]{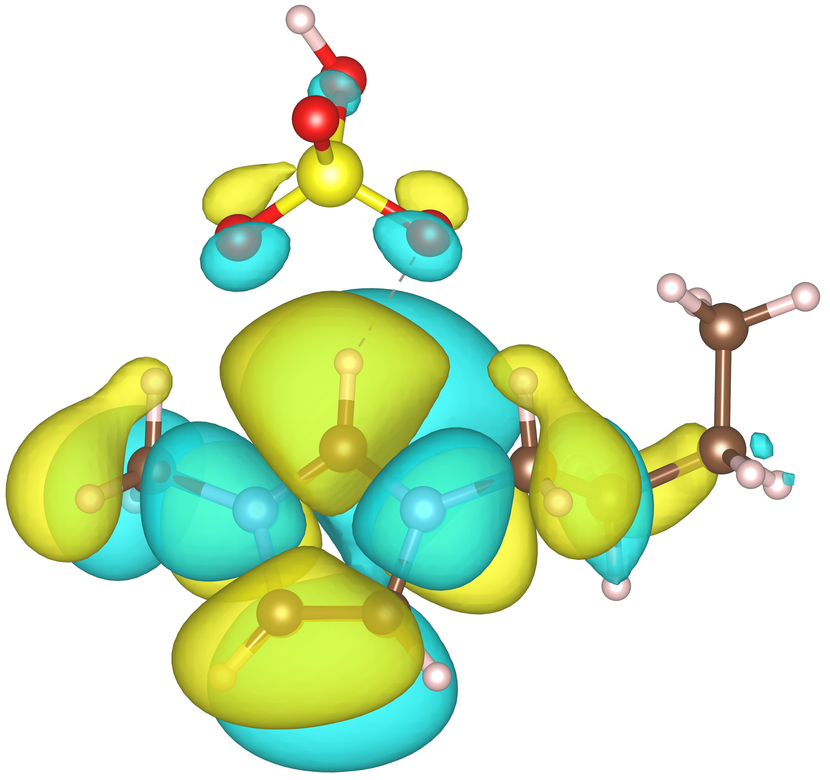}
     \includegraphics[angle=90,origin=c,width=0.32\linewidth]{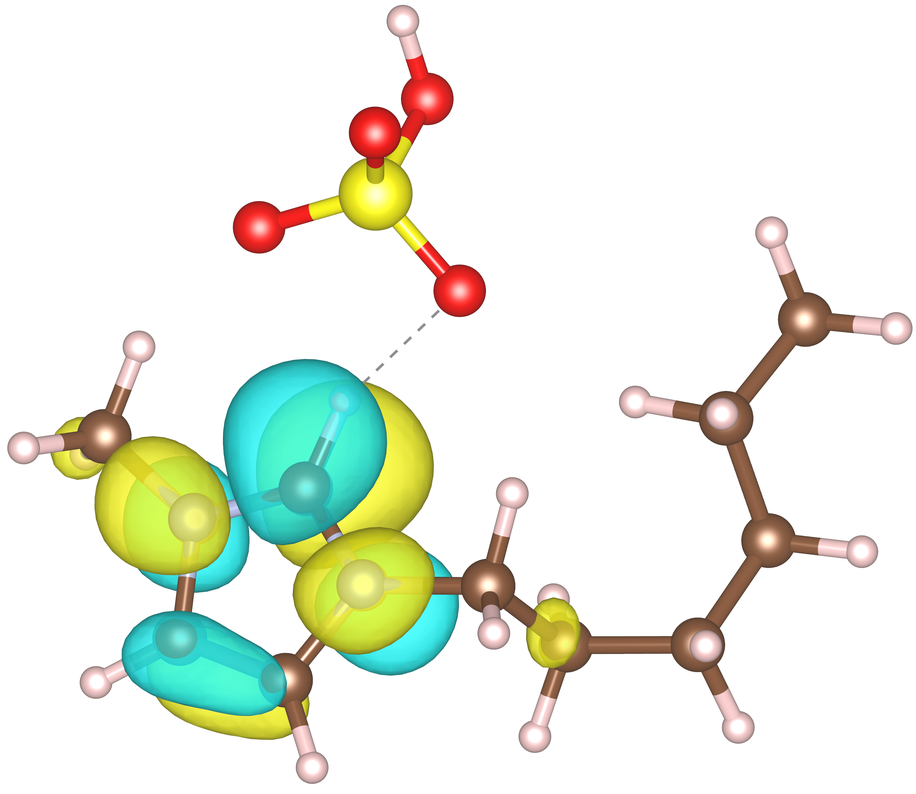}
     \includegraphics[angle=90,origin=c,width=0.32\linewidth]{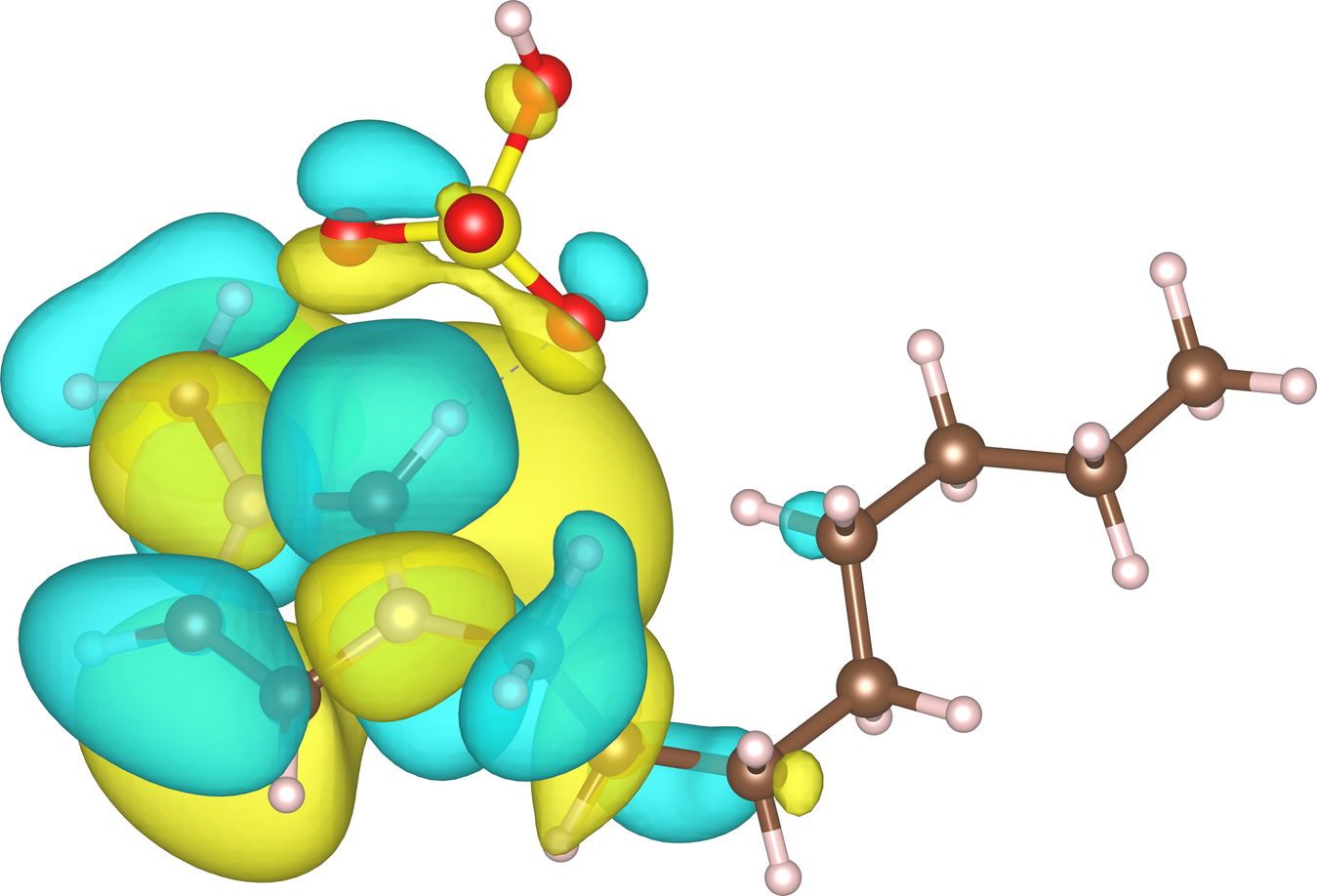}
    \hspace{-\textwidth}
    \vspace{-120pt}
    \resizebox{\textwidth}{!}{ \Large\bf
    \begin{tabular}{lll}
     \hspace{0.11\textwidth}(b) \hspace{0.11\textwidth} & \hspace{0.11\textwidth} (d) \hspace{0.11\textwidth} & \hspace{0.11\textwidth} (f)  \hspace{0.11\textwidth} \\
     && \\ && \\ && \\ && \\ && \\ && \\ && \\ && \\ && \\ && \\ && \\ && \\ && \\ && \\ && \\ && \\ && \\ && \\ && \\
    \end{tabular}}\vspace{-50pt}
    \caption{The highest occupied molecular orbital (HOMO) (\textbf{top}) and the lowest unoccupied molecular orbital (LUMO) (\textbf{bottom}) wavefunctions of 
    1-butyl-3-methylimidazolium  hydrogen sulphate \textbf{(a), (b)};
    1-hexyl-3-methylimidazolium  hydrogen sulphate \textbf{(c), (d)};
    1-octyl-3-methylimidazolium  hydrogen sulphate \textbf{(e), (f)}.
    }
    \label{fig:MIS-HOMOLUMO}
\end{figure*}

The visualization of HOMO and LUMO is presented in \cref{fig:BMIM-HOMOLUMO,fig:HMIM-HOMOLUMO,fig:OMIM-HOMOLUMO,fig:MIB-HOMOLUMO,fig:MIB-HOMOLUMO,fig:MIB-HOMOLUMO,fig:MIC-HOMOLUMO,fig:MIC-HOMOLUMO,fig:MIC-HOMOLUMO,fig:MIS-HOMOLUMO,fig:MIS-HOMOLUMO,fig:MIS-HOMOLUMO} where we present the visualizations of the real-part of the HOMO and LUMO, using the VASPkit \cite{wang2019vaspkit}.


\end{document}